\documentclass[floatfix,pre,showkeys,reprint,superscriptaddress,amsmath,amssymb,aps,longbibliography]{revtex4-1}
\usepackage[T1]{fontenc}
\usepackage{graphicx}
\usepackage{dcolumn}
\usepackage{bm}
\usepackage{xcolor}
\usepackage{mathbbol,mathrsfs}
\usepackage{mathtools}
\usepackage{overpic}
\usepackage{multirow}
\usepackage{enumitem}

\newcommand{\ve}[1]{\ensuremath{\mbox{\boldmath$#1$}}}
\newcommand{\ma}[1]{\ensuremath{\mathbb{#1}}}

\definecolor{matlabgreen}{RGB}{28,172,0}
\definecolor{matlabpurple}{RGB}{170,55,241}
\definecolor{matlabblue}{RGB}{0,0,255}
\definecolor{matlabbackground}{RGB}{255,255,255}
\definecolor{matlablinenumber}{RGB}{160,160,160}

\usepackage{listings}
\lstdefinestyle{matlaboutput}{
  basicstyle=\ttfamily\footnotesize\color{black},
  breaklines=true,
  postbreak=\mbox{\textcolor{gray}{$\hookrightarrow$}\space},
  frame=l,
  backgroundcolor=\color{white},
  rulecolor=\color{darkgray},
  keywordstyle=,
  commentstyle=,
  stringstyle=,
  identifierstyle=
}
\begin{document}
\title{Resistance tensors for aggregate particles with Stokesian dynamics}
\author{Joséphine Gissinger}
\affiliation{Aix Marseille Univ, CNRS, IUSTI, Marseille, France}
\author{Greg A. Voth}
\affiliation{Department of Physics, Wesleyan University, Middletown, CT 06459, USA}
\author{Bernhard Mehlig}
\affiliation{Department of Physics, Gothenburg University, 41296 Gothenburg, Sweden}
\author{Fabien Candelier}
\affiliation{Aix Marseille Univ, CNRS, IUSTI, Marseille, France}

\begin{abstract}
The response of particles to low-Reynolds-number flows can be compactly predicted with resistance or mobility tensors.  
However, the difficulty of obtaining accurate values for the elements of these tensors for specific geometries has held back work on particles with complex shapes. Here, we show how Stokesian dynamics can be adapted to efficiently compute the resistance and mobility tensors of rigid and flexible aggregates, including confinement by walls. We introduce \texttt{SHAPES}, an implementation of the method, and demonstrate its capabilities for complex geometries including curved fibres, chiral dipoles, interacting aggregates, and active swimmers. Aggregates are represented by assemblies of beads designed to reproduce the geometry and motion of rigid or flexible particles.  This coarse-grained description preserves the essential hydrodynamic interactions while substantially reducing computational cost. The method accurately reproduces known exact and approximate solutions, as well as experimental observations.  The ability to compute the complete resistance and mobility tensors provides new insight into how aggregate shape controls translation, rotation, and interactions with fluid-velocity gradients. Previous descriptions often relied on simplified models retaining only a few symmetry-allowed couplings. While useful, such descriptions are not always structurally stable under small perturbations of particle shape. Computing the full tensors makes it possible to draw robust conclusions and relate them to shape symmetry and hydrodynamic interactions. In particular, the method allows systematic analysis of non-Jeffery couplings to the strain rate  that arise for helicoidal aggregates. \texttt{SHAPES} therefore provides a versatile framework for studying rigid and flexible aggregates in microfluidic, biological, and environmental flows.
\end{abstract}

\maketitle
\newpage
\makeatletter
\onecolumngrid

\section{ \label{sec:intro} Introduction}
The motion of small non-spherical particles in a fluid arises in a broad range of 
scientific problems and technological applications \cite{voth2017anisotropic}. 
Among many examples are dynamics of small ice crystals in atmospheric clouds \cite{gustavsson2021effect} and the transport of small plastic particles in the atmosphere \cite{tatsii2024shape}.
Particle shape determines the angular dynamics of particles in uniform Stokes flow \cite{candelier_settling_2016,Roy_2019,ravichandran2023orientation,huseby2024helical,miara2024dynamics,maches2024settling,candelier2024torques,huseby2026bifurcations}, shear flow \cite{jeffery_motion_1922}, and in turbulence \cite{lundell2011fluid,zhao_passive_2019,parsa2012rotation,byron2015shape,voth2017anisotropic}.
For instance, the settling speed of aggregates of organic and inorganic matter is believed to be responsible for much of the material transport from the surface to the bottom of the ocean \cite{fowler1986role,iversen2023carbon}.
In biotechnology, the hydrodynamical resistance of DNA molecules confined to nanochannels determines the unfolding kinetics of the DNA \cite{alizadehheidari2015nanoconfined}.
In colloidal suspensions, particle shape and size determines both the physical and rheological properties of the suspension \cite{guazzelli2011suspension}.
\citet{witten2020review} emphasise the importance of chiral particle shapes, causing translation-rotation coupling in a quiescent fluid \cite{huseby2024helical}. 
Chirality affects particle rotation in straining flows \cite{fries_angular_2017,ishimoto2020helicoidal}, in turbulence \cite{kramel2016preferential}, and the dynamics of microswimmers with chiral flagella \cite{ishimoto2020helicoidal,zoettl2023asymmetric,jing2020chirality}. 

The dynamics in such systems is determined by the Navier-Stokes equations, with boundary conditions imposed by the moving particles. For very small particles, the problem 
simplifies for two reasons.
First, the velocity of the undisturbed flow can be linearised
around the centre of the particle:
\begin{equation} \label{eq:Vflow_intro}
    \bm{U}^{(\infty)}(\ve x) = \bm{U}_0 + \mathbb{E}^{(\infty)}
    \bm{\cdot} \bm{x} + \boldsymbol{\Omega}^{(\infty)} 
    \wedge \bm{x}
    \:,
\end{equation}
where $\ve U_0$, $\ve \Omega^{(\infty)}$ and $\ma E^{(\infty)}$ are constants.
Here, $\ve U_0$ is a spatially uniform flow velocity, $\ve \Omega^{(\infty)}$ is the undisturbed vorticity,
and $\ma E^{(\infty)}$ is the undisturbed strain rate, the symmetric part of the matrix of the undisturbed fluid-velocity gradients. 
Second, in the Stokes limit, fluid inertia 
is negligible so that the disturbance flow generated by a particle obeys the Stokes equations, satisfies a no-slip boundary condition on the particle surface, and vanishes far from the particle. The hydrodynamic force, torque, and stresslet exerted by the particle on the fluid are 
given by~\cite{happel_low_1983,kim_microhydrodynamics_1991}:
\begin{equation} 
    \begin{bmatrix}
        \bm{F} \\
        \bm{T} \\
        \mathbb{S}
    \end{bmatrix} 
    = \mu
    \begin{bmatrix}
        \mathbb{A} & \widetilde{\mathbb{{B}}} & \widetilde {\mathbb{G}} \\
        \mathbb{B} & \mathbb{C} & \widetilde{\mathbb{{H}}}\\
        \mathbb{G} & \mathbb{H} & {\mathbb{M}}
    \end{bmatrix}
    \bm{\cdot}
    \begin{bmatrix}
        \bm{v} - \bm{U}^{(\infty )}\\
        \bm{\omega} - \bm{\Omega}^{(\infty)} \\
        \ma e - \mathbb{E}^{(\infty)}
    \end{bmatrix}\,,
    \label{eq_systeme_full}
\end{equation}
where $\mu$ is the dynamic viscosity of the fluid, $\ve v$ is the velocity of the chosen particle centre, and $\ve\omega$ is the angular velocity about this centre.
Further, $\ma e$ is the strain rate induced within a deformable particle by the undisturbed flow. It characterises stretching and shearing of the particle relative to rigid-body motion, and vanishes for a rigid particle.
 The matrix in Eq.~(\ref{eq_systeme_full}) is the resistance matrix. For its blocks, we use the notation of \citet{kim_microhydrodynamics_1991}: $\ma A$, $\ma B$, \ldots .

To integrate the equations of motion of a small particle in a given fluid flow, one needs the hydrodynamic force and torque acting on the particle, $-\ve F$ and $-\ve T$. Eq.~(\ref{eq_systeme_full}) shows that this requires the resistance tensors, whose elements depend on particle shape and orientation relative to the flow. Dimensional analysis shows that the elements of the translation-resistance tensor $\ma A$ have dimensions of length, $[A_{ij}]=L$. The translation--rotation coupling tensor $\ma B$ has dimensions of length squared, $[B_{ij}]=L^2$, and so forth. The rank-3 tensor $\widetilde{\ma G}$ determines how force couples to strain, $\widetilde{\ma H}$ relates torque to strain, and the rank-4 tensor $\ma M$ connects the induced stresslet to the strain rate.
The elements of these tensors are known for simple geometries such as spheres, spheroids, and slender fibres. For a sphere, $\ma A$, $\ma C$, and $\ma M$ are proportional to the unit tensor $\ma I$, while all other coupling tensors are zero. \citet{jeffery_motion_1922} used Eq.~(\ref{eq_systeme_full}) to derive the equation of motion for small ellipsoidal particles in shear flow. Considerable effort has since gone into determining resistance tensors for more complex shapes. Early successes are summarised in Refs.~\cite{happel_low_1983,kim_microhydrodynamics_1991}, which provide approximate expressions for the resistance tensors of spheroids, symmetric bead aggregates, propellers, impellers, and slender fibres.
For single particles of more complex shapes, symmetry analysis \cite{Bretherton:1962,brenner1964stokes,happel_low_1983,fries_angular_2017,witten2020review,collins_lord_2021,miara2024dynamics,huseby2024helical,sundberg2025fluid} can be used to constrain which elements of $\ma A$, $\ma B$, \ldots may be non-zero. This substantially reduces the number of independent parameters that must be considered, but symmetry alone does not necessarily determine the particle dynamics. For example, planar curved wires with constant curvature \cite{candelier2024torques} and the bent disk studied in Ref.~\cite{miara2024dynamics} possess the same particle-shape symmetry, yet settle very differently in a quiescent fluid: the wires approach a stable aligned state, whereas the bent disk continues to rotate.

Direct numerical solution of the Stokes equations allows, in principle,  to determine how particle shape affects resistance matrices and thus particle dynamics in the Stokes limit. Finite-element methods require a volumetric discretisation of the fluid domain with an explicit mesh of the particle boundary \cite{ferziger2002computational}. Immersed-boundary methods \cite{uhlmann2005immersed,verzicco2025introduction} and lattice-Boltzmann simulations~\cite{chen1998lattice} use a fixed fluid mesh, while the particle is represented implicitly. Slender-body theory~\cite{tornberg2004simulating} avoids volumetric meshing by reducing the particle description to its centreline. Boundary-integral methods~\cite{blawzdziewicz2007boundary} require discretisation only of the particle surface. While all these approaches can yield accurate resistance tensors, they exhibit different strengths and limitations. Mesh-based methods struggle to represent particles shapes with corners or edges, and to resolve the unsteady disturbance flow created by an asymmetrical particle translating and rotating in a fluid. These methods become more expensive at smaller (but non-zero) particle Reynolds numbers, because large system sizes are needed to resolve the Oseen length $a/{\rm Re}_p$. Slender-body theory avoids meshing problems but is designed for slender particles 
with large aspect ratios $\kappa$ (length divided by diameter). The theory is derived from first principles and is exact in the limit of $\kappa\to \infty$.  However, it tends to lose accuracy when $\kappa$ is not large enough. 

For two or more particles, or in the presence of a wall, the problem of computing hydrodynamic forces and torques becomes still harder, because one must consider how hydrodynamic interactions with a nearby particle or wall affect force, torque, and stresslet. In principle, it is straightforward to deal with this in the Stokes limit, because a linear relation like Eq.~(\ref{eq_systeme_full}) holds between all forces, torques and stresslets, and all particle velocities, angular velocities and strain rates, expressing how the dynamics of the ensemble of particles is   determined by their shape and relative orientations. But many-particle conformations tend to be asymmetric, requiring many more coefficients to be calculated.
Even for only two spheres, the exact resistance tensor is very complicated~\cite{jeffrey_forces_1984,jeffrey_calculation_1984,jeffrey1992calculation}, and there are few highly accurate numerical results for more spheres in strongly symmetric configurations~\cite{wilson_stokes_2013}.

Eq.~(\ref{eq_systeme_full}) determines force and torque on a particle, given its velocity and angular velocity. 
This \lq resistance problem\rq{} occurs, for example, for magnetic micropropellers \cite{vach_fast_2015}, where an angular velocity is prescribed, and the force is sought. 
In some situations it is useful to ask instead: imposing an external force $\ve F_{\rm ext}$ and torque $\ve T_{\rm ext}$ on the aggregate, what is its instantaneous velocity and angular velocity? This is the `mobility problem', and it is the natural question for aggregates settling under gravity in a quiescent fluid \cite{candelier_settling_2016,miara2024dynamics,huseby2024helical,einarsson2017spherical}.  The resistance and mobility problems are related, one obtains the mobility formulation  by inverting the resistance tensor in Eq.~(\ref{eq_systeme_full}):
\begin{equation}
\label{eq:mobility} 
    \begin{bmatrix}
        \bm{v} \\
        \bm{\omega}\\
    \end{bmatrix} 
    = 
     \begin{bmatrix}
       \ve U^{(\infty)} \\
        \ve \Omega^{(\infty)}
    \end{bmatrix} 
    +
    \frac{1}{\mu}
    \begin{bmatrix}
        \mathbb{a} & \widetilde{\mathbb{{b}}} \\
        \mathbb{b} & \mathbb{c} 
    \end{bmatrix}
    \bm{\cdot}
    \begin{bmatrix}
        \ve F_{\rm ext} + \widetilde{\ma{G}} : \ma E^\infty  \\
        \ve T_{\rm ext} + \widetilde{\ma{H}} : \ma E^\infty 
    \end{bmatrix}\:, \, \quad \mbox{with} \quad \begin{bmatrix}
        \mathbb{a} & \widetilde{\mathbb{{b}}} \\
        \mathbb{b} & \mathbb{c} 
    \end{bmatrix} = \begin{bmatrix}
        \mathbb{A} & \widetilde{\mathbb{{B}}} \\
        \mathbb{B} & \mathbb{C} 
    \end{bmatrix}^{-1}\,.
\end{equation}

Here, we describe an alternative way of computing resistance and mobility tensors of rigid and flexible particles with complex shapes. We consider aggregates of rigid spherical beads and use Stokesian dynamics \cite{durlofsky_dynamic_1987,brady_stokesian_1988} to represent hydrodynamic interactions between the beads that make up the aggregate particle. Imposing the kinematic constraints on the positions and relative velocities of the beads within the rigid aggregate allows extraction of the resistance tensors in Eq.~(\ref{eq_systeme_full}). 
We introduce and describe the program package \texttt{SHAPES} (Stokes Hydrodynamics for Aggregate Particles via Ensemble of Spheres)  which implements this method.
We used precursors to the \texttt{SHAPES} package 
in our own research \cite{candelier_settling_2016,collins_lord_2021,huseby2024helical,candelier2024torques}, and 
other authors have used similar approaches to calculate forces and torques on colloidal aggregates in linear Stokes flow \cite{schlauch2013comparison}. 
\texttt{SHAPES} provides a versatile and accurate tool to determine the full set of resistance tensors of aggregates of arbitrary shapes at low computational cost, provided that the beads do not touch each other.  
It allows investigation of particle-particle interactions and particle-wall interactions in Stokes flow for aggregate particles of arbitrary shapes. 
We envisage that the tool will find numerous applications in biophysics and active-matter research, where the dynamics of ensembles of particles with complex shapes is of key interest. 
The program package is available on GitHub at \url{https://github.com/JosephineGiss/SHAPES/releases/tag/v1.0.0}.

Other numerical packages implementing Stokesian Dynamics have been developed in the past.
The \texttt{HYDROLIB} library was  developed by \citet{hinsen_hydrolib_1995}. The code 
is based on the Cartesian multipole-expansion technique \cite{cichocki_friction_1994} which extends the formulation of \citet{durlofsky_dynamic_1987} through a higher-order multipole expansion. 
To the best of our knowledge, this library is also the only implementation that allows the computation of drag coefficients for rigid clusters of spheres \cite{cichocki_stokes_1995}. The code includes pairwise lubrication corrections,  but it does not account for shear flows, or for confining walls.
The \texttt{RYUON} package \cite{ryon2006} 
simulates the dynamics of freely moving spherical particles in linear Stokes flow under periodic boundary conditions. In the code, near-field lubrication corrections between pairs of spheres can be included using analytical expressions derived by Jeffrey and Onishi~\cite{jeffrey_calculation_1984}, and improved numerical schemes proposed by Ichiki~\cite{ichiki_improvement_2002} enhance both the accuracy and efficiency of the multipole expansion by extending it to higher orders and employing iterative solvers.
More recently, a Python-based implementation of Stokesian Dynamics for suspension has been released by \citet{townsend_stokesian_2024}.
It also includes pairwise lubrication corrections and extends the method to viscoelastic suspensions by introducing bead-spring dumbbells that model polymer-induced fluid elasticity. Very recently, Cheng \textit{et al.}~\cite{cheng2026} presented a computational package that computes rank-2 mobility tensors for particles represented by a triangulated surface in an STL file using the method of Stokeslets distributed across the surface. 

\begin{figure*}
\begin{overpic}[height=8cm]{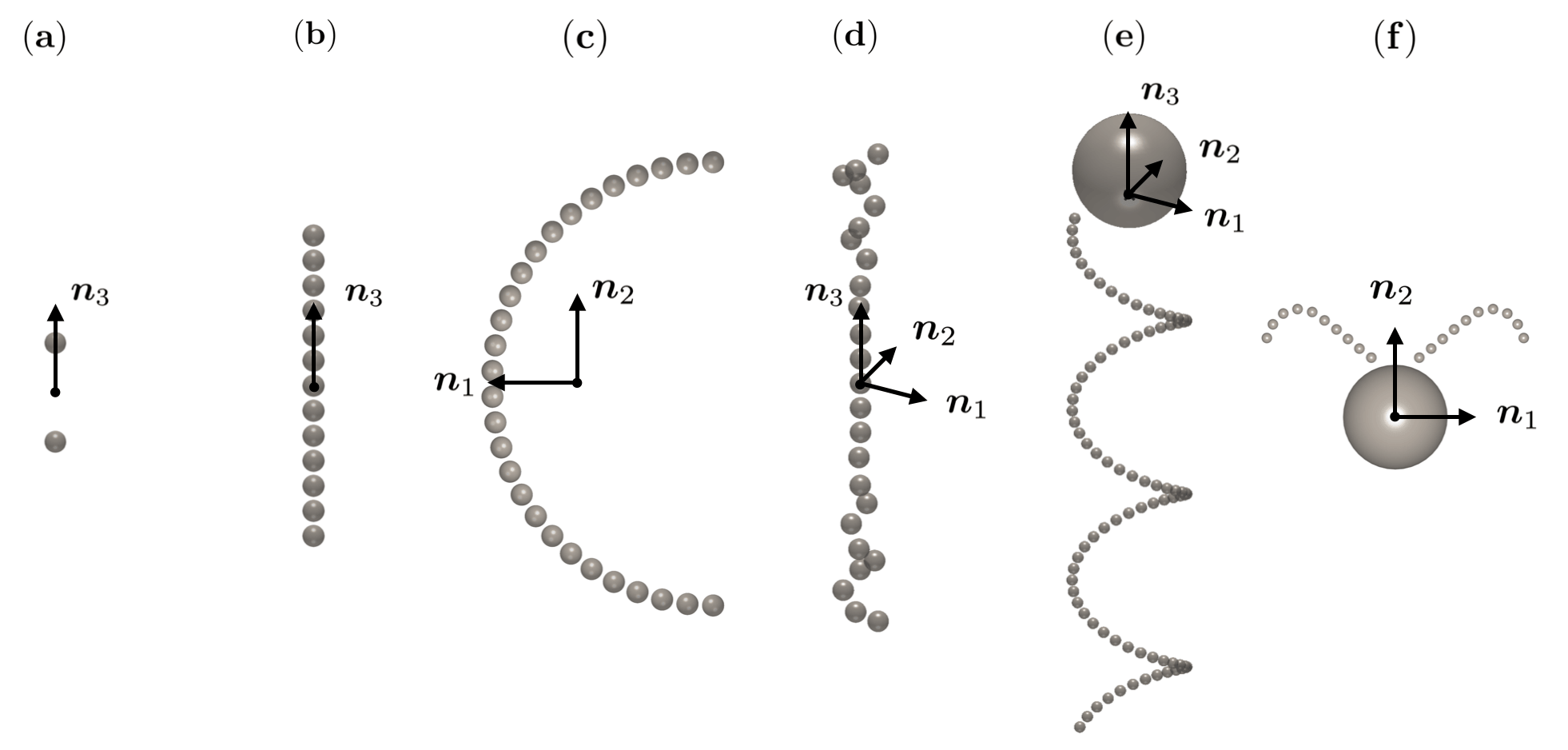}
\end{overpic}
\caption{\label{fig:particle_summary}
Examples of aggregates considered in this article. 
The origin of the frame attached to the aggregate, $\ve n_1, \ve n_2,\ve n_3$, is located at the centre of mass of the aggregate.
({\bf a}) Dumbbell. 
({\bf b}) Straight fibre. 
({\bf c}) Curved planar fibre with constant radius of curvature. 
({\bf d}) Chiral dipole~\cite{kramel2016preferential}. The aggregate consists of two helical parts, one right-handed, one left handed. 
({\bf e}) Bead with helical flagella \cite{ishimoto2020helicoidal}.  
({\bf f}) Puller \cite{geyer_cell-body_2013}.
All aggregates were visualised using \texttt{SHAPES}.
}
\end{figure*}

In the present article, we describe our implementation \texttt{SHAPES}, 
outline past and future applications, and discuss strengths and weaknesses of the approach. 
The remainder of the article is organised as follows. 
In Section~\ref{sec:background}, we review the theoretical background, the Stokesian dynamics method \cite{durlofsky_dynamic_1987}. 
Section~\ref{sec:aggregates} describes how this method  is applied to compute resistance tensors of aggregate particles. 
Section~\ref{sec:program} introduces the \texttt{SHAPES} program package and its implementation; a brief tutorial is found in Appendix \ref{appendix:tutorial}. 
In Section \ref{sec:examples}, we discuss some applications of the method to particles of different shapes (see Fig. \ref{fig:particle_summary}). Outlook and conclusions are summarised in Section~\ref{sec:conclusions}.

\section{Background}
\label{sec:background}
In this Section, we briefly describe 
the Stokesian dynamics method \cite{durlofsky_dynamic_1987} which computes hydrodynamic interactions between freely moving spherical beads in Stokes flow.
It  forms the basis for the \texttt{SHAPES} program package described in the present article. We reproduce here the essential framework required for our developments, and 
derive the expressions for the polydisperse case.

The disturbance flow field produced by a set of spherical beads at any point in space in Stokes flow is given by:
\begin{equation}
   \bm{w}(\bm{x}) 
    = -\frac{1}{8\pi\mu} \sum_{\alpha \, = \, 1}^{N} \int_{\mathcal{S}_\alpha} \mathbb{J}\left(\bm{x} - \bm{y}_{\alpha}\right) \bm{\cdot} \bm{f}(\bm{y}_{\alpha}) \mathrm{d} S \: ,
    \label{eq:disturbance_flow_integral}
\end{equation}
\noindent where $\mathbb{J}$ is the Oseen tensor, the Green function of Stokes' equation:
\begin{equation}
    \mathbb{J} \left(\bm{r}\right) = \frac{1}{r}\mathbb{I} + \frac{1}{r^3}\bm{r}\otimes\bm{r} \: ,
\end{equation}
\noindent and where $\bm{y}_\alpha$ is a point on the surface of bead $\alpha$.
Because the surface integral (\ref{eq:disturbance_flow_integral}) is computationally expensive when beads are in close proximity, the authors performed a Taylor expansion of the Oseen tensor $\mathbb{J}$ about the centre of the spherical beads, allowing the disturbance flow $\bm{w}$ to be expressed as a linear combination of the forces $\ve F_\alpha$, torques $\ve T_\alpha$, and stresslets $\ma S_\alpha$ exerted by beads $\alpha = 1,\ldots,N$ upon the fluid:
\begin{eqnarray} \label{eq:flow1}
    &&\bm{w} (\bm{x}) =
    \frac{1}{8\pi\mu} \sum_{\alpha \, = \, 1}^{N} \left[ \left(1 + \frac{a_\alpha^2}{6}\Delta \right)\mathbb{J} \left( \bm{x} - \bm{x}_\alpha \right) \bm{\cdot} \ve F_\alpha  + \: \mathbb{R}( \bm{x} - \bm{x}_\alpha ) \bm{\cdot} \ve T_\alpha  - \left(1 + \frac{a_\alpha^2}{10}\Delta \right) \hspace{-1pt} \mathbb{K} \left( \bm{x} - \bm{x}_\alpha \right)\bm{:}\mathbb{S}_{\alpha}  \right] \:.
\end{eqnarray}
Note that the minus sign preceding the tensor $\mathbb{K}( \bm{x}  )$ is missing in \citet{durlofsky_dynamic_1987}, a typographical error there.
In Eq.~(\ref{eq:flow1}), $\bm{x}_\alpha$ denotes the centre of 
the spherical bead $\alpha$. Its radius is denoted
by $a_\alpha$ and  $\Delta$ is the Laplace operator. The rotlet $\mathbb{R}( \bm{x})$ and the rank-3 tensor $\mathbb{K}( \bm{x}  )$ are derivatives of the Green function. Their elements are given by:
\begin{align}
\label{eq:2}
R_{ij} &= \frac{1}{4} \varepsilon_{\ell kj} \left( \frac{\partial J_{i\ell}}{\partial x_k} - \frac{\partial J_{ik}}{\partial x_\ell} \right) \quad\mbox{and}\quad   K_{ijk} = \frac{1}{2}\left( \frac{\partial J_{ij}}{\partial x_k} + \frac{\partial J_{ik}}{\partial x_j} \right) \,.
\end{align}
Here, $\varepsilon_{ijk}$ are the elements of the completely asymmetric rank-3 tensor, the Levi-Civita tensor. Forces, torques and stresslets exerted upon the fluid by a given bead depend on the disturbance flow generated by the other beads, 
unknown at this point. The formulation (\ref{eq:flow1}) is therefore implicit.
To move forward,
\citet{durlofsky_dynamic_1987} used Faxén’s laws \cite{faxen_bewegung_1923} which provide  hydrodynamic relations for a bead $\alpha$ moving in a disturbance flow:
\begin{subequations} \label{eq:Faxen}
    \begin{align}
        \bm{v}_{\alpha} - \bm{U}^{(\infty)} _\alpha &= \frac{\bm{F}_\alpha}{6 \pi \mu a_\alpha} + \left(1 + \frac{a_\alpha^2}{6}\Delta \right) \bm{w} (\bm{x}_\alpha) \:, \label{eq:Faxén1}
        \\
        \boldsymbol{\omega}_{\alpha} - \boldsymbol{\Omega}^{(\infty)}
         &= \frac{\bm{T}_\alpha}{8 \pi \mu a_\alpha^3} + \frac{1}{2}[\boldsymbol{\nabla} \wedge \bm{w}(\bm{x}_\alpha) ] \: ,
        \\
        - \mathbb{E}^{(\infty)} &= \frac{\mathbb{S}_{\alpha}}{\frac{20}{3} \pi \mu a_\alpha^3} + \Big(1 + \frac{a_\alpha^2}{10}\Delta \Big)\left[\frac{1}{2}\nabla \bm{w}(\bm{x}_\alpha) + \frac{1}{2} \nabla \bm{w}^{\sf T}(\bm{x}_\alpha) \right] \: .
    \end{align}
\end{subequations}
The disturbance flow field $\bm{w}$ experienced by a given bead $\alpha$ is induced by all other beads:
\begin{eqnarray} \label{eq:flow1_bis}
    &&\bm{w} (\bm{x}_{\alpha}) =
    \frac{1}{8\pi\mu} \sum_{\substack{\beta \, = \, 1 \\ \beta \, \neq \, \alpha}}^{N} \left[ \left(1 + \frac{a_\beta^2}{6}\Delta \right)\mathbb{J} \left( \bm{x}_\alpha - \bm{x}_\beta \right) \bm{\cdot} \ve F_\beta  + \: \mathbb{R}( \bm{x}_\alpha - \bm{x}_\beta ) \bm{\cdot} \ve T_\beta  - \left(1 + \frac{a_\beta^2}{10}\Delta \right) \hspace{-1pt} \mathbb{K} \left( \bm{x}_\alpha - \bm{x}_\beta \right)\bm{:}\mathbb{S}_{\beta}  \right] \:.
\end{eqnarray}
The undisturbed flow $\ve U^{(\infty)}(\ve x_\alpha) $ at the position
$\ve x_\alpha$ of bead $\alpha$ 
is denoted by $\ve U^{(\infty)}_\alpha $. 
From these relations, one can extract a linear system relating the relative velocities of the beads to their forces, torques and stresslets through the grand mobility matrix $\mathscr{M}$:
\begin{equation} \label{eq:bigMobMat}
    \begin{bmatrix}
         \bm{v}_1 - \bm{U}^{(\infty)}_1 \\[-4pt]
         \vdots \\[1pt]
         \bm{v}_N - \bm{U}^{(\infty)}_N \\[1pt]
         \boldsymbol{\omega}_1 - \boldsymbol{\Omega}^{(\infty)} \\[-4pt]
         \vdots \\[1pt]
         \boldsymbol{\omega}_N - \boldsymbol{\Omega}^{(\infty)} \\[1pt]
         - \mathbb{E}^{(\infty)}\\[-4pt]
         \vdots \\[1pt]
         - \mathbb{E}^{(\infty )}
    \end{bmatrix}
    = \dfrac{1}{\mu}
    \left[
    \begin{array}{ccc}
        \\[-5pt]
        \mathscr{M}^{(\mathrm{VF})} &
        \mathscr{M}^{(\mathrm{VT})} & 
        \mathscr{M}^{(\mathrm{VS})} \\[20pt]
        \mathscr {M}^{(\mathrm{\Omega F})} & \mathscr{M}^{(\mathrm{\Omega T})} &
        \mathscr{M}^{(\mathrm{\Omega S})} \\[20pt]
        \mathscr{M}^{(\mathrm{E F})} &\mathscr{M}^{(\mathrm{E T})} &
        \mathscr{M}^{(\mathrm{E S})} \\[10pt]
    \end{array}
    \right]
    \bm{\cdot}
    \begin{bmatrix}
        \bm{F}_1 \\[-4pt]
        \vdots \\[1pt]
        \bm{F}_N \\[1pt]
        \boldsymbol{T}_1 \\[-4pt]
        \vdots \\[1pt]
        \boldsymbol{T}_N \\[1pt]
        \mathbb{S}_1 \\[-4pt]
        \vdots\\[1pt]
        \mathbb{S}_N
    \end{bmatrix}
    \;.
\end{equation}
Here, products with $\bm{F}_\alpha$ and $\boldsymbol{T}_\alpha$ are simple contractions, while products with $\mathbb{S}_\alpha$ are double contractions. 
Each block is itself an $N\times N$ array of submatrices.
The block ${\mathscr M}^{(\mathrm{VF})}$, for instance, encodes the coupling between particle forces and particle translational velocity,
and its submatrix 
${{\mathcal{M}}}^{(\mathrm{VF})}_{\alpha\beta}$  characterises the coupling of the force applied by bead $\beta$ upon the translational velocity of bead $\alpha$.  
In this case, the submatrix  ${{\mathcal{M}}}^{(\mathrm{VF})}_{\alpha\beta}$ is a rank-2 tensor obtained using Eqs.~\eqref{eq:Faxén1} and \eqref{eq:flow1_bis}. Its expression is given by:
\begin{equation}
\label{eq:MVFab}
    {{\mathcal{M}}}^{(\mathrm{VF})}_{\alpha\beta} =
    \begin{cases}\dfrac{1}{8 \pi \mu} \Big(1 + \dfrac{a_\alpha^2}{6}\Delta \Big) \Big(1 + \dfrac{a_\beta^2}{6}\Delta \Big)\mathbb{J}(\bm{x}_\alpha - \bm{x}_\beta)  & \mbox{for $\alpha\neq\beta$},\\[4mm]
    \dfrac{\mathbb{I}}{6 \pi \mu a_\alpha} &
    \mbox{for $\alpha=\beta$}.
    \end{cases}
\end{equation}
The grand mobility matrix in Eq.
\eqref{eq:bigMobMat} is a far-field approximation of the interactions between particles since it is derived from the truncated Taylor expansion in Eq. \eqref{eq:flow1}; for this reason it is often denoted $\mathscr{M}^{(\infty)}$ in the literature.
According to \citet{durlofsky_dynamic_1987}, this leads to an error of order 
$\mathscr{O}\big((a/r)^{6}\big)$, where $r$ is the center-to-center bead spacing and $a$ is the bead radius. 
As a consequence, contact between beads (such as in the experiments by \citet{maches2024settling}) cannot be modelled using the approximations 
described here (see Section~\ref{sec:lubrication}). 
The inverse of the grand mobility matrix 
yields the grand resistance tensor $\mathscr{R}$:
\begin{equation} \label{eq:bigResMat}
    \begin{bmatrix}
        \bm{F}_1 \\[-4pt]
        \vdots \\[1pt]
        \bm{F}_N \\[1pt]
        \boldsymbol{T}_1 \\[-4pt]
        \vdots \\[1pt]
        \boldsymbol{T}_N \\[1pt]
        \mathbb{S}_1 \\[-4pt]
        \vdots\\[1pt]
        \mathbb{S}_N
    \end{bmatrix}
    = \mu
    \left[
    \begin{array}{ccc}
    \\[-5pt]
    \mathscr{R}^{(\text{FV})} & \mathscr{R}^{(\text{F}\Omega)} & \mathscr{R}^{(\text{FE})} \\[20pt]
    \mathscr{R}^{(\text{TV})} & \mathscr{R}^{(\text{T}\Omega)} & \mathscr{R}^{(\text{TE})} \\[20pt]
    \mathscr{R}^{(\text{SV})} & \mathscr{R}^{(\text{S}\Omega)} & \mathscr{R}^{(\text{SE})} \\[10pt]
    \end{array}
    \right]
    \bm{\cdot}
    \begin{bmatrix}
        \bm{v}_1 - \bm{U}^{(\infty)}_1 \\[-4pt]
         \vdots \\[1pt]
         \bm{v}_N - \bm{U}^{(\infty)}_N \\[1pt]
         \boldsymbol{\omega}_1 - \boldsymbol{\Omega}^{(\infty)} \\[-4pt]
         \vdots \\[1pt]
         \boldsymbol{\omega}_N - \boldsymbol{\Omega}^{(\infty)}\\[1pt]
         - \mathbb{E}^{(\infty)} \\[-4pt]
         \vdots \\[1pt]
         - \mathbb{E}^{(\infty)}
    \end{bmatrix}
    \:.
\end{equation}
As for the mobility matrix, the submatrices of the $\mathscr{R}^{(\mathrm{FV})}$ block are denoted by $\mathcal{R}^{(\mathrm{FV})}_{\alpha\beta}$, and the same notation applies to the other blocks.
In the next Section, we show how to extract the resistance matrix of a rigid aggregate from Eq.~(\ref{eq:bigResMat}).

\section{Resistance tensors of aggregate particles}
\label{sec:aggregates}

In this Section, we show how to compute resistance tensors of rigid aggregates of beads from Eq.~(\ref{eq:bigResMat}),  including hydrodynamic interactions between multiple aggregates (Section~\ref{sec:method_multiple_particles}), deformable and active aggregates (Section~\ref{sec:da}), and  wall effects (Section \ref{sec:wall}).

\subsection{Resistance matrix for a single rigid aggregate}
\label{sec:method_single_particle}

We adapt the Stokesian dynamics algorithm to represent rigid aggregates of spherical
beads.
Suppose that the aggregate moves with angular velocity $\bm{\omega}$ and with translational velocity $\bm{v}$ defined at a reference point belonging to the aggregate.
Then the velocity $\ve v_\alpha$ and angular velocity $\ve \omega_\alpha$ of bead $\alpha$ follow rigid-body kinematics: 
\begin{align}
\label{eq:velocity_rigid_body}
    \bm{v}_{\alpha} &= \bm{v} + \boldsymbol{\omega} \wedge \bm{X}_{\alpha}\,,\quad 
    \boldsymbol{\omega}_{\alpha} = \boldsymbol{\omega} \:,
\end{align}
where $\bm{X}_{\alpha} = \bm{x}_\alpha - \bm{x}_\text{ref}$ is the position of bead $\alpha$ relative to the chosen reference point. 
This reference point can be any point belonging to the aggregate. Common choices are the centre of mass, the centre of resistance (where the coupling tensor $\mathbb{B}$ is symmetric), or the centre of mobility (where $\ma b$ is symmetric). 
Here, the centre of mass is used as the reference point unless stated otherwise.
The undisturbed flow velocity expressed at the centre of bead $\alpha$ can be decomposed as a first-order expansion about the reference point of the aggregate:
\begin{equation} \label{eq:Vflow}
    \bm{U}^{(\infty)}_\alpha = \bm{U}^{(\infty)}_\text{ref} + \mathbb{E}^\infty 
    \bm{\cdot} \bm{X}_{\alpha} + \boldsymbol{\Omega}^\infty 
    \wedge \bm{X}_\alpha
    \:.
\end{equation}
The total force, torque and stresslet exerted by the aggregate particle upon the fluid are obtained by summing the contributions of each bead:
\begin{subequations} \label{eq:FTSparticle}
\begin{eqnarray}
    \bm{F} &&= \sum_{\alpha \: = \: 1}^{N} \bm{F}_{\alpha} 
    \:,
    \\
    \boldsymbol{T} &&= \sum_{\alpha \: = \: 1}^{N} \boldsymbol{T}_{\alpha} + \sum_{\alpha \: = \: 1}^{N} 
    \bm{X}_\alpha \wedge \bm{F}_{\alpha}
    \:,
    \\
    \mathbb{S} &&= 
    \sum_{\alpha \: = \: 1}^{N} \mathbb{S}_{\alpha} + \sum_{\alpha \: = \: 1}^{N} \left[ \frac{1}{2} \bm{X}_\alpha \otimes \bm{F}_\alpha + \frac{1}{2} \bm{F}_\alpha \otimes \bm{X}_\alpha - \frac{1}{3} \left( \bm{X}_\alpha \bm{\cdot} \bm{F}_\alpha \right) \,\mathbb{I} \right]\,.
\end{eqnarray}
\end{subequations}
We now use Eqs.~(\ref{eq:velocity_rigid_body}--\ref{eq:FTSparticle}) to extract
the resistance tensors of
the rigid aggregate from the $N$–bead grand resistance tensor
(\ref{eq:bigResMat}): 
\begin{align} \label{eq:ABGtensors}
    \mathbb{A} &= \sum_{\alpha \: = \: 1}^{N} \sum_{\beta \: = \: 1}^{N} \mathcal{R}^{(\mathrm{FV})}_{\alpha\beta}\,,
    \quad
 \widetilde{\mathbb{B}} = \sum_{\alpha \: = \: 1}^{N} \sum_{\beta \: = \: 1}^{N} \left( \mathcal{R}^{(\mathrm{F\Omega})}_{\alpha\beta} - \mathcal{R}^{(\mathrm{FV})}_{\alpha\beta} \bm{\cdot}  \hat{\mathbb{X}}_\beta \right)\,,\quad 
  \widetilde  {\mathbb{G}} = \sum_{\alpha \: = \: 1}^{N} \sum_{\beta \: = \: 1}^{N} \left( \mathcal{R}^{(\mathrm{FE})}_{\alpha\beta} + \mathcal{R}^{(\mathrm{FV})}_{\alpha\beta} \otimes \bm{X}_\beta \right)
    \:.
\end{align}
Here $\hat{\mathbb{X}}_\alpha$ denotes the antisymmetric matrix associated with the vector $\bm{X}_\alpha$, with components 
$[ \mathbb{\hat{X}}_{\alpha} ]_{ij} = \varepsilon_{ikj} [\bm{X}_{\alpha}]_k$,
and the tensorial product $\bm{A}\otimes \bm{v}$ between a rank-2 tensor $\bm{A}$ and a vector $\bm{v}$ is defined by
$ [\bm{A}\otimes \bm{v}]_{ijk} = A_{ij}v_k$.
With the same approach, the resistance tensors $\mathbb{B}$, $\mathbb{C}$ and $\widetilde{\mathbb{H}}$ are obtained from the torque $\bm{T}$ of the rigid body:
\begin{subequations} \label{eq:BCHtensors}
\begin{eqnarray} 
    \mathbb{B} &&= \sum_{\alpha \: = \: 1}^{N} \sum_{\beta \: = \: 1}^{N} \left( \mathcal{R}^{(\mathrm{TV})}_{\alpha\beta} + \hat{\mathbb{X}}_\alpha \bm{\cdot} \mathcal{R}^{(\mathrm{FV})}_{\alpha\beta} \right) 
    \:,
    \\
    \mathbb{C} &&= \sum_{\alpha \: = \: 1}^{N} \sum_{\beta \: = \: 1}^{N} \left( \mathcal{R}^{(\mathrm{T\Omega})}_{\alpha\beta} - \mathcal{R}^{(\mathrm{TV})}_{\alpha\beta} \bm{\cdot} \hat{\mathbb{X}}_\beta + \hat{\mathbb{X}}_\alpha \bm{\cdot} \mathcal{R}^{(\mathrm{F\Omega})}_{\alpha\beta}  - \hat{\mathbb{X}}_\alpha \bm{\cdot} \mathcal{R}^{(\mathrm{FV})}_{\alpha\beta} \bm{\cdot}  \hat{\mathbb{X}}_\beta \right)
    \:,
    \\
    \widetilde{\mathbb{H}} &&= \sum_{\alpha \: = \: 1}^{N} \sum_{\beta \: = \: 1}^{N} \left( \mathcal{R}^{(\mathrm{TE})}_{\alpha\beta} + \mathcal{R}^{(\mathrm{TV})}_{\alpha\beta} \otimes \bm{X}_\beta + \hat{\mathbb{X}}_\alpha \bm{\cdot} \mathcal{R}^{(\mathrm{FE})}_{\alpha\beta} + \hat{\mathbb{X}}_\alpha \bm{\cdot} \mathcal{R}^{(\mathrm{FV})}_{\alpha\beta} \otimes \bm{X}_\beta \right)
    \:.
\end{eqnarray}
\end{subequations}
Finally, from the stresslet we identify resistance tensors $\mathbb{G}, \mathbb{H}$ and $\mathbb{M}$. For convenience, their expression are given in index notation:
\begin{subequations}  \label{eq:GHMtensors}
\begin{eqnarray}
    G_{ijk} && = \sum_{\alpha \: = \: 1}^{N} \sum_{\beta \: = \: 1}^{N} \left( \left[\mathcal{R}^{(\mathrm{SV})}_{\alpha\beta}\right]_{ijk} 
    + \frac{1}{2} \Big[\bm{X}_\alpha \Big]_i \left[\mathcal{R}^{(\mathrm{FV})}_{\alpha\beta}\right]_{jk} +\frac{1}{2} \Big[\bm{X}_\alpha \Big]_j \left[\mathcal{R}^{(\mathrm{FV})}_{\alpha\beta}\right]_{ik} 
    -\frac{1}{3} \delta_{ij} \Big[\bm{X}_\alpha \Big]_\ell  \left[\mathcal{R}^{(\mathrm{FV})}_{\alpha\beta} \right]_{\ell k} \right) 
    \: ,
    \\
    H_{ijk} &&= \sum_{\alpha \: = \: 1}^{N} \sum_{\beta \: = \: 1}^{N} \left( \left[\mathcal{R}^{(\mathrm{S\Omega})}_{\alpha\beta} \right]_{ijk} - \left[\mathcal{R}^{(\mathrm{SV})}_{\alpha\beta} \right]_{ij\ell} \Big[ \mathbb{\hat{X}}_{\beta} \Big]_{\ell k} + \frac{1}{2} \Big[\bm{X}_\alpha \Big]_i \left[\mathcal{R}^{(\mathrm{F\Omega})}_{\alpha\beta} \right]_{jk} 
    - \frac{1}{2} \Big[\bm{X}_\alpha \Big]_i \left[\mathcal{R}^{(\mathrm{FV})}_{\alpha\beta} \right]_{j\ell} \Big[ \mathbb{\hat{X}}_{\beta} \Big]_{\ell k} \right. \nonumber \\
    &&   + \frac{1}{2} \Big[\bm{X}_\alpha \Big]_j \left[\mathcal{R}^{(\mathrm{F\Omega})}_{\alpha\beta} \right]_{ik} 
    - \frac{1}{2} \Big[\bm{X}_\alpha \Big]_j \left[\mathcal{R}^{(\mathrm{FV})}_{\alpha\beta} \right]_{i\ell} \Big[ \mathbb{\hat{X}}_{\beta} \Big]_{\ell k}  
    - \frac{1}{3} \delta_{ij} \Big[\bm{X}_\alpha \Big]_\ell \left[\mathcal{R}^{(\mathrm{F\Omega})}_{\alpha\beta} \right]_{\ell k}  \nonumber \\
    &&  \left. 
    + \frac{1}{3} \delta_{ij} \Big[\bm{X}_\alpha \Big]_\ell \left[\mathcal{R}^{(\mathrm{FV})}_{\alpha\beta} \right]_{\ell m} \Big[ \mathbb{\hat{X}}_{\beta} \Big]_{m k} \right)
    \: ,
    \\
    M_{ijkl} &&= \sum_{\alpha \: = \: 1}^{N} \sum_{\beta \: = \: 1}^{N} \left( \left[\mathcal{R}^{(\mathrm{SE})}_{\alpha\beta} \right]_{ijk\ell} + \left[\mathcal{R}^{(\mathrm{SV})}_{\alpha\beta} \right]_{ijk} \Big[\bm{X}_\beta \Big]_\ell \right. + \frac{1}{2} \Big[\bm{X}_\alpha \Big]_i \left[\mathcal{R}^{(\mathrm{FE})}_{\alpha\beta} \right]_{jk\ell} 
    + \frac{1}{2} \Big[\bm{X}_\alpha \Big]_i \left[\mathcal{R}^{(\mathrm{FV})}_{\alpha\beta} \right]_{jk} \Big[\bm{X}_\beta \Big]_\ell \nonumber \\
    &&  + \frac{1}{2} \Big[\bm{X}_\alpha \Big]_j \left[\mathcal{R}^{(\mathrm{FE})}_{\alpha\beta} \right]_{ik\ell} 
    + \frac{1}{2} \Big[\bm{X}_\alpha \Big]_j \left[\mathcal{R}^{(\mathrm{FV})}_{\alpha\beta} \right]_{ik} \Big[\bm{X}_\beta \Big]_\ell   - \frac{1}{3} \delta_{ij} \Big[\bm{X}_\alpha \Big]_m \left[\mathcal{R}^{(\mathrm{FE})}_{\alpha\beta} \right]_{mk\ell} \nonumber \\
    && \left.
    - \frac{1}{3} \delta_{ij} \Big[\bm{X}_\alpha \Big]_m \left[\mathcal{R}^{(\mathrm{FV})}_{\alpha\beta} \right]_{mk} \Big[\bm{X}_\beta \Big]_\ell \right)
    \: .
\end{eqnarray}
\end{subequations}
By virtue of the reciprocal theorem \cite{lorentz1907} and the reversibility of Stokes flows, the resistance tensors of the aggregate particle must satisfy the following symmetry relations \cite{kim_microhydrodynamics_1991}:
\begin{align}
\refstepcounter{equation}
    A_{ij} &= A_{ji} \:, & B_{ij} &= \widetilde{B}_{ji} \:, & C_{ij} &= C_{ji} \:, \tag{\theequation a,b,c} \\
    G_{ijk} &= \widetilde{G}_{kij} \:, & H_{ijk} &= \widetilde{H}_{kij} \:, & M_{ijk\ell} &= M_{k\ell ij} \:. \tag{\theequation d,e,f}
\end{align}
A subtlety is that
direct computation may 
yield results for $\ma G$, $\ma H$, and $\ma M$
that do not possess these
symmetries, because
the method does not explicitly represent
the fact that
 the strain-rate tensor $\mathbb{E}^\infty$ is symmetric and traceless.
However,
any  contributions
to  $\ma G$, $\ma H$, and $\ma M$ that break these symmetries, vanish upon double contraction with the traceless, symmetric strain-rate tensor.
To bring out the required symmetries of the physically relevant parts
of $\ma G$, $\ma H$, and $\ma M$, we symmetrise
them as follows:
\begin{subequations}
\label{eq:sym_GHM}
\begin{align}
    \widetilde{\mathbb{G}}_{ijk}^{\text{sym}} &= \frac{1}{2}\left( \widetilde{\mathbb{G}}_{ijk}+ \widetilde{\mathbb{G}}_{ikj} \right) - \frac{1}{3}\widetilde{\mathbb{G}}_{i\ell\ell}\delta_{jk} \:, \\
    \widetilde{\mathbb{H}}_{ijk}^{\text{sym}} &= \frac{1}{2}\left( \widetilde{\mathbb{H}}_{ijk}+ \widetilde{\mathbb{H}}_{ikj} \right) - \frac{1}{3}\widetilde{\mathbb{H}}_{i\ell\ell}\delta_{jk} \:, \\
    \widetilde{\mathbb{M}}_{ijk\ell}^{\text{sym}} &= \frac{1}{2}\left( \widetilde{\mathbb{M}}_{ijk\ell}+ \widetilde{\mathbb{M}}_{ij\ell k} \right) - \frac{1}{3}\widetilde{\mathbb{M}}_{ijmm}\delta_{k\ell} \:.
\end{align}
\end{subequations}
Applying these operations ensures that the global resistance matrix is symmetric and traceless, as it must be. 
When expressed in the particle’s body-fixed reference frame, the resistance tensors remain constant over time. 
Consequently, for an isolated particle in an unbounded fluid, the resistance matrix needs to be computed only once, 
after which the full rigid-body dynamics can be obtained at a negligible numerical cost. Some examples are given in Section \ref{sec:examples}.

\subsection{Hydrodynamic interactions between multiple aggregates}
\label{sec:method_multiple_particles}

The method naturally extends to account for hydrodynamic interactions among several rigid particles $\mathcal{P}$. Let us consider a system of $M$ particles, modeled altogether by $N$ beads. For each bead, velocities are constrained according to the rigid body to which it belongs.
For example, the resistance matrix for two particles can be written as:
\begin{equation}
    \begin{bmatrix} 
        \bm{F}_\mathit{1} \\[1pt]
        \bm{F}_\mathit{2} \\[1pt]
        \bm{T}_\mathit{1} \\[1pt]
        \bm{T}_\mathit{2} \\[1pt]
        \mathbb{S}_\mathit{1} \\[1pt]
        \mathbb{S}_\mathit{2}
    \end{bmatrix}
    = \mu
    \begin{bmatrix}
        \mathbb{A}_{\mathit{11}} & \mathbb{A}_{\mathit{12}} & \widetilde{\mathbb{B}}_{\mathit{11}} & \widetilde{\mathbb{B}}_{\mathit{12}} & \widetilde{\mathbb{G}}_{\mathit{11}} & \widetilde{\mathbb{G}}_{\mathit{12}}\\
        \mathbb{A}_{\mathit{21}} & \mathbb{A}_{\mathit{22}} & \widetilde{\mathbb{B}}_{\mathit{21}} & \widetilde{\mathbb{B}}_{\mathit{22}} & \widetilde{\mathbb{G}}_{\mathit{21}} & \widetilde{\mathbb{G}}_{\mathit{22}}\\
        \mathbb{B}_{\mathit{11}} & \mathbb{B}_{\mathit{12}} & \mathbb{C}_{\mathit{11}} & \mathbb{C}_{\mathit{12}} & \widetilde{\mathbb{H}}_{\mathit{11}} & \widetilde{\mathbb{H}}_{\mathit{12}} \\
        \mathbb{B}_{\mathit{21}} & \mathbb{B}_{\mathit{22}} & \mathbb{C}_{_\mathit{21}} & \mathbb{C}_{\mathit{22}} & \widetilde{\mathbb{H}}_{\mathit{21}} & \widetilde{\mathbb{H}}_{\mathit{22}} \\
        \mathbb{G}_{_\mathit{11}} & \mathbb{G}_{\mathit{12}} & \mathbb{H}_{\mathit{11}} & \mathbb{H}_{\mathit{12}} & \mathbb{M}_{\mathit{11}} & \mathbb{M}_{\mathit{12}} \\
        \mathbb{G}_{_\mathit{21}} & \mathbb{G}_{\mathit{22}} & \mathbb{H}_{\mathit{21}} & \mathbb{H}_{_\mathit{22}} & \mathbb{M}_{\mathit{21}} & \mathbb{M}_{\mathit{22}}
    \end{bmatrix}
    \bm{\cdot}
    \begin{bmatrix}
        \bm{v}_\mathit{1} - \bm{U}^{(\infty)}_\mathit{1} \\[1pt]
        \bm{v}_\mathit{2} - \bm{U}^{(\infty)}_\mathit{2} \\[1pt]
        \bm{\omega}_\mathit{1} - \bm{\Omega}^{(\infty)} \\[1pt]
        \bm{\omega}_\mathit{2} - \bm{\Omega}^{(\infty)} \\[1pt]
        - \mathbb{E}^{(\infty)} \\[1pt]
        - \mathbb{E}^{(\infty)}
    \end{bmatrix} \: ,
    \label{eq:system}
\end{equation}
where italic subscripts indicate the particle index.
Resistance tensors have the same expressions as in the single-particle case, except that the sums are now restricted to the beads belonging to a given particle. For example:
\begin{align}
    \mathbb{A}_{AB} &= \sum_{\alpha \: \in \: \mathcal{P}^A} \: \sum_{\beta \: \in \: \mathcal{P}^B} \mathcal{R}^{(\mathrm{FV)}}_{\alpha\beta}
    \:.
\end{align}
The sets $\mathcal{P}^A$ and $\mathcal{P}^B$ respectively contain the indices of the beads of particles A and B.
When multiple rigid particles interact, the position of the  beads evolves over time and the relative orientation of the particles 
strongly 
affects the hydrodynamic coupling. 
As a result, the global resistance matrix must be updated at each time step to account for the new geometry of the system.

\subsection{Flexible and active aggregates}
\label{sec:da}

\begin{figure}[t]
    \centering
    \includegraphics[width=0.5\linewidth]{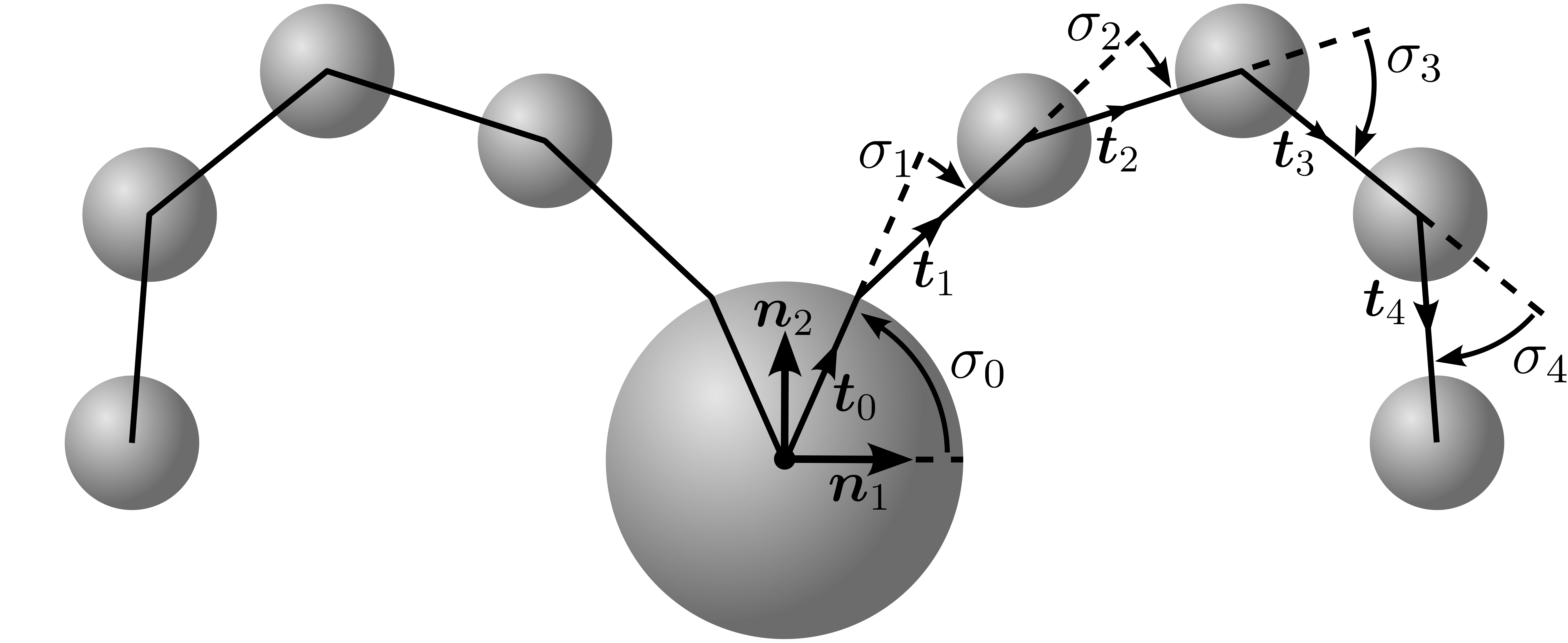}
    \caption{
  Sketch of a microswimmer with two flexible flagella, illustrating how the flagella and the swimming gait are parameterised. The mobile frame is attached to the swimmer's body, at the centre of the largest bead. An angle $\sigma_0$ determines the position of the attachment point of each flagellum. The orientations of the vectors $\bm{t}_\alpha$ are parameterised by the angles~$\sigma_\alpha$. 
  }
\label{fig:swimmer_sketch}
\end{figure}

Up to this point, we focused  on rigid particles, although the method itself is not restricted to this case.
In practice, the method presented here is much more general. 
If the relative motion among the  beads is prescribed, 
the hydrodynamic problem can be formulated as an equivalent resistance problem for a single body.
The construction of the resistance tensors follows the same procedure as for a rigid particle.
The only difference lies in the velocity field imposed on each bead 
which now corresponds to the prescribed kinematics of the deformable or active structure.
This approach is illustrated by the microswimmer \textit{Chlamydomonas reinhardtii} in Section \ref{subsec:swimmer}, where each flagellum consists of a sequence of beads whose relative motions are prescribed to reproduce the beating kinematics. 
The motion of the flagella is prescribed through time-dependent angular laws $\sigma_\alpha(t)$ together with their time derivatives $\dot{\sigma}_\alpha(t)$. 
In that context, force and torque exerted by such an aggregate upon the fluid can be cast in the form:
\begin{equation} \label{eq:swimmer_FT}
    \begin{bmatrix}
        \bm{F} \\
        \bm{T}
    \end{bmatrix}
    =
    \begin{bmatrix}
        \mathbb{A} & \widetilde{\mathbb{B}} \\
        \mathbb{B} & \mathbb{C}
    \end{bmatrix}
    \cdot
    \begin{bmatrix}
        \bm{v} \\
        \bm{\omega}
    \end{bmatrix}
    +
    \begin{bmatrix}
        \ve F_a(\boldsymbol{\sigma}, \dot{\boldsymbol{\sigma}}) \\
        \ve T_a(\boldsymbol{\sigma}, \dot{\boldsymbol{\sigma}})
    \end{bmatrix}\,,
\end{equation}
where $\boldsymbol{\sigma} = (\sigma_1,\ldots,\sigma_N)^{\sf T}$, and $\dot{\boldsymbol{\sigma}} = (\dot{\sigma}_1,\ldots,\dot{\sigma}_N)^{\sf T}$. 
The parametrisation of the swimmer is represented in Fig.~\ref{fig:swimmer_sketch}. The largest bead of the swimmer, modeling the swimmer's body, has translational velocity $\bm{v}$ and angular velocity $\boldsymbol{\omega}$.
In this case, translational velocity $\bm{v}_\alpha$ and the angular velocity $\bm{\omega}_\alpha$ of bead $\alpha$ in the laboratory frame are given by:
\begin{subequations} \label{eq:swimmer_velocities}
\begin{eqnarray} 
    \boldsymbol{\omega}_{\alpha} &=& \boldsymbol{\omega} + \displaystyle \sum_{\beta \: \in \: f}^{\alpha} \Dot{\sigma}_{\beta}   \bm{n}_{3} \:, \label{eq:swimmer_omega} \\
    \bm{v}_\alpha &=& \bm{v} + \bm{\omega} \wedge \bm{X}_\alpha  + \ell \sum_{\beta \: \in \: f}^{\alpha} \boldsymbol{\omega}_\beta  \wedge \bm{t}_\beta  - \frac{\ell}{2} \omega_\alpha \wedge \bm{t}_\alpha \:,\label{eq:swimmer_v}
\end{eqnarray} 
\end{subequations} 
where 
\begin{equation}
\refstepcounter{equation}
    \bm{t}_0 = \cos{\sigma_0} \: \bm{n}_{1} + \sin{\sigma_0}  \: \bm{n}_{2} \:,
    \quad \mbox{and} \quad 
    \bm{t}_\alpha = \cos\left({\sigma_0 + \sum_{\beta \: \in \: f}^\alpha \sigma_\beta} \right) \: \bm{n}_{1} + \sin\left({\sigma_0 + \sum_{\beta \: \in \: f}^\alpha \sigma_\beta} \right) \: \bm{n}_{2} \:.
    \tag{\theequation a,b}
\end{equation}
The set $f$ contains the indices of the beads that form the same flagellum as bead $\alpha$.
By substituting the velocities given in Eqs. \eqref{eq:swimmer_velocities} into Eq. (\ref{eq:swimmer_FT}), and collecting the terms proportional to $\ve v$ and $\boldsymbol{\omega}$, the forcing terms $\ve F_a(\boldsymbol{\sigma}, \dot{\boldsymbol{\sigma}})$ and $\ve T_a(\boldsymbol{\sigma}, \dot{\boldsymbol{\sigma}})$, which account for the swimmer’s active propulsion induced by internal flagellar motion, can be written as:
\begin{subequations}
\begin{align}
\refstepcounter{equation}
    \ve F_a(\boldsymbol{\sigma}, \dot{\boldsymbol{\sigma}}) &= \sum_\alpha \sum_\beta \left\{ \mathcal{R}^{(\mathrm{FV})}_{\alpha\beta} \cdot \left( \ell \sum_{A \: \in \: f}^{\beta} \left[ \sum_{B \: \in \: f}^{A} \Dot{\sigma}_{B}  \: \bm{n}_{3} \right] \wedge \bm{t}_A - \frac{\ell}{2} \left[ \sum_{A \: \in \: f}^\beta \Dot{\sigma}_{A}  \: \bm{n}_{3} \right] \wedge \bm{t}_\beta \right) + \mathcal{R}^{\mathrm{(F\Omega)}}_{\alpha\beta} \cdot \left[ \sum_{A \: \in \: f}^{\beta} \Dot{\sigma}_{A}  \: \bm{n}_{3} \right] \right\} \tag{\theequation}\\
    \ve T_a (\boldsymbol{\sigma}, \dot{\boldsymbol{\sigma}}) &= \sum_\alpha \sum_\beta \left\{ \left( \mathcal{R}^{(\mathrm{TV})}_{\alpha\beta} + \mathbb{\hat{X}}_\alpha  \cdot \mathcal{R}^{(\mathrm{FV})}_{\alpha\beta} \right) \cdot \left( \ell \sum_{A \: \in \: f}^\beta \left[ \sum_{B \: \in \: f}^{A} \Dot{\sigma}_{B}  \: \bm{n}_{3} \right] \wedge \bm{t}_A - \frac{\ell}{2} \left[ \sum_{A \: \in \: f}^\beta \Dot{\sigma}_{A}  \: \bm{n}_{3} \right] \wedge \bm{t}_\beta \right) \right. \notag \\
    & \left.~~~~~~~~~~~~~~+ ~\left(\mathcal{R}^{\mathrm{(T\Omega)}}_{\alpha\beta} + \hat{\mathbb{X}}_\alpha  \cdot \mathcal{R}^{\mathrm{(F\Omega)}}_{\alpha\beta}\right) \cdot \left[ \sum_{A \: \in \: f}^\beta \Dot{\sigma}_{A}  \: \bm{n}_{3} \right] \right\} \:.
\end{align}
\end{subequations}
The swimming behavior of such an active particle, together with the characteristics of the disturbance flow generated over one swimming cycle, is computed in Section~\ref{subsec:swimmer}.

\subsection{Hydrodynamic interaction with a  wall}
\label{sec:wall}
We now show how to compute the resistance matrix of a rigid aggregate moving near a planar wall using the \texttt{SHAPES} method. 
\citet{durlofsky_dynamic_1989} and \citet{bossis_stokesian_1991}  accounted for wall effects using heuristic approaches. The strengths and weaknesses of their method are discussed by  \citet{swan_simulation_2007} who developed a precise theory for wall effects using Blake’s image system \cite{blake_note_1971}.
The no-slip boundary condition at the wall is satisfied by placing image singularities on the opposite side of the wall. Using this method, \citet{swan_simulation_2007} computed how a wall affects the  rotation of a nearby pair of spheres. 
Here we include wall effects in a different way, exploiting the inherent mathematical symmetry of the Stokes equations near a planar boundary \cite[p.~87]{happel_low_1983}. The advantage of this method is that the resulting set
of equations is very close to the original Stokesian-dynamics formulation, albeit with different tensors.
Also, the method applies to any Stokes flow, and does not require to decompose the flow into a sum of singularities.
We decompose the disturbance flow into two contributions: the flow generated by the aggregate in the absence of the wall $\bm{w}$ and the reflected flow induced by the wall denoted by $\bm{w}^*$.
\begin{equation}
    \bm{w}'(\bm{x}) = \bm{w}(\bm{x}) + \bm{w}^*(\bm{x}) \:.
\end{equation}
Taken together, these two contributions must satisfy the no-slip boundary condition at the 
wall located at $x=0$:
\begin{equation} \label{eq:wall_boundary_condition}
    \bm{w}^*(\bm{x}) + \bm{w}(\bm{x}) = 0 ~~ \text{at the wall\,.}
\end{equation}
The reflected flow $\bm{w}^*$ is constructed by applying two successive linear transformations to the flow $\bm{w}$ generated in the absence of the wall. 
Each transformation is defined so that the resulting field remains a valid solution of the Stokes equation.
Writing $\bm{w}=[w_x,w_y,w_z]^{\sf T}$ in Cartesian components, the first transformation is given by:
\begin{equation}
    \widetilde{\bm{w}} = 
    \begin{bmatrix}
        w_x \\
        - w_y \\
        - w_z
    \end{bmatrix}
    - 2x
    \begin{bmatrix}
        \partial w_x / \partial x \\
        \partial w_x / \partial y \\
        \partial w_x / \partial z
    \end{bmatrix}
    + x^2 
    \begin{bmatrix}
        \Delta w_x  \\
        \Delta w_y  \\
        \Delta w_z 
    \end{bmatrix}
    \:.
\end{equation}
Here, $\Delta$ denotes the Laplace operator, as in Eq.~\eqref{eq:flow1}. 
The vector field $\widetilde{\bm{w}}$ satisfies the following boundary condition on the wall: 
\begin{equation}
    \widetilde{\bm{w}}_x = w_x \: , ~ \widetilde{\bm{w}}_y = - w_y \: , ~ 
    \widetilde{\bm{w}}_z = - w_z \:, \quad \text{at } x = 0 \:.
\end{equation}
The second transformation is a mirror reflection of $\widetilde{\bm{w}}$ with respect to the plane $x=0$:
\begin{subequations}
\begin{align}
    w_x^*(x, y, z) &= - \widetilde{w}_x(-x, y, z)
    \:,\\
    w_y^*(x, y, z) &= \widetilde{w}_y(-x, y, z)
    \:,\\
    w_z^*(x, y, z) &= \widetilde{w}_z(-x, y, z)
    \:.
\end{align}
\end{subequations}
Finally, $\bm{w}^*$ satisfies Eq.~\eqref{eq:wall_boundary_condition}
so that the 
total disturbance flow $\bm{w}'(\bm{x})$ fulfills the no-slip boundary condition on the wall.
The reflected disturbance flow induced by the wall can then be written 
as:
\begin{eqnarray} \label{eq:wall_tensors}
    &&\bm{w}^* (\bm{x}) = \frac{1}{8\pi\mu} \sum_{\alpha \, = \, 1}^{N} \left[ \mathbb{J}^{*}(\bm{x}, \bm{x}_\alpha, a_\alpha)  \bm{\cdot} \bm{F}_\alpha  + \mathbb{R}^{*}(\bm{x}, \bm{x}_\alpha, a_\alpha) \bm{\cdot} \bm{T}_\alpha - \mathbb{K}^{*}(\bm{x}, \bm{x}_\alpha, a_\alpha) \bm{:} \mathbb{S}_\alpha \right] 
    \label{Durlofsky_wall} 
    \:.
\end{eqnarray}
where $\mathbb{J}^{*}$, $\mathbb{R}^{*}$ and $\mathbb{K}^{*}$ are the wall tensors obtained by applying the transformations to Eq.~(\ref{eq:flow1}). 
We note that, in contrast with Eq.~(\ref{eq:flow1_bis}), a given bead $\alpha$ will experience the reflections $\bm{w}^*$ of its own disturbance flow on the wall.
We now 
use Faxén's laws (Eq. \eqref{eq:Faxen}) on the total disturbance flow $\bm{w}+\ve w^\ast$. 
The flow experienced by a given bead $\alpha$ is composed of the contribution $\bm{w}$ generated by all other beads in the absence of the wall, 
together with the reflected flow $\bm{w}^*$ generated by all beads including $\alpha$.
The resulting grand mobility matrix now accounts for the presence of the wall. The submatrix ${{\mathcal{M}}}^{(\mathrm{VF})}_{\alpha\beta}$, for example, 
is given by:
\begin{equation}
\label{eq:MVFab_wall}
   {{\mathcal{M}}}^{(\mathrm{VF})}_{\alpha\beta}=  \begin{cases}\dfrac{1}{8 \pi \mu} \left(1 + \dfrac{a_\alpha^2}{6}\Delta \right) \left(1 + \dfrac{a_\beta^2}{6}\Delta \right)\mathbb{J}(\bm{x}_\alpha - \bm{x}_\beta) + \dfrac{1}{8 \pi \mu} \left(1 + \dfrac{a_\alpha^2}{6}\Delta \right) \mathbb{J}^{*} \left( \bm{x}_\alpha, \bm{x}_\beta, a_\beta \right) & \mbox{for $\alpha\neq \beta$\,,}\\
   \dfrac{\mathbb{I}}{6\pi \mu a_\alpha} + \left(1 + \dfrac{a_\alpha^2}{6}\Delta \right)\left[ \dfrac{1}{8\pi\mu} \mathbb{J}^{*} \left( \bm{x}_\alpha, \bm{x}_\alpha, a_\alpha \right) \right]&\mbox{for $\alpha=\beta$}\,.
   \end{cases} 
\end{equation}
Compared with Eq.~(\ref{eq:MVFab}), an additional contribution appears in both cases ($\alpha\neq \beta$ and $\alpha=\beta$) due to the interaction of the beads with the wall.
In particular, the extra term for $\alpha \neq \beta$  
arises from the self-interaction of the bead with its own flow reflected by the wall.
The remaining submatrices are obtained analogously from Faxén’s laws.
The inversion of the mobility matrix provides the global resistance matrix for the system of $N$ spheres, now including the effect of a planar  wall. 
To obtain the effective resistance matrix for one or several rigid particles, we proceed as described in the previous subsections. 

As in the unbounded case, the resulting matrices must satisfy the required symmetry conditions. 
Any residual asymmetry in the tensors $\mathbb{G}$, $\mathbb{H}$, and $\mathbb{M}$ is corrected using  Eqs.~(\ref{eq:sym_GHM}). 
This lack of symmetry simply reflects that the tensors were constructed without explicitly enforcing the symmetric and traceless nature of the strain-rate tensor $\mathbb{E}^\infty$. However, it does not affect the computed hydrodynamic forces when contracted with the velocity field.

\subsection{Lubrication corrections}
\label{sec:lubrication}

\begin{figure}[t]
    \centering
    \hspace{-1.9cm}
    \includegraphics[width=\textwidth - 1.9cm]{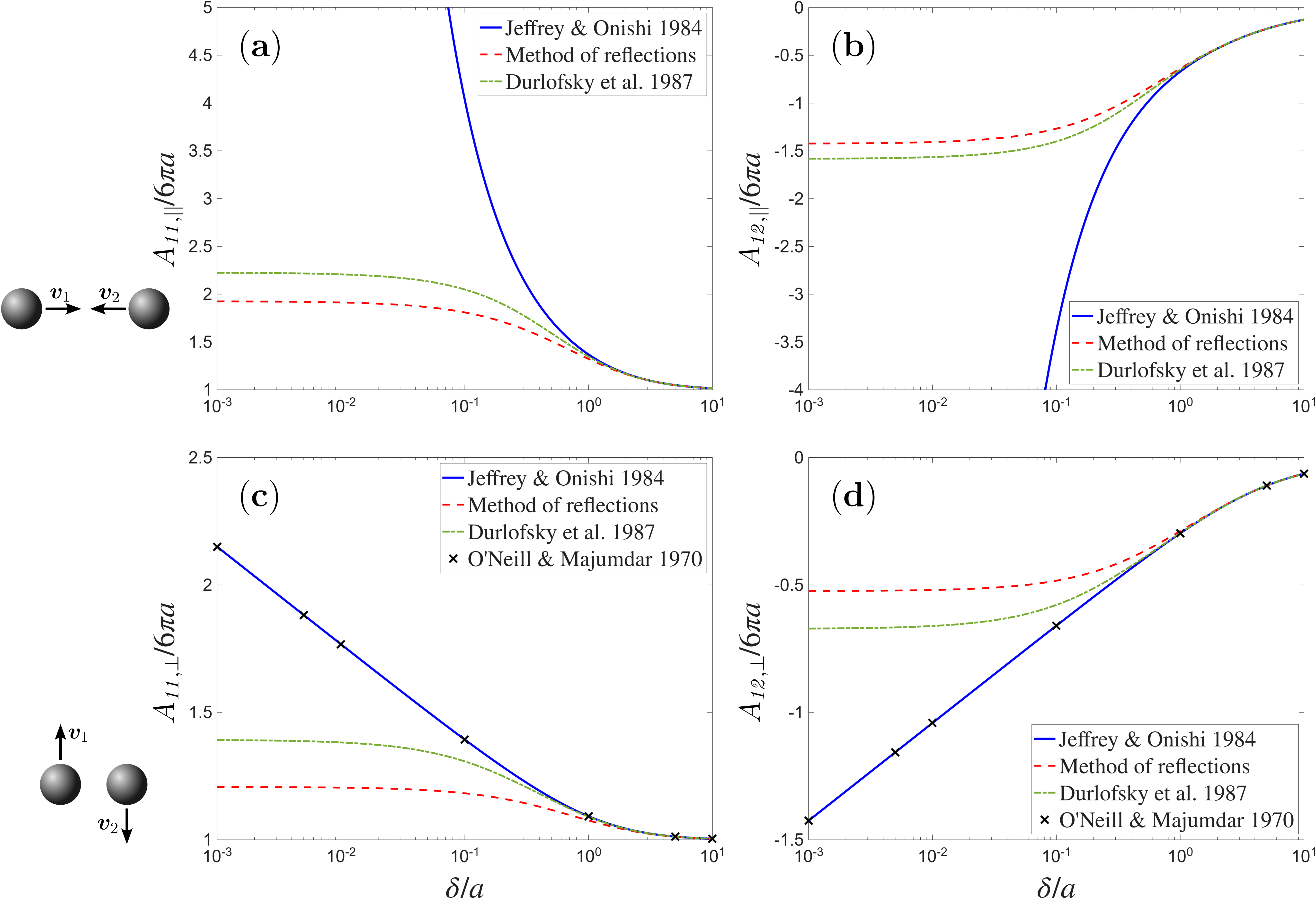}
    \caption{Resistance tensors $A_{AB}$ from Eq.~\eqref{eq:system} for two hydrodynamically interacting beads of equal radius $a$, as functions of the non-dimensional interfacial distance $\delta/a$ between them.
    Exact solutions were derived by \citet{jeffrey_calculation_1984}. \citet{oneill_asymmetrical_1970} tabulated some of these results. The method of reflection is described in Ref.~\cite{happel_low_1983}. 
    }
    \label{fig:lubrication}
\end{figure}

The Stokesian dynamics method relies on a far-field expansion of the hydrodynamic interactions between beads, leading to large errors when beads are too close to each other. 
To illustrate this,  
we consider two identical beads of radius $a$ separated by a centre-to-centre distance $r$. 
The exact hydrodynamic resistance of two interacting beads of arbitrary size in Stokes flow was derived by Jeffrey and Onishi \cite{jeffrey_calculation_1984, jeffrey_forces_1984, jeffrey1992calculation}. 
Note that the seminal paper \cite{jeffrey_calculation_1984} contains a number of errors, as identified by \citet{townsend_generating_2023}.
Exact results for unequal spheres were also available through tabulated solutions \cite{oneill_asymmetrical_1970}, while earlier approximations relied on the method of reflections \cite{happel_low_1983}. 
We examine elements of the grand resistance matrix defined in Eq.~\eqref{eq:system}, focusing on the $\ma A$ tensors for convenience.
Fig.~\ref{fig:lubrication} shows the force $\bm{F}_1$, exerted by the first bead on the fluid, as a function of the non-dimensional interfacial distance $\delta/a$, considering relative translations parallel and perpendicular to the line of centres. All methods agree at large separations, whereas at small values of $\delta$, Stokesian dynamics \cite{durlofsky_dynamic_1987} improves upon the method of reflections but still substantially deviates from the exact solution.

To partially compensate for this limitation, \citet{durlofsky_dynamic_1987} proposed a correction procedure to approximately reintroduce lubrication interactions in a pairwise additive manner.
Using the authors' notation, the approach consists in reconstructing the two-body resistance matrix $\mathscr{R}_{2B}$ 
built from the exact analytical results for the hydrodynamic interactions between two 
spheres \cite{jeffrey_calculation_1984,jeffrey1992calculation}.
The final approximation of the resistance matrix thus reads:
\begin{equation}
\label{eq:lubrication}
    \mathscr{R} = (\mathscr{M}^{(\infty)})^{-1} + \mathscr{R}_{2B} - \mathscr{R}_{2B}^{(\infty)},
\end{equation}
where $\mathscr{R}_{2B}^{(\infty)}$ is obtained by inverting a two-body mobility matrix truncated at the same order 
as $\mathscr{M}^{(\infty)}$.

For dilute suspensions where close interactions remain pairwise and transient, this pairwise additive correction is accurate and 
widely adopted in the literature \cite{brady_stokesian_1988, ladd_hydro_1989, bossis_hydrodynamic_1991, cichocki_friction_1994, townsend_frictional_2017, fiore_fast_2019}.
However, lubrication theory is only relevant when two rigid surfaces move relative to each other, 
producing large velocity gradients within the narrow fluid gap between them \cite{kim_microhydrodynamics_1991}. 
Here, aggregates are either strictly rigid or deformable through prescribed internal motions so that lubrication effects should remain minor.  
From a computational standpoint, lubrication corrections also significantly increase the numerical cost, particularly when an aggregate is represented by a large number of beads. 
Furthermore, for configurations involving three or more spheres in close proximity, pairwise superposition of lubrication corrections is an approximation and lacks a rigorous theoretical foundation.

\begin{figure}
    \centering
    \includegraphics[width=\textwidth/2 - 2cm]{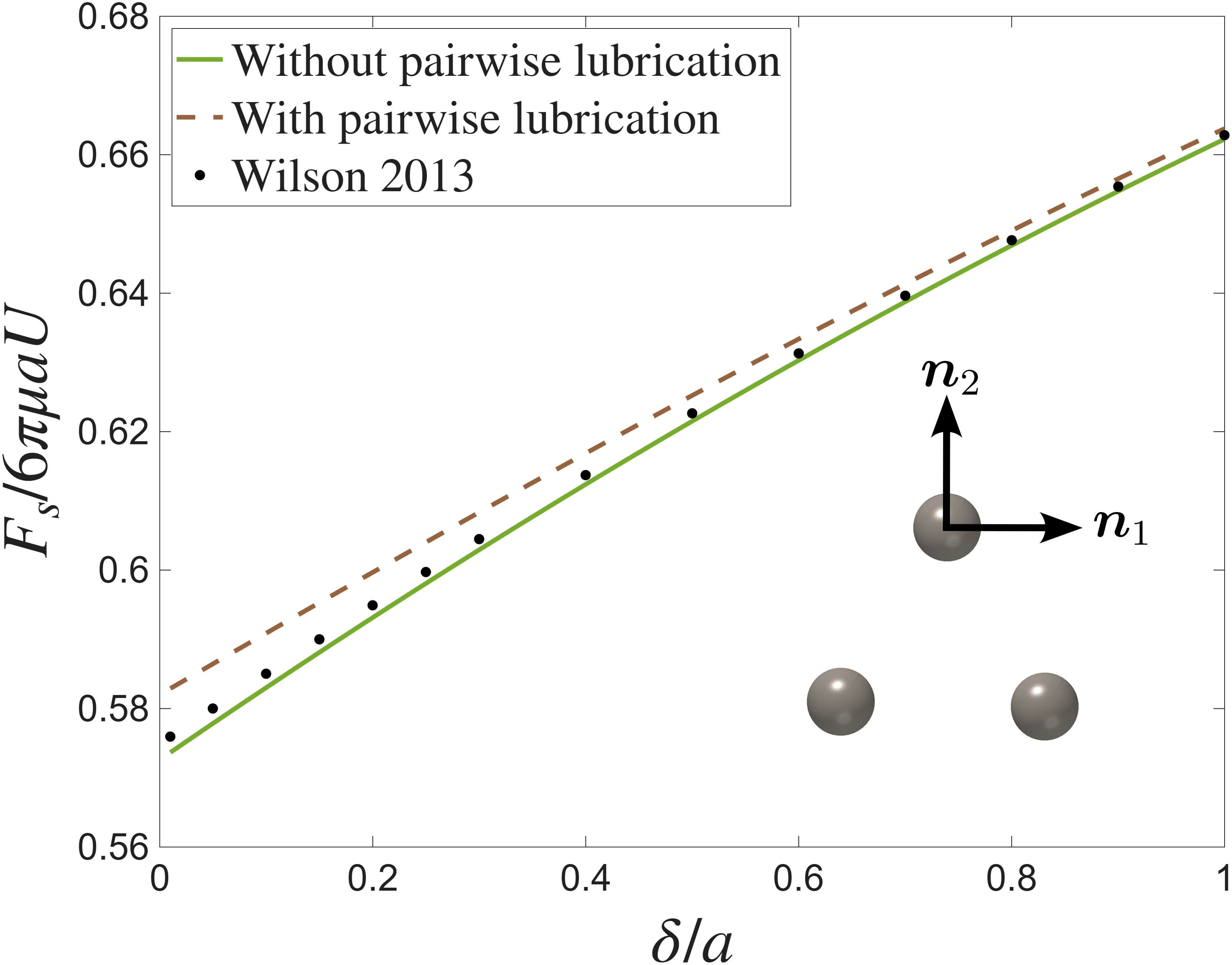} 
    \caption{
    Three beads are arranged in a triangle of side $r$ and move with the same speed $v$ in the direction $\bm{n}_3=\ve n_1 \wedge \ve n_2$, out of the plane. 
    Shown is the non-dimensional force exerted by one of the  beads upon the fluid as a function of the non-dimensional interfacial distance. The two lines are computed using Stokesian dynamics \cite{durlofsky_dynamic_1987}, without pairwise lubrication (solid line), and with pairwise lubrication using Eq.~\eqref{eq:lubrication} (dashed line). Black dots are numerical results obtained by~\citet{wilson_stokes_2013}.}
    \label{fig:triangle_lubrication}
\end{figure}

We consider three spheres of radius $a$, in close proximity, forming an equilateral triangle of side $r$.  
Fig.~\ref{fig:triangle_lubrication} shows the force exerted by each sphere on the fluid when they are moving perpendicular to their plane of centres with a velocity $\bm{v} = v\,\bm{n}_3$. 
We compare the Stokesian-dynamics results with numerical results obtained by \citet{wilson_stokes_2013}, using a method specifically designed to accurately resolve the hydrodynamic interactions of three closely spaced spheres. Fig. \ref{fig:triangle_lubrication} shows that 
 higher accuracy is 
consistently
achieved without lubrication corrections for all interfacial distances.
For the three-sphere configuration considered here, however, the discrepancies remain moderate, and both approaches provide acceptable results for practical purposes. 
The situation changes markedly for more densely packed geometries, where each sphere interacts with multiple close neighbours. 
In such cases, 
pairwise lubrication corrections can lead to substantial errors.
Avoiding these corrections is therefore desirable for both efficiency and robustness of the method.

\section{Description of the program package}
\label{sec:program}
In this Section, we explain how the program package \texttt{SHAPES} is structured and what it can do.  A tutorial on how to get started with the  most essential features of the packaged is found in Appendix \ref{appendix:tutorial}.

The method is implemented in an object-oriented \textsc{Matlab} package called \texttt{SHAPES}.
A comprehensive guide to object-oriented programming in \textsc{Matlab} is available in the  documentation provided by MathWorks \cite{matlab_oop}.
The  code is structured around building blocks called \emph{classes}. Each class defines a set of properties (i.e.,  data) and methods (i.e.,  functions that operate on these properties).
This architecture facilitates the construction, modification, and extension of particle models by organizing physical properties and numerical operations within well-defined classes.
Objects are instances of these classes with their own set of properties, they can call the methods defined by their class.
\texttt{SHAPES} supports simulation of bead aggregates in three-dimensional space, for an unbounded fluid, or in the presence of a planar wall. 
Aggregates can be placed in an externally imposed linear background flow, as defined in Eq.~\eqref{eq:Vflow_intro}. 
It should be noted that lubrication corrections are voluntarily omitted in the present implementation, as discussed in Section \ref{sec:lubrication}.

\begin{figure}[t]
\includegraphics[width=17.5cm]{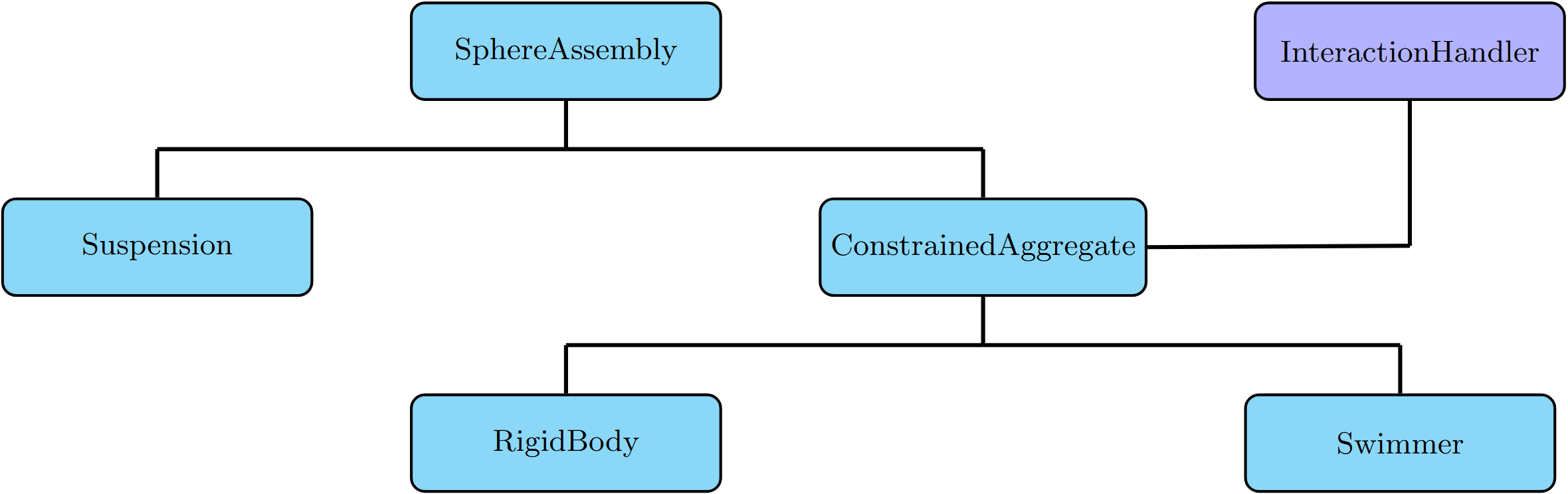} 
\caption{
Organisation chart of classes in \texttt{SHAPES} program package. The \texttt{Suspension} class is used to model hydrodynamic interactions between freely moving beads, as introduced by \citet{durlofsky_dynamic_1987}. In the present work, we show examples of \texttt{RigidBoby} and \texttt{Swimmer} objects. The \texttt{InteractionHandler} class models hydrodynamic interactions between multiple objects. 
}
\end{figure}

A single base class, \texttt{SphereAssembly}, 
encapsulates the geometric and kinematic description of a collection of spheres, together with generic functionalities 
such as visualization.
Within this class, all quantities derived from the $\mathbb{J}$, $\mathbb{R}$, and $\mathbb{K}$ tensors from Eq. \eqref{eq:flow1}
are computed (see Eqs. from Section \ref{sec:background}), including wall-effect tensors from Eq. (\ref{eq:wall_tensors}).
In particular, it constructs the full mobility and resistance matrices relating the $N$ spheres of the system [Eq. (\ref{eq:bigMobMat})].
All other classes extend this base structure through inheritance.

The \texttt{Suspension} class implements the classical Stokesian Dynamics formulation \cite{durlofsky_dynamic_1987} and is designed to compute the motion of freely moving spherical particles with different radii.
It provides the basic dynamical relations for unbounded suspensions of spheres without kinematic constraints.
The content of this class does not introduce new methodological developments and is not discussed in this paper, but is included to provide a complete and coherent implementation of the Stokesian Dynamics method.

In the \texttt{ConstrainedAggregate} class, the relative motion of the beads is prescribed.
The aggregate is treated as a single dynamical entity, with well-defined centre of mass and resistance and mobility tensors.
Centre of resistance and centre of mobility can also be computed.
Resistance tensors are calculated using Eqs.~(\ref{eq:ABGtensors}--\ref{eq:GHMtensors}) and are valid for both rigid and deformable aggregates (provided that their deformation is known). 
The beads are therefore no longer treated as independent components. 
There are currently two subclasses of \texttt{ConstrainedAggregate}: the \texttt{RigidBody} subclass and the \texttt{Swimmer} subclass. 
The main difference between them is the way the aggregates move.

The \texttt{RigidBody} subclass describes rigid aggregates: bead velocities follow the rigid-body kinematic relations given in Eqs.~\eqref{eq:velocity_rigid_body}.
The \texttt{Swimmer} subclass, by contrast, is designed to model active particles propelled by prescribed internal motions, such as flagellar beating. Relative velocities are imposed according to the swimmer kinematics [Eqs.~\eqref{eq:swimmer_velocities}].
The distinction between these subclasses therefore lies in how the relative velocities of the beads are specified.

Finally, the \texttt{InteractionHandler} class 
allows to describe multiple interacting aggregates, which can be \texttt{RigidBody} or \texttt{Swimmer} types.
With this, one can compute the resistance tensors of the multi-particle system  using Eq.~\eqref{eq:system}.
Dynamic simulations of such systems are also enabled. 
The time evolution of the system is computed using the Runge–Kutta–Fehlberg scheme, and built-in visualization tools allow both static rendering of aggregate geometries and generation of animations illustrating their dynamics.

\section{Examples}
\label{sec:examples}

We now illustrate the range of applications of the \texttt{SHAPES} package, 
highlighting how \texttt{SHAPES} may help to solve research questions regarding the dynamics of aggregates in Stokes flows.
We compare results for different aggregate shapes  
to earlier theoretical results and/or experimental data from the literature.
 We begin with the dumbbell [Fig.~\ref{fig:particle_summary}({\bf a})], a standard model for studying lubrication effects. We then consider rods [Fig.~\ref{fig:particle_summary}({\bf b})], commonly described using slender-body theory, before turning to curved slender geometries [Fig.~\ref{fig:particle_summary}({\bf c})]. 
Next, we apply \texttt{SHAPES} to compute the resistance tensors $\ma  G$ and $\ma H$
in shear flow, for a chiral dipole [Fig.~\ref{fig:particle_summary}({\bf d})]
and for an aggregate made out of a spherical head and a helical flagella [Fig.~\ref{fig:particle_summary}({\bf e})].
Then we examine hydrodynamic interactions between rigid aggregates.  Finally we consider an active aggregate (puller) 
[Fig.~\ref{fig:particle_summary}({\bf f})]. We determine how its gait affects the disturbance flow, and how it interacts with a wall.

\begin{figure}[t]
    \centering
    \includegraphics[width=\textwidth - 3.7cm]{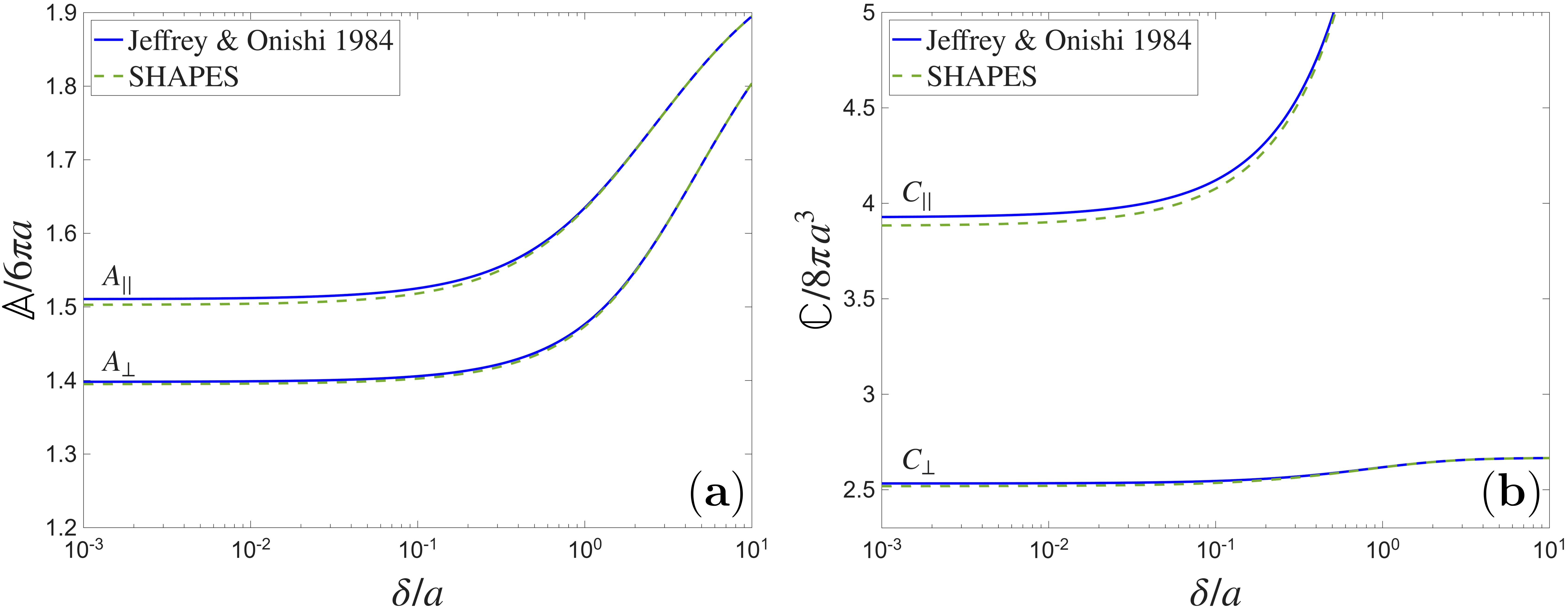}
   \caption{
    Resistance tensors $\mathbb{A}$ and $\mathbb{C}$ of a dumbbell made of two equal beads of radius $a$, as functions of the  non-dimensional interfacial distance between the two beads. Solid curves are exact analytical results derived from the exact results of \citet{jeffrey_calculation_1984}.
    }
    \label{fig:dumbbell_tensors}
\end{figure}

\subsection{Dumbbell}
As a simple test case, we apply the method introduced in Section~\ref{sec:method_single_particle} to compute the resistance tensors of a dumbbell, an aggregate consisting of two equal beads of radius $a$, as shown in Fig.~\ref{fig:particle_summary}({\bf a}). The distance between the bead centres is $r$, so that the minimum surface-to-surface (interfacial) distance is $\delta=r-2a$. Fig.~\ref{fig:dumbbell_tensors} compares the resistance tensors obtained when the grand resistance matrix is computed either exactly \cite{jeffrey_calculation_1984} or using \texttt{SHAPES}. The results show that \texttt{SHAPES} accurately reproduces the exact resistance tensors for all values of $\delta/a$ considered.

The good agreement is noteworthy because the exact calculation contains lubrication corrections in individual elements of the grand resistance matrix, as seen in Fig.~\ref{fig:lubrication}. 
This result indicates that lubrication corrections of opposite sign cancel when the resistance tensors of the aggregate are computed
using Eqs.~(\ref{eq:ABGtensors}--\ref{eq:GHMtensors}).
As a result, large pairwise lubrication corrections have only a weak effect on the resistance tensors of the  aggregate. This explains why the tensors obtained from the exact solution and from \texttt{SHAPES} differ only minimally. The results described in this Section therefore support the conclusions of Section~\ref{sec:lubrication}: explicit lubrication corrections are unnecessary for accurately determining the resistance tensors of rigid aggregates. 

\begin{figure}[p]
    \centering
    \includegraphics[width=\textwidth - 3.8cm]{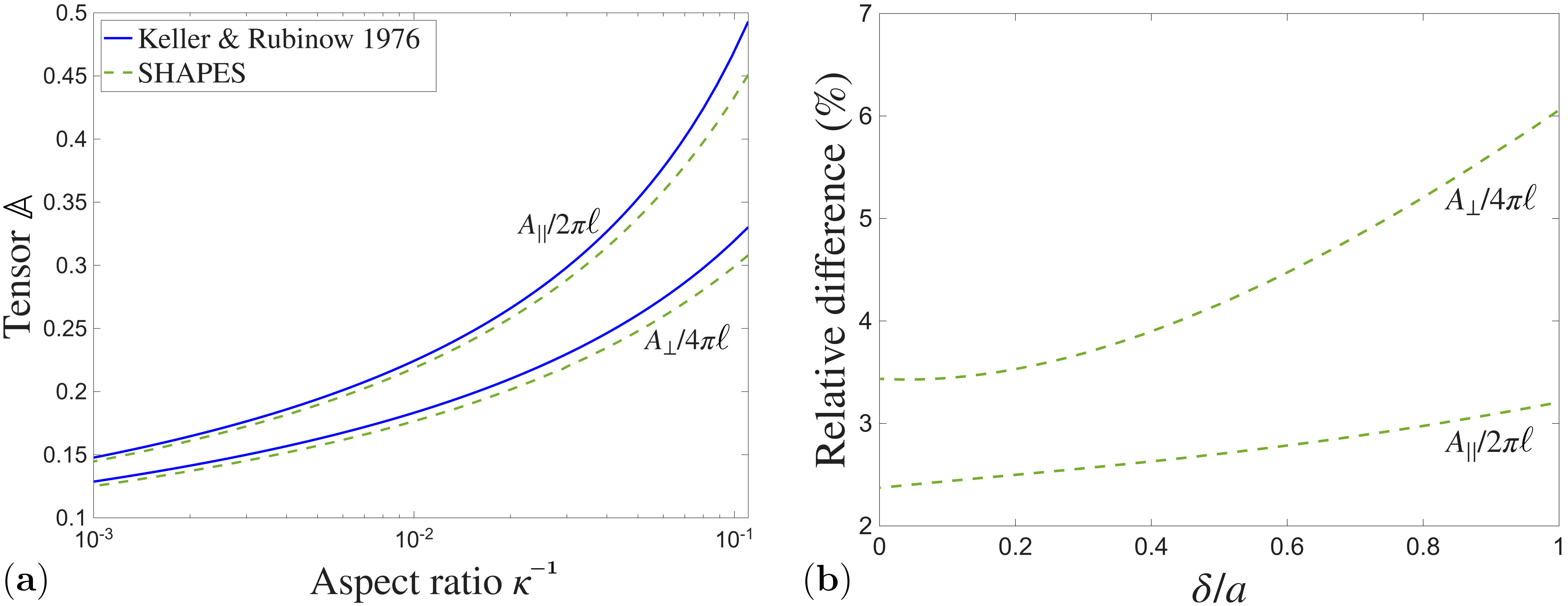}
    \caption{(\textbf{a}) Resistance tensor $\mathbb{A}$ for a cylinder of  aspect ratio $\kappa = \ell/2a$. Here, $A_\parallel = A_{33}$ and $A_\perp = A_{11} = A_{22}$ in the coordinate system shown in   Fig.~\ref{fig:particle_summary}({\bf b}).  
    Results from \texttt{SHAPES} closely match slender-body theory \cite{keller_slender_1976}. 
    The interfacial distance between neighbouring beads in the chain is
    $\delta = 0.3$.
    (\textbf{b}) Relative difference of \texttt{SHAPES} with respect to slender-body theory \cite{keller_slender_1976} 
    as a function of  bead spacing 
    for a rod with aspect ratio $\kappa = 100$.
    For larger values of $\delta/a$,  fewer beads are needed to form the straight chain. As a consequence, the elements of the drag tensors are smaller. This leads to larger differences 
    when compared with slender-body theory.
    ~~~~~~~~~~~~~~~~~~~~~~~~~~~~~~~~~~~~~~~~~~~~~~~~~~~~~~~~~~~~~~~~~~~~~~~~~~~~~~~~~~~~~~~~~~~~~~~~~~~~~~~~~~~~~~~~~~~~~~~~
    }
    \label{fig:Acylinder}
\end{figure}

\begin{figure}
    \centering
    \includegraphics[width=\textwidth - 3.8cm]{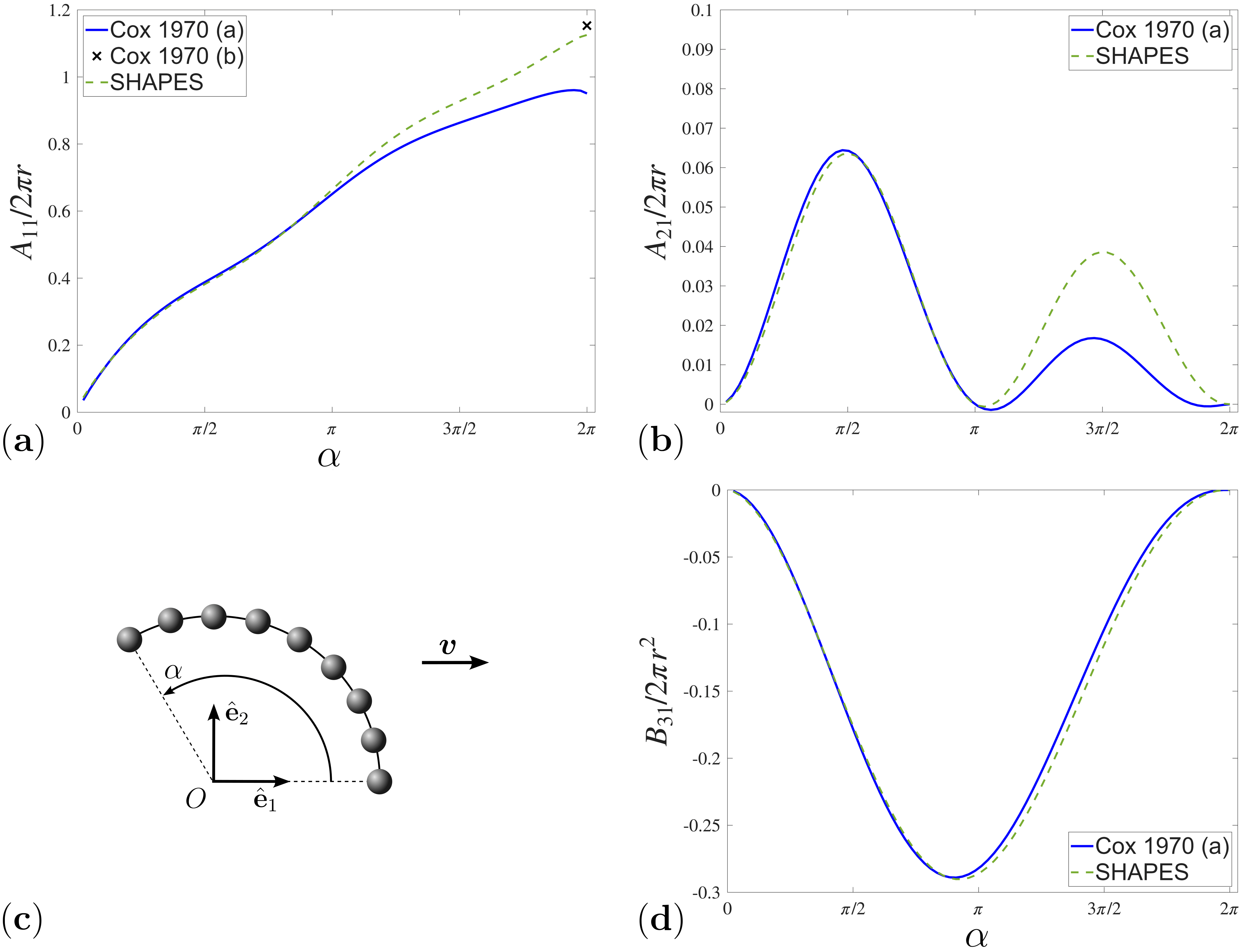}
    \caption{
    Non-dimensional elements of resistance tensors $\mathbb{A}$ and $\mathbb{B}$ for a curved fibre translating with velocity $\bm{v}$. The body length $\ell$ is set according to an angle $\alpha = \ell/R$, where $R$ is the radius of curvature of the body. Here, as an exception, 
    resistance tensors are computed from the origin $O$ of the laboratory frame.
    Blue lines [Cox 1970 (a)] are computed following Cox's general theory \cite{cox_motion_1970} while the black cross labeled Cox 1970 (b) in panel (\textbf{a}) is obtained from the special case of a  ring corresponding to $\alpha = 2\pi$,  see Eq.~(7.36)~in Ref. \cite{cox_motion_1970}.
    ~~~~~~~~~~~~~~~~~~~~~~~~~~~~~~~~~~~~~~~~~~~~~~~~~~~~~~~~~~~~~~~~~~~~~~~~~~~~~~~~~~~~~~~~~~~~~~~~~~~~~~~~~~~~~~~~~~~~~~~~~~~~~~~~~~~~~~
    }
\label{fig:Fcurved_rod}
\end{figure}

\subsection{Rod}
We approximate a cylindrical rod of length $\ell$ and radius $a$ by a straight chain of beads [Fig.~\ref{fig:particle_summary}({\bf b})]. Fig.~\ref{fig:Acylinder}(\textbf{a}) shows the resistance coefficients obtained using \texttt{SHAPES} for chains of different aspect ratios. 
To validate the numerical method, we compare these results with slender-body theory. \citet{cox_motion_1970} investigated the force acting on slender bodies in an unbounded fluid using an asymptotic expansion.
\citet{keller_slender_1976} extended this analysis to higher order: 
\begin{subequations}
    \begin{align}
        A_\parallel &= 2\pi\ell \left\{ \left[\ln{\frac{\ell}{a}} - \frac{3}{2} +\ln{2} - \left(1 - \dfrac{\pi^2}{12} \right)\Big/\ln{\frac{\ell}{a}} \right]^{-1} + \mathscr{O}\left( \frac{1}{(\ln{\ell/a})^4} \right) \right\} \:,
        \\
        A_\perp &= 4\pi\ell \left\{ \left[\ln{\frac{\ell}{a}} - \frac{1}{2} +\ln{2} - \left(1 - \dfrac{\pi^2}{12} \right)\Big/\ln{\frac{\ell}{a}} \right]^{-1} + \mathscr{O}\left( \frac{1}{(\ln{\ell/a})^4} \right) \right\} \:.
    \end{align}
\end{subequations} 
Fig.~\ref{fig:Acylinder}(\textbf{a}) demonstrates good agreement between these asymptotic predictions and the \texttt{SHAPES} results, although the difference increases with lower aspect ratio $\kappa = \ell/2a$. This reflects the asymptotic nature of the slender-body approximation, which becomes accurate only in the limit $\kappa\to\infty$. The increasing discrepancy therefore suggests limitations of the asymptotic approximation rather than a deterioration of the numerical method implemented in \texttt{SHAPES}.

A natural question is how the bead spacing along the rod influences the accuracy of \texttt{SHAPES}. 
For a fixed aspect ratio~$\kappa$, increasing the spacing between neighbouring beads reduces the total number of beads required. 
Fig.~\ref{fig:Acylinder}(\textbf{b}) shows the relative difference as a function of the interfacial distance $\delta$ between neighbouring beads for a chain with $\kappa=100$. 
Increasing the bead spacing increases the difference, while decreasing the spacing reduces it. 
Good agreement is maintained even when the interfacial distance between neighbouring beads is of the order of one bead radius ($\delta/a=1$). 
If the beads are placed too far apart, however, the aggregate no longer accurately represents the target geometry. 
Notably, the differences remain modest even for closely spaced beads, despite the absence of lubrication corrections.

\subsection{Planar curved fibre with constant radius of curvature 
}  
In the two previous examples, the aggregates are sufficiently 
symmetric to ensure that the tensor $\ma B$ vanishes \cite{happel_low_1983,sundberg2025fluid}. An example of a particle with a simple shape but exhibiting translation–rotation coupling is a curved planar fibre with constant radius of curvature \cite{candelier2024torques}.
As a model for this particle, we consider the aggregate shown in Fig.~\ref{fig:Fcurved_rod}({\bf c}) moving with a velocity $\bm{v} = v\,\hat{\bf{e}}_1$ in a fluid at rest. 
Eq. \eqref{eq_systeme_full} implies:
\begin{equation}
\refstepcounter{equation}
    F_i = A_{ij} v_j \quad \text{and} \quad T_i = B_{ij} v_j \:. \tag{\theequation a,b}
\end{equation}
Fig.~\ref{fig:Fcurved_rod} compares the tensor components $A_{11}$ (\textbf{a}), $A_{21}$ (\textbf{b}) and $B_{31}$ (\textbf{d}), obtained from Cox's asymptotic expansion \cite{cox_motion_1970} and from the \texttt{SHAPES} method, as a function of $\alpha$, i.e. as the arc length $\ell = R\alpha$ of the fibre increases. 
Note that, unlike tensor $\mathbb{A}$, the tensor $\mathbb{B}$ depends on the chosen centre. 
Here, all tensor were  computed from the origin $O$ of the laboratory coordinate system $\hat{\bf e}_i$.
We are doing this as an exception (instead of using the centre of mass as our reference) so that we can more easily compare our results with those of \citet{cox_motion_1970}.

Both methods agree well for $\alpha\leq\pi$, but progressively diverge for larger values of $\alpha$. This discrepancy is expected because Cox's slender-body theory neglects hydrodynamic interactions between different parts of the body. When $\alpha>\pi$, the arc bends back toward itself and different sections of the fibre come into close proximity. 
In this regime, the disturbance flow generated by one segment significantly modifies the local flow experienced by other segments. 
These interactions are explicitly included in \texttt{SHAPES}, but not in the asymptotic theory. 
We observe that tensor $\mathbb{B}$ seems to be less sensitive to these long-range hydrodynamic interactions than tensor $\mathbb{A}$.
Since Cox's approximation \cite{cox_motion_1970} is also less accurate than the higher-order slender-body theory \cite{keller_slender_1976} discussed in the previous Section, larger discrepancies are expected here.

\subsection{Aggregates
in shear flow
} 
\label{sec:shear}

\begin{figure}[b]
   \centering
   \includegraphics[width=0.3\linewidth]{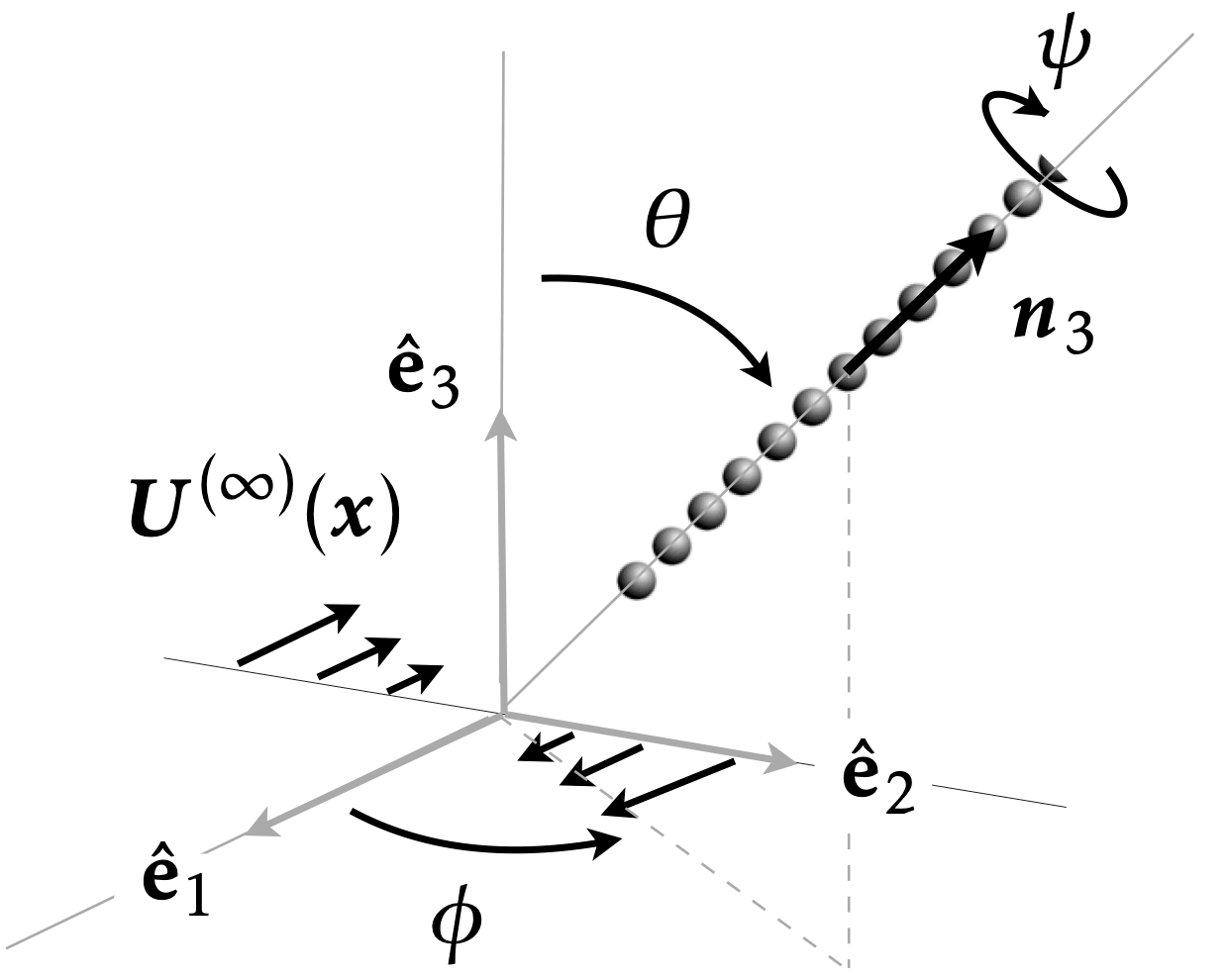}
   \caption{Linear chain in  shear flow. The flow–shear plane is spanned by $\hat{\bf e}_1$ and $\hat{\bf e}_2$. Vorticity is oriented along $-\hat{\bf e}_3$. The direction of the aggregate axis  $\ve n_3$ is parameterised by the angles $(\phi,\theta)$, corresponding to the $ZYZ$ Euler-angle convention.}
   \label{fig_Euler_angle}
\end{figure}
Up to this point, the examples concerned the tensors $\ma A$, $\ma B$ and $\ma C$ for aggregate particles in a uniform flow.
Now we turn to particles in shear flow [Fig.~\ref{fig_Euler_angle}]. We have:
\begin{equation}
\label{eq:simpleshear} 
\ve U^{(\infty)} = \dot{\gamma} \,x_2\, \hat{\bf e}_1, \quad \mbox{ so that } \quad  \ma E^{(\infty)} = 
\frac{\dot{\gamma}}{2} ( \hat{\bf e}_1 \otimes \hat{\bf e}_2 + \hat{\bf  e}_2 \otimes \hat{\bf  e}_1) \quad \mbox{ and } \quad \ve \Omega^{(\infty)} =   -\frac{\dot{\gamma}}{2} \hat{\bf  e}_3\,.
\end{equation}
The coupling of the shear $\ma E^{(\infty)}$ to force and torque is
determined by the tensors $\widetilde{\ma H}$ and $\widetilde{\ma G}$ in Eq.~(\ref{eq:ABGtensors}). The dynamics of non-spherical particles and aggregates in shear flow is of interest in many scientific problems \cite{jeffery_motion_1922,Bretherton:1962,happel_low_1983,kim_microhydrodynamics_1991,kramel2016preferential,fries_angular_2017,ishimoto2020helicoidal,zoettl2023asymmetric,jing2020chirality}. In the Stokes regime, the dynamics of a neutrally buoyant aggregate particle
in  shear flow follows from 
Eq.~(\ref{eq:mobility}) upon setting
force and torque to zero. This leads to the equations of motion:
\begin{subequations}\label{eq:dynamics_bis}
\begin{align}
\dot{\ve x}  &= \ve v\,, &\quad \ve v &= \ve U^\infty +\mu^{-1} \ma T : \mathbb{E}^{(\infty)}\,, \tag{\theequation a,b} \label{eq:dynamics_ab} \\
\dot{\ve n}_\alpha &= \ve \omega \wedge \ve n_\alpha\,, &\quad
\boldsymbol{\omega} &= \boldsymbol{\Omega}^{(\infty)} + \mu^{-1}\ma L : \mathbb{E}^{(\infty)} \:, \tag{\theequation c,d} \label{eq:dynamics_cd}
\end{align}
\end{subequations}
for the aggregate centre and the angular dynamics around it. Eq.~(\ref{eq:mobility}) shows
that $\ma T = \ma a \cdot \widetilde{\ma G}+\widetilde{\ma b} \cdot\widetilde{\ma H}$, and 
$\ma L = \ma b \cdot\widetilde{\ma G}+{\ma c}  \cdot\widetilde{\ma H}$.
By definition, $\ma L$ and $\ma T$  are symmetric and traceless with respect to their second and third indices. 
We parameterise the angular dynamics as in Ref.~\cite{einarsson2016tumbling}, denoting 
 by $\phi$ the angle between $\hat{\bf e}_1$ and the orthogonal projection of the vector $\ve n_3$ onto the horizontal plane
 spanned by $\hat{\bf e}_1$ and $\hat{\bf  e}_2$, and by $\theta$ the angle between $\hat {\bf e}_3$ and $\ve n_3$.  Finally, $\psi$ denotes the spin angle (Figure~\ref{fig_Euler_angle}).  These choices correspond to the $ZYZ$ Euler-angle convention. 
With this parameterisation, 
the angular equation of motion takes the form: 
\begin{subequations}\label{eq:angular_motion_shear_flow}
\begin{align}
\dot{\phi} &= -\frac{\dot{\gamma}}{2}+ \dot{\gamma} L_{231}\cos(2\phi)
- \dot{\gamma} L_{232}\sin(2\phi)\cos (\theta) + \dot\gamma  F_\phi(\phi,\theta,\psi;\ma L) \,,\label{eq_phi_dot} \\
\dot{\theta} &= - \frac{\dot{\gamma}}{2}L_{132}\sin(2\theta)\sin(2\phi)+\dot{\gamma} L_{131}\sin\theta\cos (2\phi)+ \dot\gamma F_\theta(\phi,\theta,\psi;\ma L)
\,,\label{eq_theta_dot}\\
\dot{\psi} &=
-\dot\gamma\Bigg[
(L_{231} + L_{321})\cos(2\phi)\cos(\theta)
\nonumber \\
&\hspace{2.5cm}
- \Big(
\frac{L_{311} + 2L_{322} + 2L_{232}}{4}\cos(2\theta)
+ \frac{L_{232}}{2}
- \frac{3L_{311}}{4}
\Big)\sin(2\phi)
\Bigg] +\dot\gamma  F_\psi(\phi,\theta,\psi;\ma L). \label{eq_psi_dot}
\end{align}
\end{subequations}
The terms $F_\phi$ and  $F_\theta$
couple the $\phi$-$\theta$ dynamics of $\ve n_3$ to the spin
around the $\ve n_3$-axis. They are given in 
Appendix~\ref{app:coupling_terms}, together with $F_\psi$. Even if one neglects the spin, the equations of motion for $\phi$ and $\theta$ are quite complicated, with many parameters determined by particle shape. By contrast, the angular dynamics of chiral and non-chiral aggregates is often described using simplified models involving only a few parameters \cite{kramel2016preferential,fries_angular_2017,zoettl2023asymmetric,jing2020chirality}. 
 
 \subsubsection{Jeffery orbits}
 For axisymmetric aggregate particles with fore-aft symmetry, Eq.~(\ref{eq:angular_motion_shear_flow}) simplifes~\cite{jeffery_motion_1922}. 
As an illustration, consider a linear chain composed of 13 beads of radius $a$, uniformly distributed over a length $L = 30a$ (Figure~ \ref{fig_Euler_angle}).
Using \texttt{SHAPES}, one finds
that all elements of $\ma T$ vanish, and the only non-zero elements of $\ma L$ are:
\begin{equation}
    L_{132} = -L_{231} = -0.4939\,.
\end{equation}
This follows
from particle-shape symmetry~\cite{sundberg2025fluid} (prismatic symmetry $D_{nh}$ with $n > 2$ ) which constrains $\ma T$
to vanish, and $\ma L$ to take the simple form 
$\ma L = \mathbb{c}\cdot \widetilde{\mathbb{H}}$ \cite{jeffery_motion_1922}.
Particle-shape symmetry further implies that the only non-zero elements $L_{ijk}$ 
are those with three distinct indices.  Moreover, since $\ma L$ is symmetric with respect to its last two indices, one has 
$L_{132} = L_{123} = -L_{231} = -L_{213}$, and $
L_{312} = -L_{321} = 0 $. In  other words, 
Eqs.~(\ref{eq_phi_dot}) and (\ref{eq_theta_dot}) simplify to Jeffery's angular equation of motion for spheroids, in the form of Eq.~(42) in \citet{einarsson2015rotation}.

In particular, the coupling to the spin vanishes, and the angular dynamics consists of marginally stable periodic orbits (the Jeffery orbits).
The shape parameter in Jeffery's equation -- Bretherton's parameter $\Lambda$ \cite{Bretherton:1962} -- is related to $L_{132}$
by $\Lambda = 2 L_{231}$. \texttt{SHAPES} simulations
show that $L_{231} = -L_{132} \to {1}/{2}$
as ${\ell}/{a} \to \infty$, as expected
for an infinitely slender rod, 
where $\Lambda \to 1$.
We note that
coupling to spin does not vanish for aggregates with
less shape symmetry, for example for ellipsoids with $D_{2h}$ symmetry. In this case,
the phase-space of the angular dynamics is three-dimensional, which allows for chaotic tumbling  \cite{hinch1979rotation,yarin1997chaotic,einarsson2016tumbling}.

\subsubsection{Chiral dipole}
\label{sec:cds}
The chiral dipole in
Fig.~\ref{fig:particle_summary}({\bf d})
consists of two helical parts with opposite handedness arranged in such a way that the aggregate is symmetric under $\ve n_3 \to -\ve n_3$. 
As a consequence, the aggregate 
is not chiral. However, 
it has a non-zero chiral-dipole moment \cite{kramel2016preferential},
with chiral-dipole vector 
 along~$-\ve n_3$.
\citet{kramel2016preferential} measured the alignment and preferential spinning of chiral dipoles in turbulence. 

\texttt{SHAPES} allows computation of the 
tensors ($\ma L$ and $\ma T$) that determine alignment and spin in response to a given linear flow. \texttt{SHAPES} yields the following results for the elements of $\ma L$ for the chiral dipole shown in Fig.~\ref{fig:particle_summary}({\bf d}):
\begin{align}
\label{eq:cdL}
L_{1ij} &= 
{\footnotesize
 \begin{bmatrix}
0 & 0 & \;\;{ 0.0034} \\
0 & 0 & -0.4922 \\
\;\;{ 0.0034} & -0.4922 & 0
\end{bmatrix}}, \:
L_{2ij} = 
{\footnotesize \begin{bmatrix}
0 & 0 & \;\;0.4887 \\
0 & 0 & \;\;{ 0.0022} \\
\;\; 0.4887 & \;\;{ 0.0022} & 0
\end{bmatrix}}, \:
L_{3ij} = 
{\footnotesize \begin{bmatrix}
\;\;{ 0.0571} & \;\; 0.0338 & 0 \\
\;\;0.0338 & \;\;{ 0.0757} & \;\;0 \\
0 & \;\;0 & {-0.1328}
\end{bmatrix}}\,.
\end{align}
The structure of $\ma L$ is explained
by the fore-aft symmetry of the chiral dipole under $\ve n_3 \mapsto -\ve n_3$ \cite{fries_angular_2017,ishimoto2020helicoidal,sundberg2025fluid}. One finds that the only non-zero elements $\ma L$ are those
where the indices contain an odd number of the index 3.  The chiral dipole vector is reversed by mirroring either $\ve n_1 \mapsto -\ve n_1$ or $\ve n_2 \mapsto -\ve n_2$. Denoting by $\alpha \in \{1,2\}$ the index associated with the reflected coordinate, the elements $L_{ijk}$ change sign if the number of indices equal to $\alpha$ is even. Those with an odd number of indices equal to $\alpha$ remain unchanged.  

\citet{kramel2016preferential} approximated the angular dynamics of a chiral dipole using only two parameters: the Bretherton parameter $\Lambda$ and 
a chiral spinning coeffcient $\beta$ [their Eq.~(5)]. 
As shown in the previous Section,  $\Lambda$
is determined by the elements $L_{132}$ and $L_{231}$ which remain unchanged upon reversing the chiral dipole vector.
The parameter
$\beta$ is determined by $L_{333}$ which couples extensional strain to spin.
Here we focus on dynamics of chiral dipoles in simple shear near the log rolling orientation, by contrast, 
 and the leading effects from the chiral dipole geometry come from $L_{131}$ and $L_{232}$.
Since $|L_{123}|\approx L_{231}$
and $L_{131} \approx L_{232}$,
we start by analysing the dynamics under the simplifying assumption:
\begin{equation}
\refstepcounter{equation}
L_{132} = -L_{231}=-\tfrac{1}{2}\Lambda  \quad \text{(Jeffery)}, \qquad
L_{131} = L_{232}= {\chi/2} \quad \text{(chiral-shear coupling strength)}\,.
\label{eq_model_equi}
\end{equation}
With this choice, the coupling between the $(\phi,\theta)$-dynamics and the spin degree of freedom vanishes, leading to:
\begin{subequations}
\label{eq:eomsimple}
\begin{align}
\dot{\phi} &= \tfrac12\bigl(\Lambda
 \cos(2\phi) - 1\bigr) - \tfrac{\chi}{{2}}
 \sin(2\phi)\cos\theta, \label{eq_phi_dot_2} \\
\dot{\theta} &= \tfrac{\Lambda}{4}\sin(2\theta)\sin(2\phi) + \tfrac{\chi}{{2}} \sin\theta \cos(2\phi). \label{eq_theta_dot_2}
\end{align}
\end{subequations}
Here we have set the shear rate to unity, $\dot\gamma =1$.
Eq.~\eqref{eq_theta_dot_2} admits
two steady orientations, namely
alignment with the vorticity direction ($\theta=\pi$) and opposite
to vorticity ($\theta=0$). In both cases, the particle
continues to spin (log-rolling orbit). 
Starting from Eq.~(45) in Ref.~\cite{einarsson2015rotation}, one can compute the Lyapunov exponent of the log-rolling orbit; it turns out that it vanishes.  We contrast 
this with the fluid-inertia perturbation (non-zero shear Reynolds number) considered by \citet{einarsson2015rotation}. While the latter breaks the marginal stability of the log-rolling orbit, the chiral-dipole  perturbation does not.  The latter twists the Jeffery orbit without causing an orbit drift, while the former breaks the Jeffery orbits (except log rolling and tumbling in the flow-shear plane), resulting in 
non-zero orbit drift.
\begin{figure}[t]
\centering
   \includegraphics[width=\textwidth - 3.8cm]{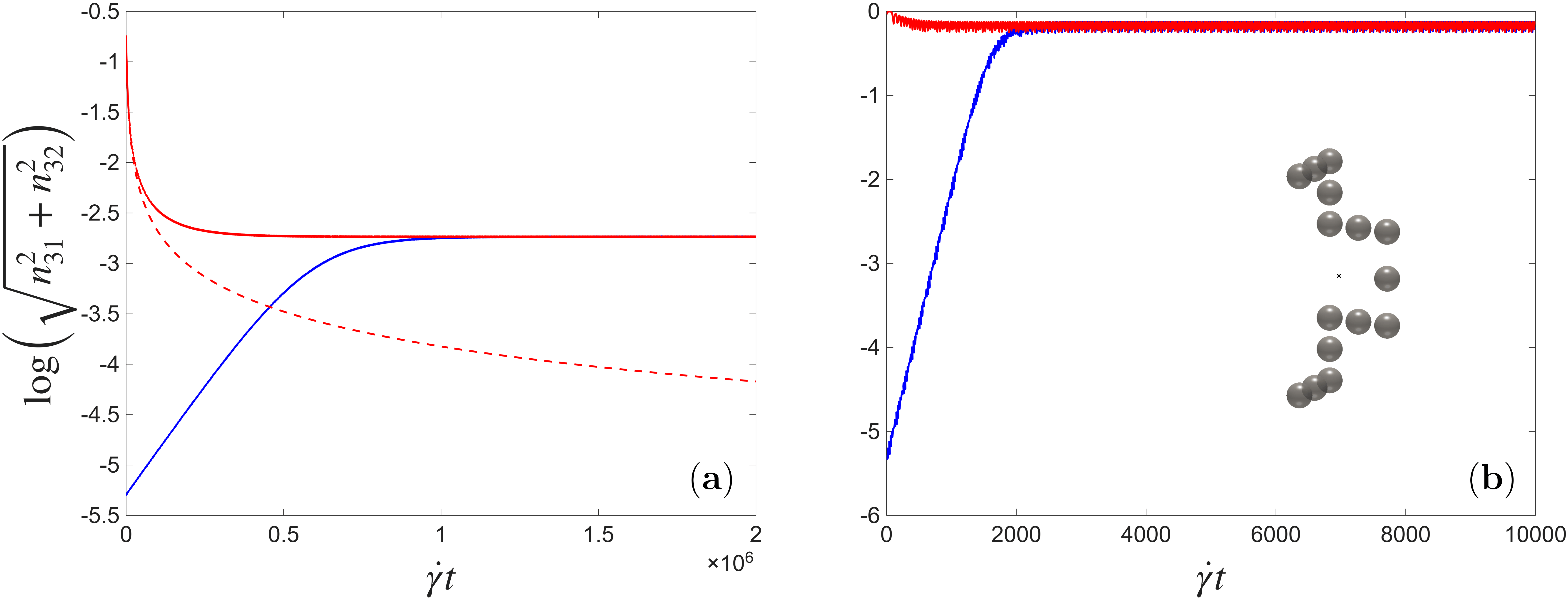}
\caption{\label{fig:chiraldipole}
Alignment of chiral dipole with vorticity in shear flow. ({\bf a}) Angular dynamics of the chiral
dipole shown in Fig.~\ref{fig:particle_summary}({\bf d}) near the log-rolling orbit $\theta=\pi$. Shown is $\log\sqrt{n_{31}^2+n_{32}^2}$ versus time ($n_{3i}=\ve n_3 \cdot \hat{\bf e}_i$) for the parameter values  from Eq.~(\ref{eq:cdL}), 
for two initial conditions (red and blue solid lines) converging to the same plateau. This indicates that there is stable (quasi-periodic) motion around $\theta=\pi$. 
The angular dynamics of the simplified model (dashed line)  slowly converges to $(0,\pi)$, confirming non-linear stability of this steady state of the simplified model.
({\bf b}) Angular dynamics near $(0,\pi)$ for the chiral dipole shown in the panel. The dynamics converges to stable (quasi-periodic) motion, but far away from log rolling.}
\end{figure}
Since log rolling has vanishing Lyapunov exponent, non-linear corrections near (anti-)alignment determine the stability. \citet{ishimoto2020helicoidal} calculated their effect for Eq.~(\ref{eq:eomsimple}).
His results imply that the dynamics converges to alignment of $\ve n_3$ with vorticity, $\theta \to \pi$, albeit very slowly. Anti-alignment of $\ve n_3$ with vorticity, by contrast, is unstable. Stability switches from anti-parallel to parallel as $\chi \to -\chi$, and the dynamics near alignment is slow for $ |\chi|\ll 1$  and $\Lambda^2+\chi^2 < 1$. The calculations summarised in Appendix~\ref{app:staban} confirm these results.

Now compare these findings with 
the dynamics of the full model, keeping all elements
in Eq.~(\ref{eq:cdL}). 
Fig.~\ref{fig:chiraldipole}({\bf a}) shows 
that alignment with vorticity is linearly unstable now. Non-linear terms still lead to a drift towards alignment. That linear and non-linear stability differ in sign is not a contradiction, because they are determined by different elements of $\ma L$.
Writing the elements of $\ma L_1$ and $\ma L_2$ as:
\begin{align}
\label{eq:defcd}
\Lambda = L_{231}-L_{132}\,,\quad 
\delta\Lambda = \tfrac12(L_{231}+L_{132})\,,\quad
\chi = L_{232}+L_{131}\,,\quad 
\delta\chi = \tfrac12 (L_{232}-L_{131})\,,
\end{align}
we find in Appendix~\ref{app:staban1} that linear stability depends on $\delta\chi$
and $\delta \Lambda$ (while
non-linear stability depends on $\chi$ and $\Lambda$).

In the full model,  linear and non-linear stability have opposite effects, leading to quasi-periodic  angular dynamics near alignment. 
Fig.~\ref{fig:chiraldipole}({\bf a}) shows how the distance from alignment with vorticity ($\theta=\pi$) evolves as a function of time. We see that the deviation reaches a plateau, consistent with quasi-periodic motion that is transversally stable. 
Since the quasi-periodic motion is quite close to alignment, the dynamics of the full model is similar to the simplified one [Fig.~\ref{fig:chiraldipole}({\bf a})]. The behaviour near $(0,0)$ is opposite, the non-linear drift is outwards, but $(0,0)$ is linearly stable (not shown). This means that the full-chiral dipole model exhibits asymmetric bistability \cite{zoettl2023asymmetric}: trajectories started inside the torus around $(0,0)$ remain there {\em ad infinitum}, but those started outside converge to the torus around $(0,\pi)$.
Trajectories started near the torus around $(0,\pi)$ remain there, regardless of whether they are started inside or outside. The simplified model, by contrast, does not show bistability. In conclusion, one needs to consider all $\ma L$-elements to deduce the correct stability of the steady states. 

On the other hand, the simplified and full model are similar to each other in at least one respect: the angular dynamics stays near log rolling for a long time.
For the full-chiral dipole model,
this requires $\delta \chi \ll \chi$. Fig.~\ref{fig:chiraldipole}({\bf b}) shows a chiral dipole where $\delta \chi \gg \chi$. The aggregate has the same symmetry as the one shown in Fig.~\ref{fig:particle_summary}({\bf d}), but its dynamics are entirely different, because linear stability is much stronger than the non-linear instability. As a consequence, the quasi-periodic torus moves far away
from $(0,\pi)$, as seen in Fig.~\ref{fig:chiraldipole}({\bf b}).
In other words, this aggregate  does not align in any way with vorticity, although it has the same shape symmetry as the one shown in Fig.~\ref{fig:particle_summary}({\bf d}).
So particle-shape symmetry alone 
is not sufficient to infer the angular dynamics, emphasising the need
of a tool like \texttt{SHAPES}. 
The fact that two different chiral parameters -- $\chi$ and $\delta\chi$ --
are needed to understand the angular dynamics (besides $\Lambda$ and $\delta\Lambda$) indicates that one needs the full $\ma L$ tensor to 
understand the angular dynamics, in general.

One may wonder why a small chiral-shear coupling causes alignment with vorticity. The reason is that the Jeffery angular dynamics in simple shear is marginally stable, so any small perturbation potentially has a strong effect. Contrast this with  elongational flow
$  \ma E^{(\infty)} 
= 
-\tfrac{1}{2}
\left(
\hat{\bf e}_1 \otimes \hat{\bf e}_1
+
\hat{\bf e}_2 \otimes \hat{\bf e}_2
\right)+\hat{\bf e}_3 \otimes \hat{\bf e}_3
$.
In this case, the angular equations of the simplified model read
$
    \dot{\phi}
=
\tfrac{3}{2}\chi\cos\theta$, 
$
\dot{\theta}
=
-\tfrac{3}{4}\Lambda\sin(2\theta)$,
and $\dot{\psi}
=\tfrac{3}{8} \beta + \tfrac{9}{8} \beta \cos(2 \theta)
-\tfrac{3}{2} \chi  \cos^2(\theta)$,
with the additional parameter $\beta = L_{333}$, setting $L_{311}=L_{322} = - \beta/2$ and all other elements of $\ma L_3$ to zero (we remind the reader
that the parameter $\beta$ does not matter in simple shear,
because the aggregate aligns parallel to $\hat{\bf e}_3$, the vorticity direction, and $E^{(\infty)}_{33}=0$). 
Now  $\ve n_3$ aligns with 
the extensional direction $\pm \hat{\bf e}_3$ with stability exponent $-\tfrac{3}{2} \Lambda$, independent of $\chi$. The parameters $\chi$ and $\beta$ determine the spinning rate when aligned. In the full model,  $L_{311}$ and $L_{322}$, and $L_{321}$ also add contributions to the spinning rate
when $\theta\neq 0$ or $\pi$. They affect the spinning rate, but they do not change the stability exponent.  
In summary, in an elongational flow, a weak chiral dipole affects only spinning, but not alignment. 

To conclude, we discuss the centre of mass dynamics of the chiral dipole in simple shear, determined by the tensor $\ma T$. \texttt{SHAPES} yields:
\begin{align} 
\label{eq:cdT}
{T}_{1ij}& = 
{\footnotesize \begin{bmatrix}
\;\;0.0190 & -0.0016 & 0 \\
-0.0016 & \;\;0.0115 & 0 \\
0 & 0 & -0.0305
\end{bmatrix}}, \:
{T}_{2ij} = 
{\footnotesize \begin{bmatrix}
-0.0073 & \;\;0.0030 & 0 \\
\;\;0.0030 & -0.0063 & 0 \\
0 & 0 & \;\;0.0136
\end{bmatrix}}, \:
{T}_{3ij} = 
{\footnotesize \begin{bmatrix}
0& 0 & \;\;0.0067 \\
0 & 0 & -0.0039 \\
\;\;0.0067 & -0.0039 & \;\;0
\end{bmatrix}}\,.
\end{align}
Particle-shape symmetry explains the structure:  the only non-zero components of $\ma T$ are those whose indices contain an even number of 3 (including zero occurrences). 
It follows that in simple shear, a chiral dipole aligned with vorticity does not experience a drift
in the $\pm\hat{\bf e}_3$-direction, because
$T_{311}= T_{322}=T_{312}=T_{321}=0$. 

\subsubsection{Spherical head with chiral flagella}
\label{sec:zoettls}
In the previous Section, we discussed different models and aggregates with the same particle-shape symmetry, but very different angular dynamics. 
Now we consider an aggregate without
any point-group shape symmetry [Fig.~\ref{fig:particle_summary}({\bf e})], and show that its dynamics is similar to that of the chiral dipole, despite the different symmetry.
\citet{zoettl2023asymmetric} 
analysed the angular
dynamics of aggregates made
out of a large spherical head attached to a helical flagella, and reported alignment with vorticity and asymmetric bistability of the angular dynamics in their model simulations and experiments. To describe the angular dynamics of their particle, \citet{zoettl2023asymmetric} used a simplified model with only two parameters. Their model is equivalent to the simplified chiral-dipole model discussed in the previous Section, which does not exhibit true bistability.

To understand the angular dynamics of the aggregate shown in Fig.~\ref{fig:particle_summary}({\bf e}), we use 
\texttt{SHAPES} to compute the mobility tensors.
We denote the radius of the spherical head by
$a$, and model the helical flagella as a chain of smaller beads of radius $a/10$, arranged along a circular helix of radius $a$ and of total length $10a$, corresponding to three full turns.
\texttt{SHAPES} yields the following elements of the tensor $\ma L$:
\begin{equation}
\label{eq:zoettl}
\begin{aligned}
L_{1ij} = 
{\footnotesize
\begin{bmatrix}
\phantom- 0.0020 & -0.0049 & \phantom- 0.0092 \\
-0.0049 & \phantom- 0.0250 & -0.4616 \\
\phantom- 0.0092 & -0.4616 & -0.0270
\end{bmatrix}}\,, \:
L_{2ij} = 
\footnotesize{\begin{bmatrix}
\phantom- 0.0176 & -0.0122 & \phantom- 0.4573 \\
-0.0122 & \phantom- 0.0062 & \phantom- 0.0098 \\
\phantom- 0.4573 & \phantom- 0.0098 & -0.0238
\end{bmatrix}}\,, \:
L_{3ij} = 
{\footnotesize\begin{bmatrix}
0.0596 & 0.0100 & \phantom- 0.0136 \\
0.0100 & 0.0646 & \phantom- 0.0011 \\
0.0136 & 0.0011 & -0.1242
\end{bmatrix}}\,.
\end{aligned}
\end{equation}
 If one keeps only:
\begin{equation}
\label{eq:zchi}
    L_{132} =  -L_{231}= -\tfrac{1}{2}\Lambda\quad \mbox{(Jeffery)}\,,\quad L_{131}=L_{323} = \tfrac{1}{2}\chi \quad\mbox{(chiral shear coupling strength)}\,,
\end{equation}
one obtains the simplified model 
discussed in the previous Section.
The analysis of this model showed that only one of the steady states of the angular dynamics is stable, the other unstable.
  
\begin{figure}
\includegraphics[width=0.35\linewidth]{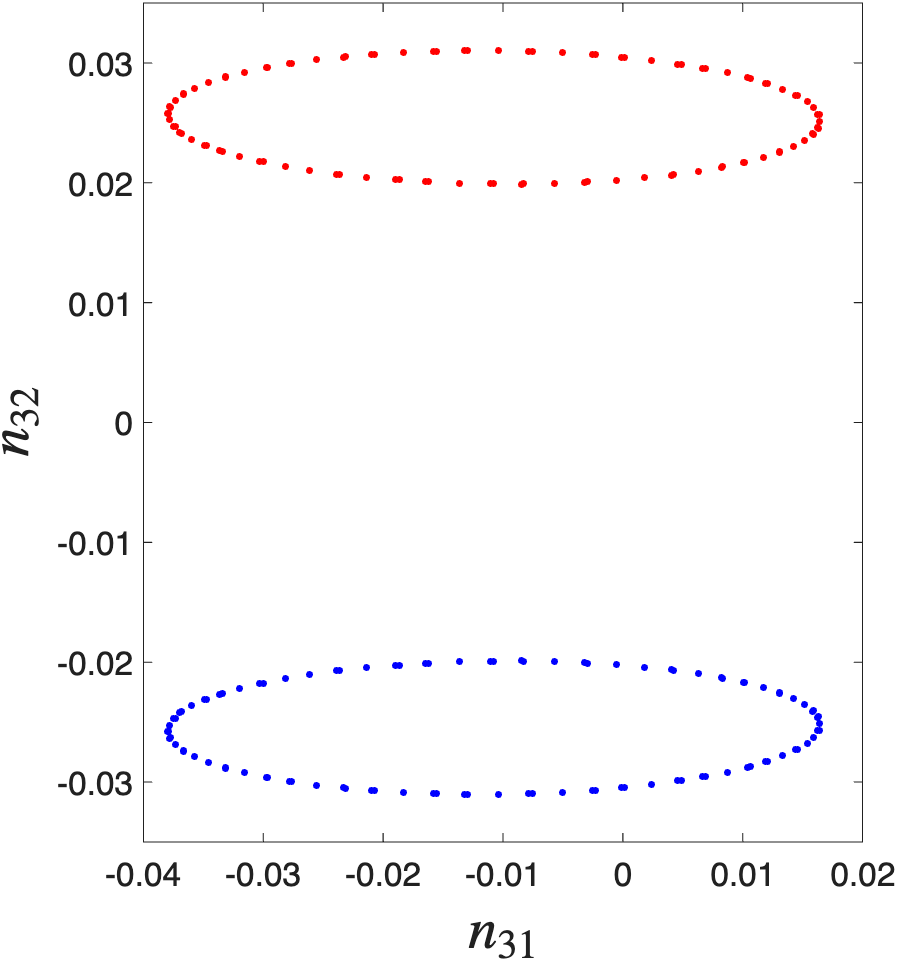}
\caption{\label{fig:z} 
Quasi-periodic tori near log rolling for the aggregate shown in
Fig.~\ref{fig:particle_summary}({\bf e}).
Shown are  Poincar\'e surface-of-sections in the $n_{31}-n_{32}$
plane ($n_{3i}=\ve n_3 \cdot \hat{\bf e}_i$), obtained for spin angles
$\psi(t) = 2\pi k$ for $k \in \ma{N}$ (see Appendix~\ref{app:staban} for the definition of $\psi$). Blue points show the transversally stable torus near the south pole whereas red points show 
the transversally unstable torus near the north pole.}
\end{figure}

Now consider the effect of the non-zero elements of $\ma L$ in Eq.~(\ref{eq:zoettl}). Compared with Eq.~(\ref{eq:cdL}) for the chiral dipole, there are more non-zero elements in (\ref{eq:zoettl}), because 
the aggregate does not have fore-aft symmetry. Moreover, 
the chiral coupling $\chi $ is of the same order as the other non-Jeffery elements in $\ma L$. So the latter cannot be neglected. In particular,  coupling to spin matters. 
The first consequence of these observations is that alignment with vorticity is no longer a steady state. 
Instead, the dynamics close to alignment and anti-alignment is quasi periodic (close because the disturbing elements of $\ma L$ are small).
 Numerical simulations of the full model with Eq.~(\ref{eq:zoettl}) show that the quasi-periodic motion near log rolling is similar to that of the full-chiral dipole model: there is a stable quasi-periodic torus near $(0,\pi)$ and an unstable one near $(0,0)$. The latter was hard to find using numerical simulations, so we reversed the time to find it. 
The result is shown in Fig.~\ref{fig:z} which illustrates quasi-periodic angular dynamics near log rolling.
The figure depicts the quasi-periodic tori near the north and the south pole. 
We note that the tori are reflected in
$n_{32}$; this follows from
the symmetry of the equations of motion for the simple shear (\ref{eq:simpleshear}) under  $[n_{31},n_{32},n_{33}]\to[n_{31},-n_{32},-n_{33}]$ and time-reversal symmetry. 
Although the tori are deformed compared with the chiral dipole,
and do not include the poles, they have the same linear stabilities as for the chiral dipole. Therefore, the model (\ref{eq:zoettl}) exhibits true bistability, just like the chiral dipole, but unlike the simplified model.

\citet{marcos2009separation} used resistive-force theory to show that alignment with vorticity leads to a drift in the vorticity direction for aggregates with helical shapes. This drift 
is determined by the elements
of the tensor $\ma T$ in Eq.~(\ref{eq:dynamics_bis}b).
For the aggregate from Fig.~\ref{fig:particle_summary}({\bf e}), 
\texttt{SHAPES} yields:
\begin{equation}
\begin{aligned}
{T}_{1ij} = 
{\footnotesize\begin{bmatrix}
\phantom- 0.0398 & -0.0318 & -0.1530 \\
-0.0318 & \phantom- 0.0060 & -0.0278 \\
-0.1530 &-0.0278 & -0.0458
\end{bmatrix}}\,, \:
{T}_{2ij} = 
{\footnotesize \begin{bmatrix}
0.0067 & \phantom- 0.0127 & \phantom- 0.0333 \\
0.0127 & -0.0606 & -0.1434 \\
0.0333 & -0.1434 & \phantom- 0.0539
\end{bmatrix}}\,, \:
{T}_{3ij} = 
{\footnotesize\begin{bmatrix}
\phantom- 1.3043 & 0.0012 & -0.0070 \\
\phantom- 0.0012 & 1.2966 & \phantom- 0.0320 \\
-0.0070 & 0.0320 & -2.6009
\end{bmatrix}}\,.
\end{aligned}
\end{equation}
Now all elements of $\ma T$ are  non-zero.
This produces an oscillatory drift if one
keeps the aggregate perfectly (anti-) aligned with vorticity. Since the spin is not perfectly uniform, 
this is expected to cause a net centre of mass drift when averaging over $\psi$. An additional contribution to the drift is expected because the aggregate is not perfectly aligned, as discussed above (see Fig.~\ref{fig:z}). 
\texttt{SHAPES} explains why
the drift velocity is small, simply because the relevant elements of $\ma L$ and $\ma T$ are small, since the helical aggregate is slender. A  practical consequence is that the aggregates need to travel a long way in a channel flow in order to separate them 
by this mechanism
\cite{marcos2009separation,eichhorn2010microfluidic,meinhardt2012separation,bogunovic2012chiral,marcos_analysis_2014}.
In principle, \texttt{SHAPES} could be used to search for aggregate shapes which maximize the drift. Along a different line of investigation, 
\citet{jing2020chirality} considered helical swimmers
essentially like the aggregate shown in Fig.~\ref{fig:particle_summary}({\bf e}), but with a swimming velocity $v_s$ imposed along the $\ve n_3$-direction. 
In this case, there is an active contribution to the drift of order $v_s$ that can be much larger than the passive drift discussed above \cite{jing2020chirality}.

 We conclude this Section with three  general remarks. First, we found  that one must keep the small elements in $\ma L$ to reach correct qualitative conclusions regarding the angular dynamics. 
 Moreover, 
 the discussion of the elongational flow shows that the angular dynamics in different flows is governed by different elements of $\ma L$. 
Second, in the past, the dynamics of helical particles has been discussed in terms of idealised models \cite{fries_angular_2017,ishimoto2020helicoidal,zoettl2023asymmetric,jing2020chirality} with only few parameters (like the simplified model discussed above).
We have seen here that the dynamics of the simplified and the full model can be qualitatively very different, so one needs to know all elements of $\ma L$ to understand the angular dynamics of aggregates in shear flow, in general.
Third,  for isotropic helicoids,  translation-rotation coupling  is caused by hydrodynamic interactions between non-chiral vanes, explaining why this coupling tends to be small \cite{collins_lord_2021}. A similar reasoning may explain why the chiral couplings of the aggregates studied in this Section are all rather small. This raises the question of how to find aggregate shapes with large chiral couplings. Here a tool like \texttt{SHAPES} can be immensely useful. 

\subsection{Interactions between rigid aggregates}

\begin{figure}[b]
    \centering
    \includegraphics[width=\textwidth]{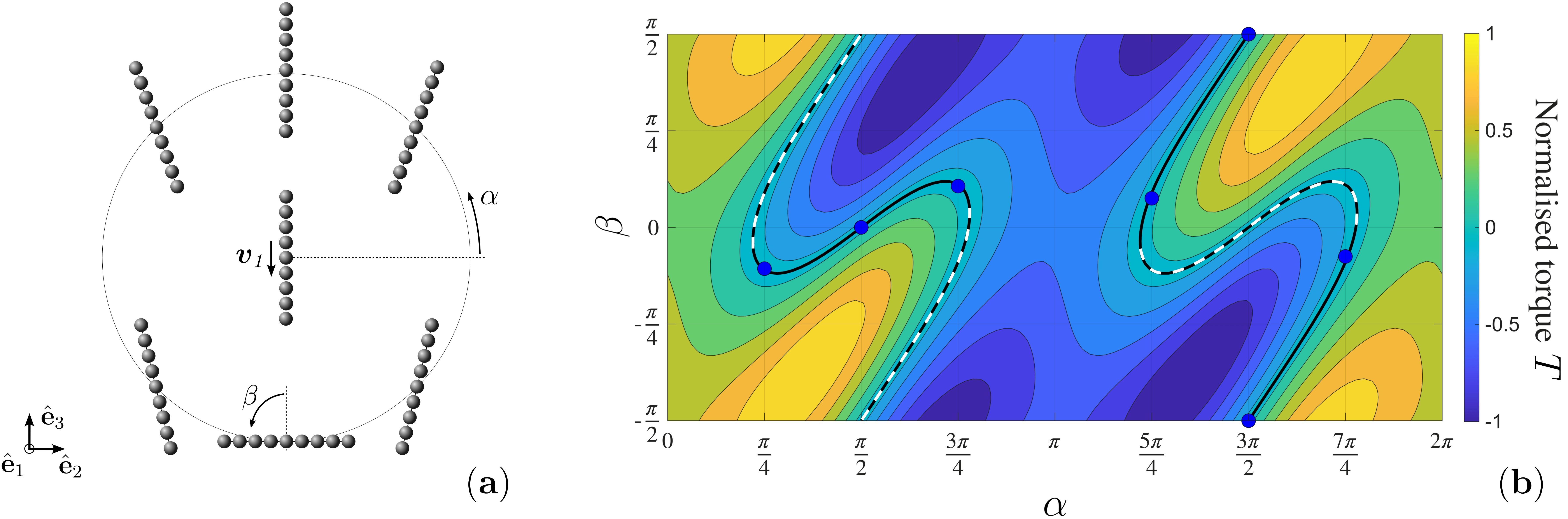}
    \caption{
  Hydrodynamic interactions between two rods of length $\ell$ and radius $a$ in a quiescent fluid. 
  The first rod, at the center settles vertically with velocity $\bm{v}_\mathit{1}$. The second rod is at rest, and placed on a circle of radius $2\ell$, its relative position and orientation to the first rod is described by two angles, $\alpha$ and $\beta$.
  (\textbf{a}) Six stable steady orientations of the second rod. These stable equilibrium orientations are shown as blue points in panel (\textbf{b}). 
  (\textbf{b}) 
  Contour lines of the torque $T = -\bm{T}_\mathit{2} \cdot \hat {\bf e}_1$ acting on the second rod, normalised by its maximum value. 
  Thick black lines indicate equilibrium orientations $T=0$. Stable equilibria are shown as solid black lines, whereas unstable equilibria are shown as dotted black lines. }
  \label{fig:phase_portrait}
\end{figure}

We illustrate how \texttt{SHAPES} handles interactions
between rigid aggregates by looking at  the mechanism
that causes an instability to number-density
fluctuations in an ensemble of slender particles
settling in a quiescent fluid. \citet{koch_instabilities_1989}
showed that the instability is caused by hydrodynamic interactions between the settling particles which leads to preferential alignment.
A slender particle settling vertically tends to turn particles below so that they settle horizontally, increasing their drag. As a consequence, the upper particle catches up with the lower one. The elegant analysis of \citet{koch_instabilities_1989} assumes that the two particles are sufficiently far apart so that the lower one sees the disturbance caused by the upper one as a Stokeslet, and that the backreaction upon the upper one can be neglected. 

\texttt{SHAPES} allows computation of the precise interaction, without relying on these assumptions. We now show that the interaction comes out as predicted by \citet{koch_instabilities_1989}, although the downstream region where rods align horizontally is somewhat narrower than shown in Fig.~1 of \citet{koch_instabilities_1989}.
We examine the hydrodynamic interactions between two rods, each composed of 9 spheres of radius \(a\) spaced such that the total length is \(\ell = 20 a\). We use numbers in italics to refer to these two rods. 
The first rod, denoted by the index $\mathit{1}$, is oriented vertically.
It moves with downward velocity $\ve v_\mathit{1} = - \hat{\bf e}_3$. 
The second rod, denoted by the index $\mathit{2}$, is at rest and its center of mass is placed such that $\ve x_\mathit{2} - \ve x_\mathit{1}= 2\ell (\hat{\bf e}_2 \cos\alpha  + \hat {\bf e}_3 \sin\alpha )$, with $\alpha \in [0,2\pi]$ and 
where $\bm{x}_\mathit{1}$ and $\bm{x}_\mathit{2}$ are the centre of mass of each rod.
For each value of $\alpha$, we denote by $\beta$ the angle formed by the rod axis with $\hat{\bf e}_3$.
From Eq.~(\ref{eq:system}), the hydrodynamic torque experienced by the second rod is given by:
\begin{equation}
\label{eq:shaqfeh_torque}
-\bm{T}_\mathit{2} =- \mathbb{B}_\mathit{21} \cdot \ve v_\mathit{1} \:.
\end{equation}
We used \texttt{SHAPES} to compute the elements of $\ma B_\mathit{21}$. In Fig.~\ref{fig:phase_portrait}({\bf b}), the torque acting on the second rod, $T = -\bm{T}_\mathit{2} \cdot \hat {\bf e}_1$, is shown as a function of $\alpha$ and $\beta$.  
The black lines show steady states
$(\alpha^\ast,\beta^\ast)$, locations of zero torque. Stable steady states for which:
\begin{equation}
\frac{\partial T}{\partial \beta}\Big|_{\alpha^\ast,\beta^\ast}
<0
\end{equation}
are shown as solid black lines, while unstable ones by dashed black lines. Six stable steady orientations of the second rod are shown as points for specified values of $\alpha$. They correspond to the arrangements in panel ({\bf a}). For example, Fig.~\ref{fig:phase_portrait}({\bf b}) shows that when rod $\mathit{2}$ is downstream of rod $\mathit{1}$, near $\alpha = 3\pi/2$, the stable orientation is horizontal. Thus \texttt{SHAPES} recovers the alignment mechanism proposed by 
\citet{koch_instabilities_1989}.

However, the downstream region, in which the horizontal orientation is selected, is narrower than in the idealised far-field picture: in our calculation, strong horizontal alignment occurs for:
\begin{equation}
    \alpha \approx 3\pi/2 \pm \pi/6\,,
\end{equation}
for the chosen parameters.
This suggests that finite-size hydrodynamic interactions and higher-order flow contributions can reduce the angular range over which pair interactions reinforce density fluctuations, potentially weakening the  growth rate of the concentration instability,  compared 
with the theory of \cite{koch_instabilities_1989}. 
This is consistent with later simulations and experiments~\cite{mackaplow1998shaqfeh,metzger2005streamers,metzger2007experimental,saintillan2006stratification}, which confirm the instability of sedimenting fibre suspensions but indicate that form and strength of the instability
differs in detail from the predictions of the idealised model. 
Since horizontal alignment only occurs in a narrow down-stream region in our case, one may expect the  amplification mechanism to be weaker and more anisotropic. 

\subsection{Microswimmers}  
\label{subsec:swimmer}

\begin{figure}[b]
    \centering
    \includegraphics[width=16cm]{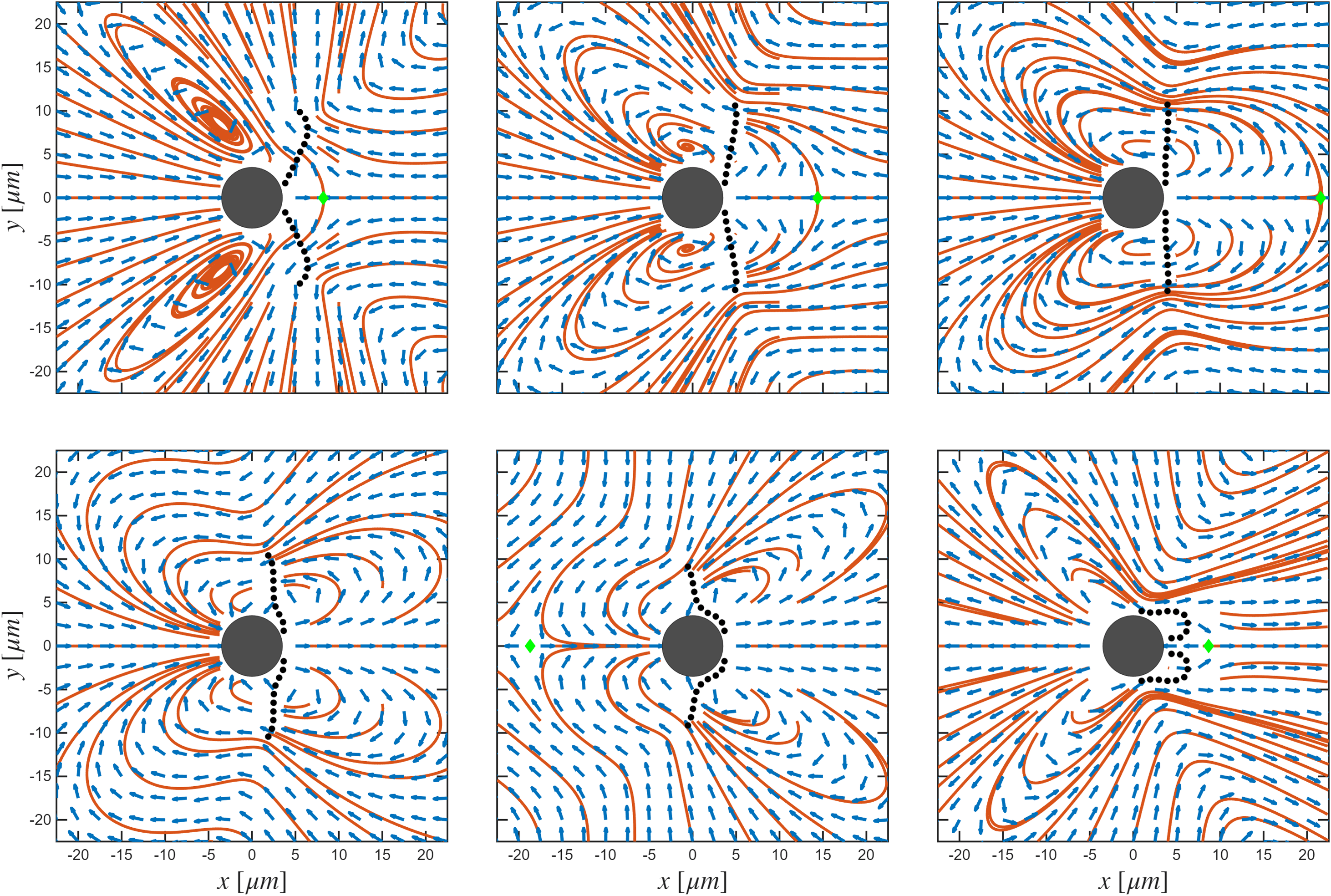}
    \caption{
    The motion of the microalga \textit{Chlamydomonas reinhardtii} over a swimming cycle generates a disturbance flow field. There is no external forces applied in this simulation.
    The normalized velocity field is represented by blue arrows. Streamlines of the disturbance flow are shown in orange. Green diamonds denote stagnation points.
    }
    \label{fig:swimmer_flow}
\end{figure}

This last example serves two purposes.  First, we demonstrate that \texttt{SHAPES} provides realistic disturbance flows for active swimmers using prescribed experimental gaits. Second, we illustrate how the method can be used to compute hydrodynamic interactions between a swimmer and a nearby wall.

To make the first point, we consider  a simplified model of the microalga \textit{Chlamydomonas reinhardtii}.  This microorganism is a  puller \cite{lauga_hydrodynamics_2009}, generating propulsion by pulling fluid inward along its swimming direction. 
Our model for the swimmer consists of a large central bead representing the cell body and two flagella that are attached at to the body as shown in  Figs.~\ref{fig:particle_summary}({\bf f}) and \ref{fig:swimmer_sketch}. Each flagellum is represented as a chain of $N$ beads. 
The swimming gait is parameterised by a vector $\ve \sigma(t)$ of time-dependent relative angles $\sigma_\alpha(t)$ ($\alpha = 1,\ldots,N$) between neighbouring beads in a given flagellum.

To model realistic flagellar kinematics, we reconstructed a sequence of flagellar shapes from the experimental snapshots reported by~\citet{geyer_cell-body_2013}. 
Since the temporal resolution of the experimental measurements is low, 
we reconstructed 14 representative configurations uniformly distributed over one period, and then interpolated using Fourier-series representations (based on nonlinear least-squares regression to a truncated Fourier series, available in \textsc{Matlab} toolboxes) of the angular variables $\sigma_\alpha(t)$. 
The dynamics of the flagella is mirrored and periodically continued in time. 
Using \texttt{SHAPES}, we then computed the dynamics of the swimmer and the generated disturbance flow.
Fig.~\ref{fig:swimmer_flow}
shows the disturbance-flow field
 over one swimming cycle. 
The simulations recover the main features observed experimentally by~\citet[Fig.~3]{guasto_oscillatory_2010}, including the axial symmetry of the disturbance flow around the swimmer and its temporal evolution over the beat cycle. 
In particular, the model reproduces the characteristic back-and-forth motion 
associated with non-reciprocal locomotion in Stokes flows. 
In Fig.~\ref{fig:swimmer_flow}, the first five panels display a forward-propelling flow field, whereas the final panel exhibits a total reversal in the direction of the flow arrows, signifying backward motion.
This illustrates how experimentally inspired gaits can be directly incorporated into \texttt{SHAPES} to generate realistic swimmer-induced flows.
A movie of the simulated alga swimming in an unbounded fluid is provided in Appendix E.

\begin{figure}[b]
    \centering
    \hspace{1.5cm}
\includegraphics[width=0.8\linewidth]{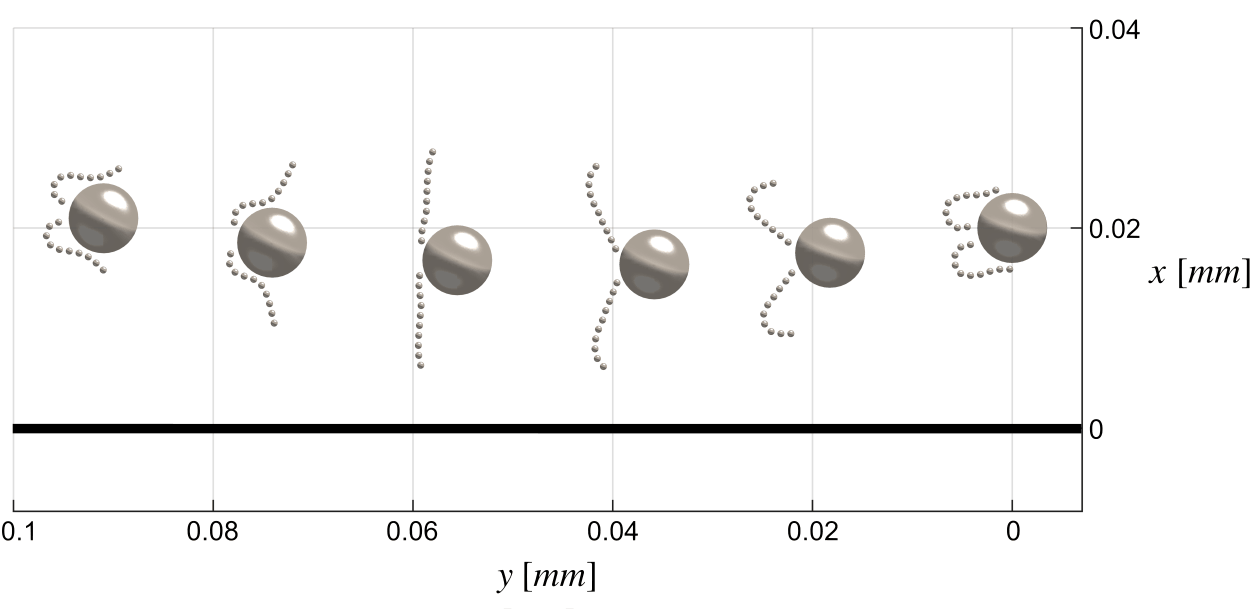}
    \caption{
    Snapshots over time of the microalga swimming near a wall located at $x=0$. 
    Initially, the alga swims towards the wall.
   As it approaches, hydrodynamic interactions cause the alga to turn and move away from the wall, illustrating the mechanism suggested by \citet{lauga_hydrodynamics_2009}.
    \texttt{SHAPES}-simulation details
    are given in the main text.}
    \label{fig:swimmer_wall}
\end{figure}

We next consider the interaction of the swimmer in a quiescent fluid near a wall, assuming no-slip boundary conditions for the disturbance flow. 
The swimmer is initially oriented toward the wall and placed sufficiently close for wall-induced hydrodynamic interactions to become significant [Fig.~\ref{fig:swimmer_wall}].
As the swimmer approaches the wall, hydrodynamic interactions reorient its swimming direction and deflect it away from the surface. This wall-induced repulsion was explained by \citet{lauga_hydrodynamics_2009}  using an idealised far-field model of a puller, in which the swimmer is represented by the leading hydrodynamic singularities generated by its force and torque-free motion. In this description, the image system associated with the no-slip wall modifies the swimmer-induced flow and produces a rotation that tends to align pullers away from the boundary. \texttt{SHAPES}  simulations capture the same qualitative mechanism, but without reducing the swimmer to its leading singularity alone. Instead, \texttt{SHAPES} evaluates the hydrodynamic interactions generated by a bead-based swimmer geometry with prescribed, experimentally inspired flagellar kinematics, including higher-order far-field contributions from the instantaneous swimmer shape and stroke. Fig.~\ref{fig:swimmer_wall} shows that this approach reproduces the expected wall-induced reorientation and repulsion.
A movie capturing the full dynamics of this swimmer-wall interaction is available in Appendix E.

This formulation extends classical far-field descriptions to more detailed swimmer models, allowing one to vary swimmer geometry, stroke kinematics, confinement, and wall geometry within a unified numerical framework. These effects have been shown to play important roles in swimmer--wall interactions, including hydrodynamic trapping and scattering, near-wall accumulation, and the dependence of surface interactions on swimmer type and gait \citep{lauga_hydrodynamics_2009}. 
The present framework has the potential to investigate the dynamics of active particles near boundaries and, for instance, to extend the analysis performed here to pusher-type swimmers \cite{spagnolie2012hydrodynamics, berke2008hydrodynamic}. 
However, when the dynamics are dominated by lubrication interactions (with a wall or another swimmer), as observed for \textit{Volvox} colonies \cite{drescher2009dancing}, 
the current model cannot accurately capture the underlying physics.
Indeed, since lubrication forces are not accounted for in the present formulation, these phenomena cannot yet be faithfully reproduced.

\section{Conclusions}
\label{sec:conclusions}
The present method and its implementation in the \texttt{SHAPES} program package provides a versatile and efficient approach for computing the resistance tensors of rigid and flexible aggregates in Stokes flows, including the confining effect of a wall. Aggregates are represented as assemblies of beads making it easier to study complex geometries and interactions between multiple particles. Hydrodynamic interactions between beads are approximately accounted for using Stokesian dynamics \cite{durlofsky_dynamic_1987}.  The resistance tensors of the full aggregate are then obtained by enforcing the kinematic constraints on the beads, effectively reducing the many-bead system to a single equivalent particle.   

To validate the \texttt{SHAPES} tool and to provide examples of its capabilities, we computed resistance tensors for the six different shapes shown in  Fig.~\ref{fig:particle_summary}:  a dumbbell where an exact solution is available, a slender rod which we compared with slender-body theory,  a curved planar fibre which is a simple shape that has translation-rotation coupling,  a chiral dipole which has elements in the strain coupling matrices that are more complex than the standard (Jeffery) couplings of spheroids,  beads with helical flagella which have been used as passive models for bacterial motion in fluids, two interacting slender rods to demonstrate how to handle multiple rigid aggregates, and an active puller which shows how \texttt{SHAPES} can be used with prescribed deformation of particles and interactions with a wall.

The  capability to obtain quantitatively accurate values for the full resistance tensors has produced several substantial new insights. \citet{collins_lord_2021} used an earlier program implementing the method to understand why the translation-rotation coupling for isotropic helicoids is not zero, but very small. \citet{huseby2024helical} determined the resistance matrices $\ma A$, $\ma B$, and $\ma C$ for helical ribbons, in order to explain the complicated dynamics of small helical ribbons settling in a quiescent fluid. \citet{candelier2024torques} used the earlier program to determine the same tensors for planar curved fibres as a starting point to explain tilted steady states of curved fibres
settling in a fluid at rest (the tilted steady states arise from a competition of Stokes and fluid-inertia torques \cite{candelier_settling_2016,Roy_2019}). 
In Section \ref{sec:examples}, we highlighted further insights obtained using \texttt{SHAPES}. 
First, we used \texttt{SHAPES} to 
analyse the mechanism giving rise to number-density fluctuations in suspensions of rods settling in a quiescent fluid. The results confirm the instability mechanism found by \citet{koch_instabilities_1989}, but
also indicate that the instability might be weaker than predicted by the idealised model of Ref.~\cite{koch_instabilities_1989}.
Second, our calculations for a mechanistic model of a microswimmer illustrate that the induced disturbance flow computed by \texttt{SHAPES} is in good agreement with experiments. 
Third, the ability of \texttt{SHAPES} to provide the full strain-coupling tensors \ma G and \ma H has provided the most substantial new insights.  We show that key results in a sequence of previous papers that explored shapes with strain coupling more complex than spheroids 
\cite{kramel2016preferential,fries_angular_2017,ishimoto2020helicoidal,zoettl2023asymmetric,jing2020chirality,marcos2009separation} can be synthesized by quantifying which of elements of the strain-rate coupling tensors matter for the angular dynamics, for different particle shapes, and in different flows. 
The analysis
shows that simplified models 
of the angular dynamics in a simple shear \cite{fries_angular_2017,ishimoto2020helicoidal,zoettl2023asymmetric,jing2020chirality} may not correctly predict when alignment with vorticity is stable. For the correct conclusion, one needs to consider all elements of the strain-rate coupling tensor. 

There is much more work to be done to develop a comprehensive understanding of how particle shape can be designed to achieve prescribed couplings to fluid-velocity gradients. Many such design questions become much more accessible with a tool like \texttt{SHAPES}. For example, one may wish to optimise translation--rotation coupling, $\ma B$, or non-Jeffery elements of $\widetilde{\ma H}$ that couple strain to angular velocity, in particular for chiral particles. Such couplings are relevant for microfluidic applications, where particles with different strain-rotation or strain-translation responses can, in principle, be separated or steered \cite{marcos2009separation,eichhorn2010microfluidic,meinhardt2012separation,bogunovic2012chiral,marcos_analysis_2014}. Since \texttt{SHAPES} computes the full resistance and mobility tensors for aggregates of arbitrary shape, including their couplings to strain, it can be used to precisely predict how particle migration and rotation depend on particle shape (and the corresponding point-group symmetries), rather than relying on highly symmetric idealised models 
that may give rise to artificial degeneracies.

Several extensions appear possible and useful. One is to include particle flexibility. In the present framework, this can be done by solving, at each time step, for the internal tensions required to enforce connectivity constraints.  
This makes it possible to study flexible fibres, worm-like chains in confinement, and deformation-induced changes of the strain--rotation couplings.  A caveat is that steric interactions and self-avoidance would be needed for dense conformations or nano-channel geometries \cite{alizadehheidari2015nanoconfined,werner2017one}.
Second, hydrodynamic interactions with walls are central in microfluidics, sedimentation near surfaces, and biological transport. Wall effects influence  how aggregates rotate, drift, align close to solid boundaries, and interact with each other. Very close to the wall, lubrication effects and steric interactions may become important. As we have shown above, current ways of including lubrication corrections \cite{durlofsky_dynamic_1987} are not reliable when applied to the present method. 
Further improvements in the modelling of near-contact interactions may therefore be of interest. 
Third, the framework could also be extended to study fragmentation in flows by introducing failure thresholds for the internal tensions within an aggregate. This might be relevant for fragile clusters, marine snow, porous aggregates, volcanic ash, atmospheric microplastics, and other irregular particles whose breakup is controlled by hydrodynamic loading. Further developments could incorporate additional interactions, such as steric repulsion, van der Waals forces, electrostatics, or permeability of porous particles. 

Finally, including weak fluid inertia would open the possibility of generalising inertial corrections known for spheroids in simple shear \cite{einarsson2015rotation} and symmetric particles in uniform flow \cite{Khayat_Cox_1989,candelier2024torques,bhowmick2024inertia,sundberg2025fluid} to aggregates of arbitrary shape in general linear flow. This would be particularly useful for larger aggregates, where  shape, inertia, and velocity-gradient coupling all contribute to the angular dynamics. A first goal would be to generalise the perturbative inertial analysis of \citet{einarsson2015rotation} from spheroids to aggregates of arbitrary shape in general linear flows. Since \texttt{SHAPES} provides the complete coupling tensors for the Stokes dynamics, it may be possible to derive systematic fluid-inertial corrections for aggregates without rotational symmetry, including chiral and irregular aggregates. Such a framework could clarify how inertia modifies strain--rotation couplings, orbit drift, and stability selection in general linear flows.

%

\appendix
\section{Tutorial}
\label{appendix:tutorial}

This tutorial introduces the simplest features of the \texttt{SHAPES} program package and focuses on rigid particles. Here we show how to instantiate a rigid aggregate using the \texttt{RigidBody} class, compute its resistance tensors and simulate its dynamics, either in an unbounded fluid or near a wall.

\subsection{Rigid-aggregate shapes}

For each aggregate, the user must choose the coordinates \texttt{xx} and radii \texttt{a} of the  beads. 
The $i$-th column of the matrix \texttt{xx} contains the coordinates of the bead \texttt{i}. The user must also ensure that beads do not touch or overlap. If all the beads have the same radius, the variable \texttt{a} can be a scalar.

\noindent
\begin{minipage}{0.56\textwidth}
\begin{lstlisting}
xx1 = [0; 0; 0];
xx2 = [5; 0; 0];
xx3 = [5; 0; 5];
xx4 = [5; 5; 5];
% Coordinates for an object made of 4 spheres
xx = [xx1, xx2, xx3, xx4];

a = [1, 1.1, 1.4, 1]; % Radii of the spheres
% a = 1; % same radius for all the spheres

% Instantiate an object
obj = classes.RigidBody(xx,a);

figure
Show(obj) % Visualiation
\end{lstlisting}
\end{minipage}
\hspace{1cm}  
\begin{minipage}{0.313\textwidth}
\includegraphics[width=\textwidth]{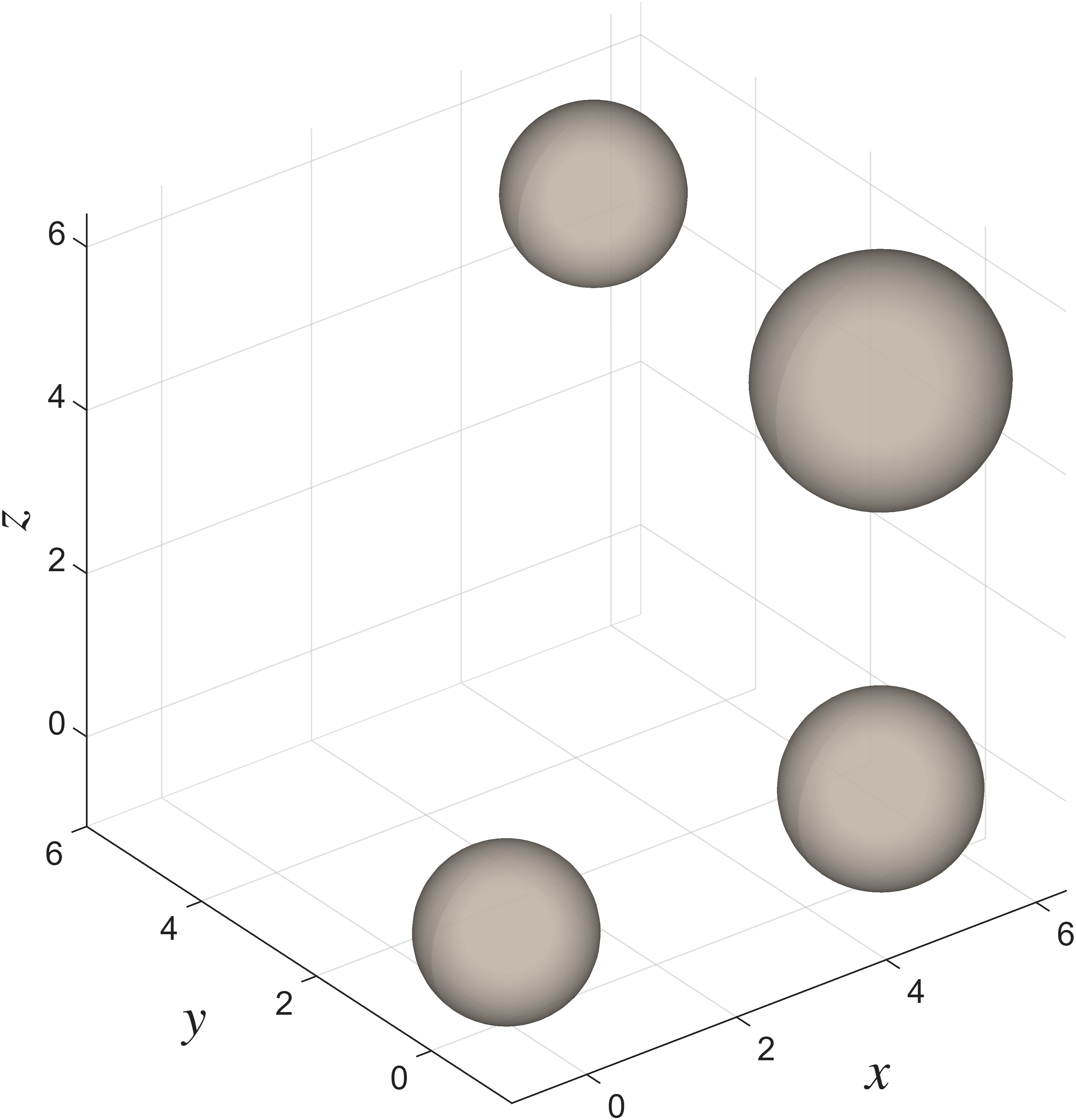}
\end{minipage}

\subsection{Resistance tensors and mobility tensors}

In \texttt{SHAPES}, one can compute resistance and mobility tensor of a given aggregate. By default, if no centre is specified, resistance tensors are computed from the centre of mass, but it can be computed from any point \texttt{xr}. Here, as an example, we use the centre of resistance as our reference point and display the tensor $\mathbb{B}$ in the command window:
\begin{lstlisting}
xr = CenterOfResistance(obj);
getResistanceTensors(obj, xr)
\end{lstlisting}
\begin{lstlisting}[style=matlaboutput]
>> obj.B
     0.7108    6.4054    0.0073
     6.4054    0.5011   -0.3446
     0.0073   -0.3446   -1.1775
\end{lstlisting}
When computing resistance tensors multiple times, only the last result is kept. You can check from which point you got your resistance tensors with \texttt{xres} property.
\begin{lstlisting}[style=matlaboutput]
>> obj.xres
     3.7920
     1.1771
     2.6990
\end{lstlisting}
You can also obtain mobility tensors and compute the centre of mobility.
\begin{lstlisting}
getMobilityTensors(obj)
xm = CenterOfMobility(obj);
\end{lstlisting}
\begin{lstlisting}[style=matlaboutput]
>> xm
     3.8077
     1.1616
     2.6991
\end{lstlisting}
Mobility tensors are always computed from the same points used for resistance tensors and you can double check which point it is by using \texttt{xmob} property. For this aggregate, the centre of mobility is slightly offset from the centre of resistance~\cite{kim_microhydrodynamics_1991}.

\subsection{Interactions with a wall}

Hydrodynamic interactions between a particle and a planar wall can also be studied. 
Creating a RigidBody object near a wall requires specifying four input arguments: the mass density as the third and the setup as the fourth.
Mass density is only useful for dynamic simulations. The setup is either 0 (no wall, by default) or 1 (wall present).
To instantiate an object near a wall, all the beads must be on the same side of the wall (which is located at x = 0), but the side itself does not matter.

\noindent
\begin{minipage}{0.5\textwidth}
\begin{lstlisting}
N = 4; % number of spheres
a = 0.1; % radius of the spheres
xx = zeros(3, N);
% coordinates of the spheres
for i = 1:N
    xx(:, i) = [0.5*i; 0; 0]; 
end
rho = 1030; % mass density of the particle
setup = 1;  % setup

obj_near_a_wall =... 
    classes.RigidBody(xx, a, rho, setup);

figure
Show(obj_near_a_wall)
getResistanceTensors(obj_near_a_wall)
\end{lstlisting}
\end{minipage}
\hspace{1cm}  
\begin{minipage}{0.35\textwidth}
\includegraphics[width=\textwidth]{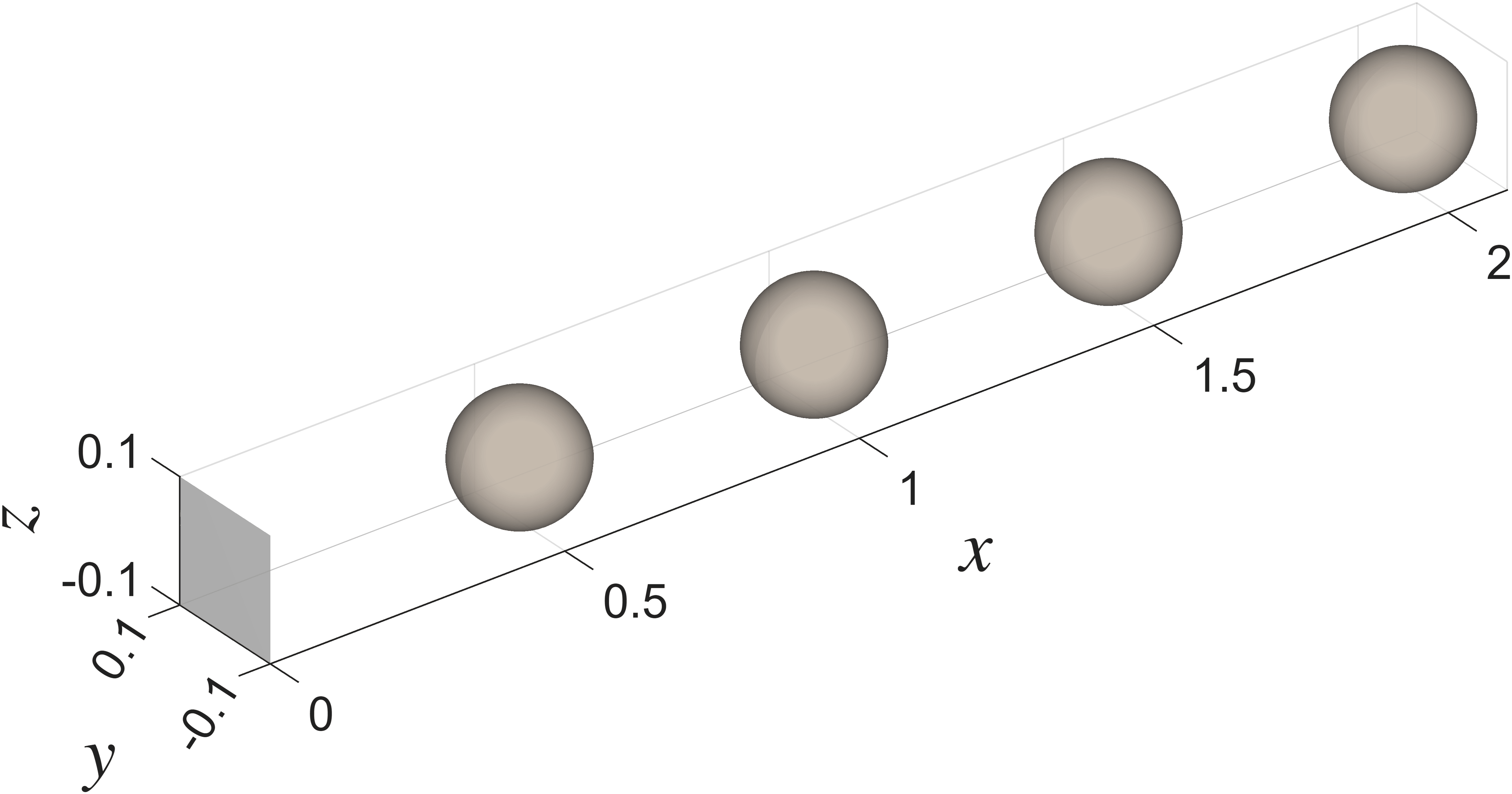}
\end{minipage}

\noindent Resistance tensors now account for the interactions with the wall. The representation of the wall here is restricted by the axis limits but the aggregate interacts with an infinite wall in the plane $x = 0$.
\begin{lstlisting}[style=matlaboutput]
>> obj_near_a_wall.B
     0         0         0
     0         0    0.2047
     0   -0.2047         0
\end{lstlisting}

\subsection{Pre-existing classes}

Several shapes are available in the \texttt{SHAPES} program package, a comprehensive list can be found in the \textit{+classes} file of the package. Each of them has their own properties. Here, we made a two-bladed propeller 
(see \citet[Figure 5-4.1]{happel_low_1983} or \citet[Figure 5.3]{kim_microhydrodynamics_1991}) using the \texttt{ScrewPropeller} subclass.
The main input arguments are:
\texttt{rad} the radius of each disk, 
\texttt{h} the distance between the center of the two disks,
\texttt{theta} the relative orientation of the two disks.

\noindent
\begin{minipage}{0.56\textwidth}
\begin{lstlisting}
rad = 12; 
h = 30;   
theta = pi/4; 
prop = classes.ScrewPropeller(rad, h, theta);

figure
Show(prop)
view(95, 15)

getResistanceTensors(prop); % about the centre of mass
\end{lstlisting}
\begin{lstlisting}[style=matlaboutput]
>> prop.B
     828.8171     0.0000     0.0000
      -0.0000  -828.8171    -0.0000
       0.0000     0.0000     0.0000
\end{lstlisting}
\end{minipage}
\hspace{3cm}
\begin{minipage}{0.35\textwidth}
\includegraphics[width=0.5\textwidth]{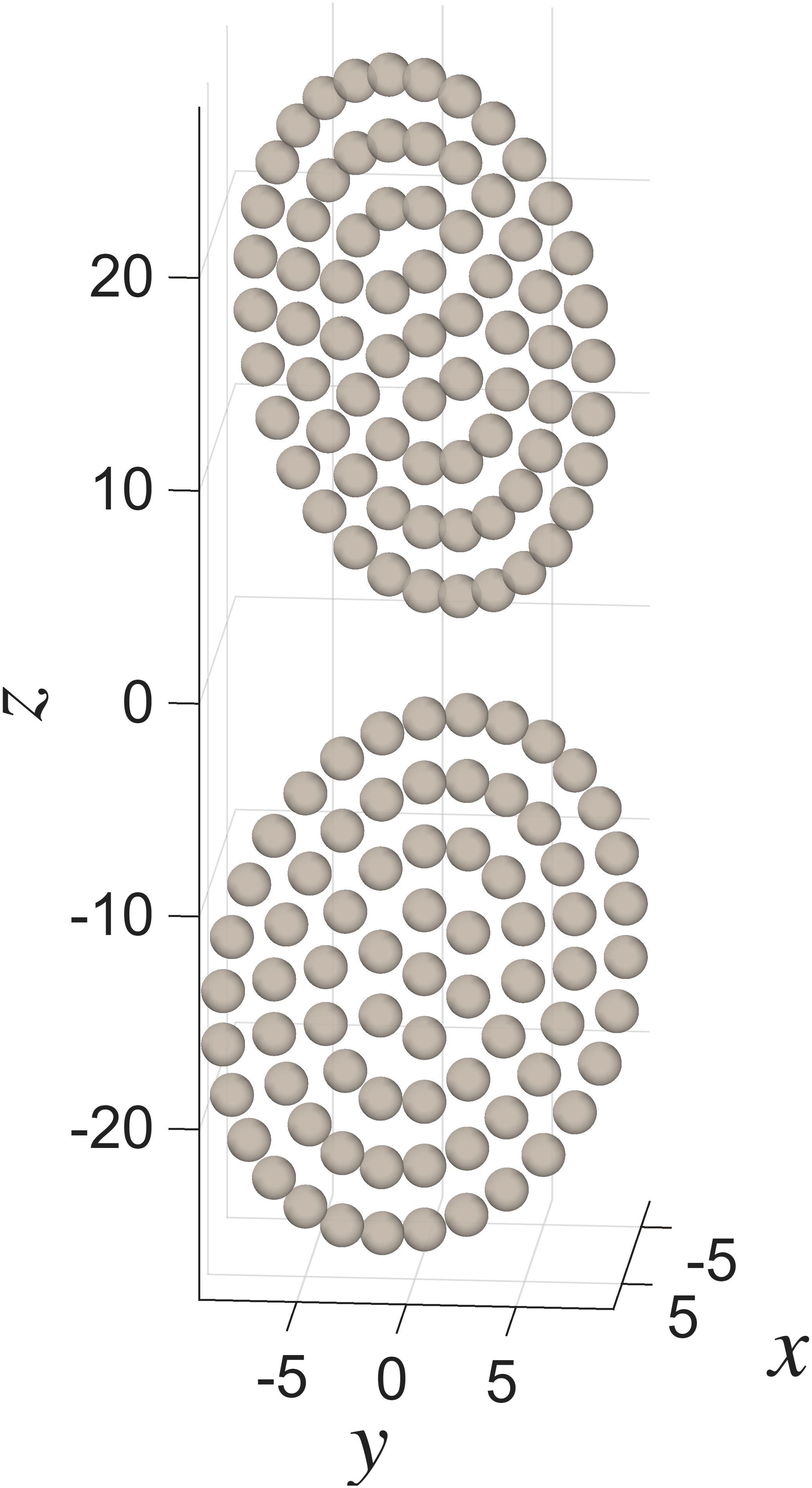}
\end{minipage}

\noindent Most classes have optional arguments useful to modify the position and orientation of the particle. 
Each particle has a default position, specified by the center of mass, and a default orientation, specified by a rotation matrix. 
The package includes auxiliary functions that are not class methods but are useful for performing certain mathematical operations. 

\noindent For instance, two functions are available to compute a rotation matrix:
\begin{itemize}
    \item \texttt{R = aux.rot\_u(u, alpha)} returns the rotation matrix corresponding to the rotation of an angle $\alpha$ around the $\bm{u}$ axis,
    \item \texttt{R = aux.rot\_euler(phi, theta, psi)} returns the rotation matrix using Euler angles \cite{goldstein_classical_2008}.
\end{itemize}
Here, we use the first option:

\noindent
\begin{minipage}{0.56\textwidth}
\begin{lstlisting}
R = aux.rot_u([1 0 0], pi/2);
prop_horizontal = classes.ScrewPropeller(rad, h,...
    theta, xG = [5;0;0], rot = R);

figure
Show(prop_horizontal)
view(95, 15)

getResistanceTensors(prop_horizontal); 
\end{lstlisting}
\begin{lstlisting}[style=matlaboutput]
>> prop_horizontal.B
     828.8171    -0.0000     0.0000
       0.0000     0.0000    -0.0000
       0.0000     0.0000  -828.8171
\end{lstlisting}
\end{minipage}
\hfill
\begin{minipage}{0.4\textwidth}
\includegraphics[width=\textwidth]{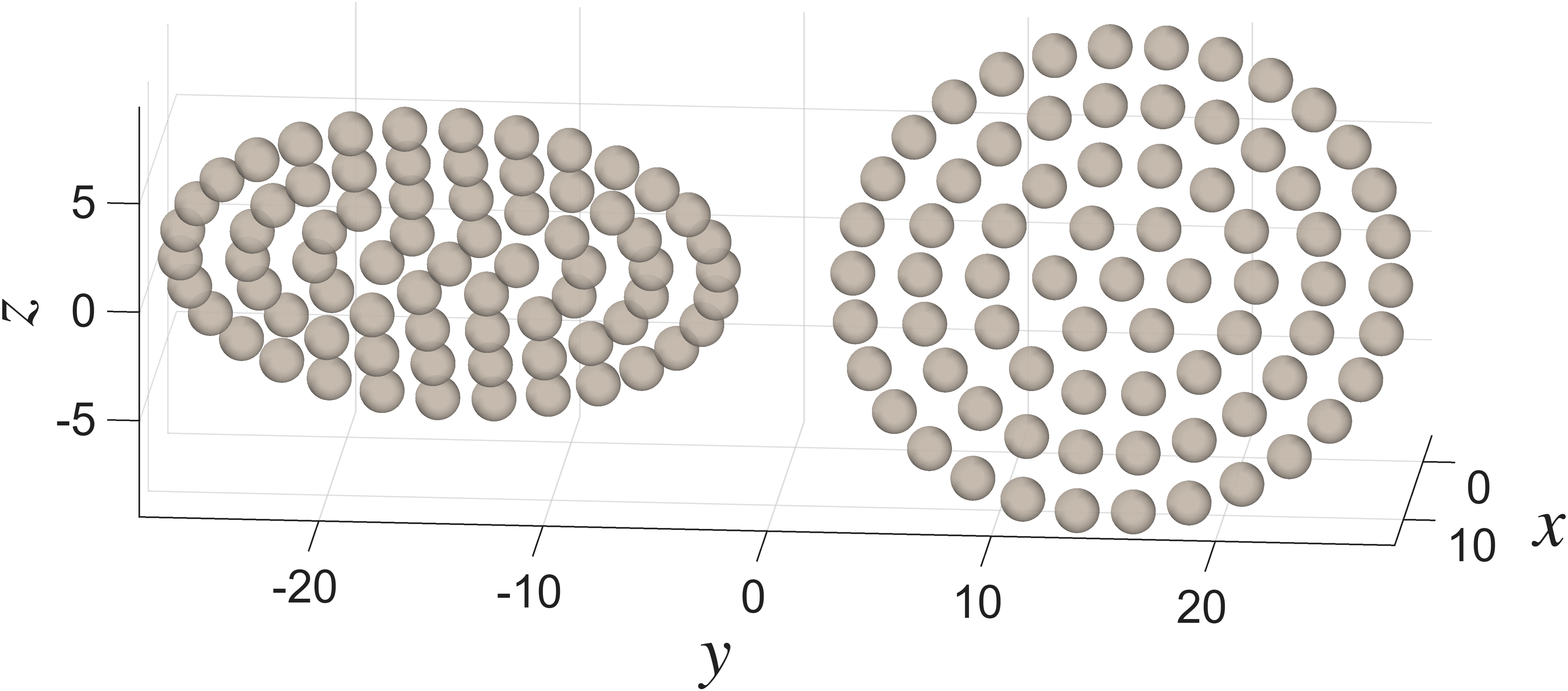}
\end{minipage}

\subsection{Aggregate dynamics} 
The translational and angular dynamics of an aggregate is calculated
in the same way for any class or subclass.
The input arguments are the aggregate \texttt{obj} in question, as well as any external forces and torques \texttt{FText}.
In addition, one can also choose different linear velocity fields, the dynamic viscosity $\mu$, as well as location and orientation of the aggregate at the beginning of the simulation.
While the internal integration time steps are automatically adapted by the solver to guarantee precision, the user-defined output time steps can be chosen arbitrarily to obtain data or generate animations without affecting the underlying resolution of the dynamics.
Here, the particles are assumed to be non-inertial. Therefore, when a buoyancy force is applied, the fluid and particle densities should ideally be close to ensure computational efficiency. For large particles or high particle-to-fluid density ratios, increasing the fluid viscosity $\mu$ reduces particle velocities and can prevent excessively small integration time steps, thereby keeping computation times reasonable.
As an example, consider a sedimenting propeller in a fluid at rest.

\begin{lstlisting}
rho_f = 0997;   % density of the fluid
rho_p = 1010;   % density of the particle
FText = [0; 0; prop.N*(4/3)*pi*prop.a{1}^3*(rho_p - rho_f)*(-9.81); 0; 0; 0]; % external forces, torques
% orientation of the particle at the start of the simulation (input arg must be a quaternion)
q_init = matrice_to_quaternion(aux.rot_u([0 0 1], theta)); 

t_step = 0.05;
t_start = 0;
t_end = 40;
[t_out,y_out] = Dynamics(prop, FText, tspan = t_start:t_step:t_end, mu = 0.1, q0 = q_init);
\end{lstlisting}
We solve for particle dynamics (Eq.~\eqref{eq:mobility} with the kinematic relations in Eqs.~(\ref{eq:dynamics_bis}a,c)) using the Runge–Kutta–Fehlberg numerical scheme. The vector \texttt{t\_out} stores the output time steps (which are chosen using the \texttt{tspan} option in the inputs). 
The first four columns of \texttt{y\_out} store quaternions (parameterisation of the rotation matrix) and the position of the centre of mass is stored in columns 5 through 7.
Animation of the motion can be made using space data over time and can be saved as a video:
\begin{lstlisting}
figure
Animation(prop, t_out, y_out, video = "yes", filename = "sedimentation_propeller",...
    orient = [90 0], framerate=1/t_step)
\end{lstlisting}
Sometimes, it may be helpful to adjust the frame rate (or the viscosity $\mu$) to speed up or slow down the animations.

\section{Coupling terms in Eqs.~(\ref{eq_phi_dot}--\ref{eq_psi_dot})}
\label{app:coupling_terms}
In Section~\ref{sec:shear}, the equations of motion for the angular dynamics of an aggregate in  shear flow, Eqs.~(\ref{eq_phi_dot}--\ref{eq_psi_dot}),  were written in compact form abbreviating the couplings between tumbling and spin by $F_\phi$, $F_\theta$, and $F_\psi$.
In this appendix we give the explicit expressions for $F_\phi$, $F_\theta$, and $F_\psi$.
These terms arise from the hydrodynamic couplings between the orientation of the aggregate,
the imposed shear flow, and the geometrical coefficients $L_{ijk}$. In this appendix
we use the short-hand notation
\begin{align}
c_\phi=\cos\phi,
\quad
s_\phi=\sin\phi,
\quad
c_\theta=\cos\theta,
\quad
s_\theta=\sin\theta,
\quad
c_\psi=\cos\psi,
\quad
s_\psi=\sin\psi\,.
\end{align}
With this notation, the coupling terms read
\begin{subequations}
\begin{align} 
F_\phi = 
& \frac{c_\psi^3}{s_\theta}
\left[ 
c_\theta c_\phi^2 (2L_{222}-4L_{121}-2L_{211}) 
+ c_\theta (2L_{121}+L_{211}-L_{222})
+ s_\phi c_\phi (c_\theta^2+1)(L_{122}+2L_{221}-L_{111}) 
\right] 
\nonumber 
\\
& + \frac{c_\psi^2 s_\psi }{s_\theta}
\left[ 
s_\phi c_\phi (c_\theta^2+1) (2L_{121}+L_{211}-L_{222}) 
+ c_\theta c_\phi^2 (2L_{122}+4L_{221}-2L_{111}) 
+ c_\theta (L_{111}-L_{122}-2L_{221}) 
\right] 
\nonumber 
\\
& + c_\psi^2 
\left[ 
(L_{132}+L_{231})(1-2c_\phi^2) 
+ 2s_\phi c_\theta c_\phi (L_{232}-L_{131}) 
\right] 
\nonumber 
\\
& + c_\psi s_\psi 
\left[ 
(L_{131}-L_{232})(1-2c_\phi^2) 
+ 2s_\phi c_\theta c_\phi (L_{132}+L_{231}) 
\right] 
\nonumber 
\\
& + \frac{c_\psi}{s_\theta} 
\left[ 
s_\phi c_\phi (L_{111}(2-c_\theta^2) + L_{122}(1-2c_\theta^2) - 2L_{221}(1+c_\theta^2))  
\right. 
\nonumber 
\\ 
& \hspace{1.2cm} 
\left. 
+  
c_\theta c_\phi^2 (2L_{121}+2L_{211}-2L_{222}) 
+ c_\theta (L_{222}-L_{121}-L_{211}) 
\right] 
\nonumber 
\\
& + \frac{s_\psi }{s_\theta}
\left[ 
c_\theta L_{221} (1-2c_\phi^2) 
+ s_\phi c_\phi (L_{211}(c_\theta^2-2) + L_{222}(2c_\theta^2-1)) 
\right] \,,
\end{align}
\begin{align}
F_\theta = 
& c_\psi^3
\Big[
c_\theta
(L_{111}-L_{122}-2L_{221}
-2c_\phi^2L_{111}
+2c_\phi^2L_{122}
+4c_\phi^2L_{221})
\hspace*{6.3cm}
\nonumber
\\
& \hspace{1.2cm}
+
s_\phi c_\phi
(2L_{121}c_\theta^2
+L_{211}c_\theta^2
-L_{222}c_\theta^2
+2L_{121}+L_{211}-L_{222})
\Big]
\nonumber
\\
& + c_\psi^2 s_\psi
\Big[
s_\phi c_\phi
(1+c_\theta^2)
(L_{111}-L_{122}-2L_{221})
+
c_\theta
(2c_\phi^2-1)
(2L_{121}+L_{211}-L_{222})
\Big]
\displaybreak[3]
\nonumber
\\
& + c_\psi^2
\Big[
s_\theta
(L_{131}-L_{232})(1-2c_\phi^2)
+
2s_\phi c_\phi s_\theta c_\theta
(L_{231}+L_{132})
\Big]
\nonumber
\\
& + c_\psi s_\psi
\Big[
2s_\theta c_\phi^2(L_{231}+L_{132})
+
2s_\theta c_\theta s_\phi c_\phi
(L_{131}-L_{232})
-
s_\theta(L_{132}+L_{231})
\Big]
\nonumber
\\
& + c_\psi
\Big[
c_\theta
(2c_\phi^2-1)
(L_{111}-L_{122}-L_{221})
+
s_\phi c_\phi
(L_{211}c_\theta^2
+2L_{222}c_\theta^2
-2L_{121}c_\theta^2
-2L_{121}-2L_{211}-L_{222})
\Big]
\nonumber
\\
& + s_\psi
\Big[
c_\theta L_{121}
+
s_\phi c_\phi c_\theta^2
(L_{111}+2L_{122})
-
2c_\theta c_\phi^2L_{121}
-
s_\phi c_\phi
(2L_{111}+L_{122})
\Big]\,,
\end{align}
\begin{align}
F_\psi = 
& \frac{c_\psi^3}{s_\theta}
\left[
s_\phi c_\phi (c_\theta^3 + c_\theta) (L_{111}-L_{122}-2L_{221})
+ 2c_\theta^2 \big( c_\phi^2(2L_{121}+L_{211}-L_{222}) - (L_{121}+\tfrac12L_{211}-\tfrac12L_{222}) \big)
\right]
\nonumber
\\
& + \frac{c_\psi^2 s_\psi}{s_\theta}
\left[
s_\phi c_\phi (c_\theta^3 + c_\theta) (L_{222}-2L_{121}-L_{211})
+ (2c_\theta^2 c_\phi^2 - c_\theta^2) (L_{111}-L_{122}-2L_{221})
\right]
\nonumber
\\
& + c_\psi^2
\Big[
c_\theta \big( 2c_\phi^2(L_{132}+2L_{321}+L_{231}) - (L_{132}+2L_{321}+L_{231}) \big)
\nonumber \\ & \hspace{1.2cm}
+ s_\phi c_\phi c_\theta^2 (2L_{131}+L_{311}-2L_{232}-L_{322}) + s_\phi c_\phi (L_{311}-L_{322})
\Big]
\nonumber
\\
& + c_\psi s_\psi
\Big[
2c_\theta c_\phi^2 (L_{131}-L_{311}+L_{322}-L_{232}) + c_\theta (L_{232}-L_{311}+L_{322}-L_{131})
\nonumber \\ & \hspace{1.2cm}
- 2s_\phi c_\phi c_\theta^2 (L_{132}+L_{231}+L_{321}) - 2L_{321}s_\phi c_\phi
\Big]
\nonumber
\\
& + \frac{c_\psi}{s_\theta}
\left[
s_\phi c_\phi c_\theta^3 (L_{111}+2L_{122}+2L_{221}-2L_{331})
+ s_\phi c_\phi c_\theta (-2L_{111}-L_{122}+2L_{221}+2L_{331})
\right. \nonumber \\ & \hspace{1.2cm} \left.
+ 2c_\theta^2 \big( c_\phi^2(L_{222}-L_{211}-L_{121}-L_{332}) + \tfrac{1}{2}(L_{121}+L_{211}-L_{222}+L_{332}) \big)
+ L_{332}(2c_\phi^2-1)
\right]
\nonumber
\\
& + \frac{s_\psi}{s_\theta}
\left[
s_\phi c_\phi c_\theta^3 (2L_{332}-2L_{222}-L_{211}) + s_\phi c_\phi c_\theta (L_{222}+2L_{211}-2L_{332})
\right. \nonumber \\ & \hspace{1.2cm} \left.
+ 2c_\theta^2 (c_\phi^2-1/2) (L_{221}-L_{331}) + L_{331}(2c_\phi^2-1)
\right]\,.
\end{align}
\end{subequations}

\section{Stability analysis for the simplified model}
\label{app:staban}
In Section \ref{sec:cds}, we discussed the angular dynamics 
of the simplified model [Eq.\eqref{eq:eomsimple}]
for a chiral dipole in shear flow. 
The model has two log-rolling steady states, alignment and anti-alignment with vorticity. 
Linear stability analysis 
shows that both are marginally stable. 
\citet{ishimoto2020helicoidal} considered
higher-order corrections and computed the non-linear stability of the two steady states.
Here we analyse the effect of the non-linear corrections in a slightly different way, and the conclusions are consistent with Ref.~\cite{ishimoto2020helicoidal}. The most convenient
Euler angles $\varphi,\vartheta$ for analysing log-rolling stability are such that log rolling is a steady state
in the $\varphi$-$\vartheta$-dynamics \cite{rosen2015numerical,zoettl2023asymmetric} (this is not the case for the standard choice $\phi,\theta$ introduced in the main text). In this Appendix, we  choose
XYZ Tait-Bryan angles 
which correspond to the rotation matrix
\begin{equation}
    \ma R = \begin{bmatrix}
     c_\vartheta c_\psi &   -c_\vartheta s_\psi  & s_\vartheta
\\
\phantom- c_\varphi s_\psi  -s_\varphi   s_\vartheta c_\psi
   & \phantom- c_\varphi c_\psi+  s_\varphi s_\vartheta   s_\psi  &s_\varphi  c_\vartheta  
\\
- s_\varphi s_\psi  - c_\varphi s_\vartheta  c_\psi  & -s_\varphi c_\psi +c_\varphi s_\vartheta  s_\psi   & c_\varphi c_\vartheta  
    \end{bmatrix}\,.
\end{equation}
Using Eqs.~(\ref{eq_model_equi})
and setting all other elements of $\ma L$ to zero, the coupling to spin vanishes and the angular dynamics becomes
\begin{subequations}
    \label{eq:ecd}
\begin{eqnarray}
\dot \varphi &= &\frac{1}{2}
(\Lambda-1)\cos\varphi\,\tan\vartheta\,
+\frac{\chi}{2}\,\sin(\varphi)\left(
-2\cos(\vartheta)+\frac{1}{\cos(\vartheta)}
\right)\,,
\\
\dot \vartheta&=&\frac{1}{2}
(1+\Lambda \cos2\vartheta)\sin\varphi
+\frac{\chi}{2}\,\sin(\vartheta)\cos(\varphi)\,,
\end{eqnarray}
\end{subequations} 
where we have set the shear rate to unity, $\dot\gamma=1$.
The log-rolling steady states
are
\begin{equation}
\begin{bmatrix}
    \varphi_\sigma^\ast\\
    \vartheta_\sigma^\ast
\end{bmatrix}
=
\begin{bmatrix}
    0\\\tfrac{\pi}{2} (1-\sigma)
\end{bmatrix}
\end{equation}
for $\sigma = \pm 1$. Expanding
Eq.~(\ref{eq:ecd}) around these steady states
to third order in  $\delta \varphi$ and $\delta \vartheta$ 
leads to
\begin{equation}
\tfrac{\rm d}{{\rm d}t} \ve X = \ma A_\sigma \ve X 
+ \ve N_\sigma(\ve X)\,,
\end{equation}
with 
    \begin{eqnarray}
 \ve X =\begin{bmatrix}
     \delta \varphi\\\delta \vartheta
    \end{bmatrix}\,,\quad
    \ma A_\sigma = 
\tfrac{1}{2}\begin{bmatrix}
-\sigma{\chi} & \Lambda-1\\
\Lambda+1& \sigma {\chi}
\end{bmatrix}\,,
\quad\ve N_\sigma = 
\begin{bmatrix}
    \tfrac{\sigma\chi}{12}\delta\varphi^3 + \tfrac{1-\Lambda}{4} \delta \varphi^2 \delta\theta + \tfrac{3\sigma\chi}{4}\delta\varphi \delta\theta^2 + \tfrac{\Lambda-1}{6}\delta\theta^3\\
    -\tfrac{\Lambda+1}{12}\delta\varphi^2
    -\tfrac{\sigma\chi}{4} \delta\varphi^2 \delta \theta -\Lambda\delta\varphi\delta\theta^2-\tfrac{\sigma\chi}{12}\delta\theta^3
    \end{bmatrix}\,.
 \end{eqnarray}
Here $\ve N_\sigma$ is the leading non-linear correction of third order
(there are no second-order contributions). 
    The eigenvalues of $\ma A_\sigma$
    determine the linear stability of the fixed point. Since
    ${\rm tr}\ma A_\sigma = 0$
    and 
    ${\rm det}\ma A_\sigma=\tfrac{1}{4}(1-\Lambda^2-\chi^2)\equiv \omega_0^2$,
   the fixed points 
are  marginally stable  for $\Lambda^2+\chi^2<1$, with eigenvalues  $\pm {\rm i} \omega_0$, where
    $\omega_0$ is  real and positive (slender body, weak chirality).
The orbits of the linearised
system are ellipses, 
\begin{equation}
\delta \varphi_0(t) = \sqrt{\frac{2H_\sigma}{1+\Lambda}}[\cos(\omega_0t +\beta)- \frac{\sigma\chi}{2\omega_0} \sin(\omega_0 t+\beta)]\\,\quad
\delta\vartheta_0(t) = \sqrt{\frac{2H_\sigma}{1+\Lambda}} {\frac{1+\Lambda}{2\omega_0}} \sin(\omega_0 t+\beta)\,.
\end{equation}
The trajectories are determined by the initial phase $\beta$
and the 
quadratic invariant \cite{arnold1980gewohnliche}
\begin{equation}
H_\sigma = \frac{1}{2}
\ve X\cdot \ma K_\sigma
\ve X\quad \mbox{with}
\quad \ma K_\sigma = \begin{bmatrix}
    1+\Lambda & \sigma\chi\\
    \sigma\chi &1-\Lambda
\end{bmatrix}\,.
\end{equation}
The function $H_\sigma$
is positive semidefinite and 
assumes a minimum $H_\sigma=0$
at the fixed point. 
In order to quantify how the non-linear dynamics behaves near the fixed point,
one monitors how $H$ changes
under the non-linear dynamics. The rate-of-change of $H$  is given by
\begin{equation}
\frac{{\rm d}}{{\rm d t}}H_\sigma
= \frac{\partial H_\sigma}{\partial \ve X}\cdot \ve N_\sigma=\ve X\cdot \ma K_\sigma \ve N_\sigma\,.
\end{equation}
Near the fixed point, the non-linear dynamics
 corresponds to a slow drift of a rapidly oscillating solution. 
Averaging over one period
of $\varphi_0$ and $\vartheta_0$ 
as described in Ref.~\cite{sanders2007averaging},
one finds
\begin{equation}
\label{eq:drift} 
\overline{\tfrac{{\rm d}}{{\rm d t}}H} = \frac{3}{8}\frac{\sigma\chi\Lambda}{\omega_0^2} H^2\,.
\end{equation}
This shows that the sign of the drift depends on sgn$(\sigma\chi\Lambda)$. 
For one of the steady states the drift is inwards, for the other one it is outwards. 
Eq.~(\ref{eq:drift}) implies that stability flips under $\chi\to -\chi$,
as expected, and that stability
also flips under $\Lambda\to -\Lambda$ (prolate to oblate and {\em vice versa}).

\section{Linear stability analysis for the chiral dipole}
\label{app:staban1}
Now consider the full chiral dipole, Eqs.~(\ref{eq:angular_motion_shear_flow}) and (\ref{eq:cdL}), in the coordinates defined above. For the full system, log rolling is 
a periodic orbit, not a fixed point, because $\dot \psi$ does not vanish in general at $(\varphi_\sigma^\ast,\vartheta_\sigma^\ast$). 
To check the linear stability of log rolling, we linearise
around $\varphi_\sigma^\ast$ and $\vartheta^\ast_\sigma$
using the definitions in Eq.~(\ref{eq:defcd}).
\begin{align*}
\Lambda = L_{231}-L_{132}\,,\quad 
\delta\Lambda = \tfrac12(L_{231}+L_{132})\,,\quad
\chi = L_{232}+L_{131}\,,\quad 
\delta\chi = \tfrac12 (L_{232}-L_{131})\,.
\end{align*}
This results in 
\begin{subequations}
\label{eq:deltaAcd}
\begin{align}
\tfrac{\rm d}{{\rm d}t} \ve X &= \ma [\ma A_\sigma+\delta\ma A_\sigma(\psi) ]\ve X\,, \quad
\quad
\delta\ma A_\sigma = \begin{bmatrix}
\sigma a(\psi) & b(\psi)\\
b(\psi) & \sigma a(\psi)
\end{bmatrix},
\quad
\begin{bmatrix} a \\b \end{bmatrix} =
\begin{bmatrix}
\phantom-
\delta\Lambda \sin2\psi +\delta\chi \cos2\psi\\
-\delta \Lambda \cos2\psi+\delta \chi\sin2\psi\end{bmatrix}\,,\\
\tfrac{\rm d}{{\rm d}t} \psi &=-\tfrac{1}{2}+ L_{321}\cos2\psi + \tfrac{1}{2}(L_{311}-L_{322})\sin 2\psi\,.
\end{align}
\end{subequations}
The dynamics of $\psi$ is autonomous to this lowest order, determined by the coefficients in the upper left
$2\times 2$ block of $\ma L_3$. Since
these elements are small, the dynamics is very close to $\dot \psi = -\tfrac{1}{2}$.
Averaging $\delta \ma A_\sigma(\psi)$ over this dynamics yields $\overline{{\rm tr}\delta \ma A_\sigma}=0$, so still marginal stability. But accounting for small deviations from uniform dynamics of $\psi$ yields
\begin{equation}
\label{eq:trav}
    \overline{{\rm tr}\delta \ma A_\sigma}
    = \sigma \delta\Lambda(L_{311}-L_{322}) +2\sigma \delta\chi L_{321}\,.
\end{equation}
For $\delta\Lambda=\delta\chi = 0$, this is consistent with the simplified model:  vanishing trace implies 
marginal linear stability, as found above. For the coefficients of
the full model, Eq.~(\ref{eq:cdL}), one finds
\begin{equation}
\Lambda =0.9809\,,\quad  \delta\Lambda=-0.0018\,,\quad \chi = 0.0056\,,\quad\delta \chi=-0.0006\,,
\end{equation} 
So Eq.~(\ref{eq:trav})
gives 
$\overline{{\rm tr}\delta \ma A_\sigma}\approx -8.77 \times 10^{-6}\sigma$, and the Lyapunov exponent is half of this value.
So now $(0,\pi)$ is linearly unstable,  but Eq.~(\ref{eq:drift}) shows that $(0,\pi)$ is non-linearly stable  in the simplified model since $\Lambda>0$ and $\chi<0$. Numerical \texttt{SHAPES} simulations show
that the non-linear stability of the full chiral-dipole model is qualitatively the same as for the simplified model: the non-linear terms give rise to a drift towards $(0,\pi)$, and away from $(0,0)$. 
So the linear and non-linear stability predictions do not agree. 
These observations are explained by the fact that 
linear and non-linear stability are
determined by different elements of $\ma L$, for example $\delta\chi$ instead of $\chi$. The consequences of this mismatch for the angular dynamics of the chiral dipole (asymmetric bistability) are discussed in the main text.

\section{Video recordings of model results for a motile microorganism}
\label{app:video}

A video recording at \url{https://youtube.com/shorts/g-Uqdcq_SUA} shows \texttt{SHAPES} simulation-results for the motile microswimmer described in the main text, Figure~\ref{fig:particle_summary}({\bf f}), in an unbounded fluid.   The video highlights the reciprocal back-and-forth motion characteristic of locomotion in the Stokes regime. The motion of the left flagellum is the mirror image of that of the right flagellum, resulting in a straight swimming trajectory.

A second video recording at \url{https://youtube.com/shorts/gnsCZO06jq4} shows how the  motile microorganism 
interacts with a wall. Hydrodynamic interactions cause the swimmer to turn
away from the wall (initial distance $20\,\mu\mathrm{m}$, initial orientation: $\ve n_2$ at an angle of $\pi/18$ with the wall). The vector $\ve n_2$ is defined in Figure~\ref{fig:particle_summary}({\bf f}) in the main text.

\end{document}